\documentclass[a4paper,12pt]{article}

\usepackage{latexsym}
\usepackage{amssymb,amsmath,amsthm}
\usepackage[pdftex]{graphicx,color}
\usepackage{bm}          
\usepackage[
	colorlinks=true,
	urlcolor=black,
	linkcolor=black,
  hypertexnames=false
]{hyperref}

\newcommand{\beq}{\begin{eqnarray*}}
\newcommand{\beqn}{\begin{eqnarray}}
\newcommand{\eeq}{\end{eqnarray*}}
\newcommand{\eeqn}{\end{eqnarray}}

\newcommand{\bmat}{\begin{pmatrix}}
\newcommand{\emat}{\end{pmatrix}}

\newcommand{\lbeq}[1]{\label{eq:#1}}
\newcommand{\req}[1]{\ref{eq:#1}}

\newcommand{\nn}{\nonumber}

\newcommand{\ds}{\displaystyle}
\newcommand{\ts}{\textstyle}
\newcommand{\la}{\langle}
\newcommand{\ra}{\rangle}
\newcommand{\pt}{\partial}
\newcommand{\pro}{\mathop\Pi}
\newcommand{\summ}{\mathop{\ts\sum}}

\newcommand{\prodd}{\mathop{\ts\prod}}
\newcommand{\up}{\uparrow}
\newcommand{\down}{\downarrow}

\newcommand{\su}{{\hbox{\tiny$\uparrow$}}}
\newcommand{\sd}{{\hbox{\tiny$\downarrow$}}}

\newcommand{\sabs}{{\hbox{\scriptsize$|$}}}

\newcommand{\vep}{\varepsilon}
\newcommand{\eps}{\epsilon}

\newcommand{\R}{{\mathbb R}}

\newcommand{\cF}{{\cal F}}

\newcommand{\sign}{{\rm sign}}

\newcommand{\pI}{{\phantom{\ds\int_I^i}}}
\newcommand{\pS}{{\phantom{\ds\sum}}}
\newcommand{\pM}{{\phantom{\ds MM}}}

\newcommand{\ti}{\tilde }
\newcommand{\G}{{|\Gamma|} }
\newcommand{\GA}{{\Gamma_{\!A}} }
\newcommand{\GB}{{\Gamma_{\!B}} }

\newcommand{\rint}{{\rm int}}

\newcommand{\lam}{{\lambda}}

\newcommand{\Idz}{ {Id_{\!\phantom{.}_{2\G}} }}

\newcommand{\Idv}{ {Id_{\!\!\phantom{.}_{4\G}} }}

\begin{document}

\pagenumbering{gobble}

$$\phantom{mm}$$
\vskip 3.5cm

\centerline{\LARGE Thermodynamics of the Fermi-Hubbard Model}  

\medskip
\bigskip
\centerline{\large through}

\smallskip
\bigskip
\centerline{\LARGE Stochastic Calculus} 

\bigskip
\centerline{\large and}

\smallskip
\bigskip
\centerline{\LARGE  Girsanov Transformation}

\bigskip
\bigskip
\bigskip
\bigskip
\bigskip
\centerline{Detlef Lehmann, Hochschule RheinMain, Faculty of Engineering } 
\centerline{Postfach 3251, 65022 Wiesbaden, Germany}

\medskip

\centerline{\footnotesize detlef.lehmann@hs-rm.de}
\centerline{\footnotesize \url{https://www.hs-rm.de/de/hochschule/personen/lehmann-detlef/}}

\vskip 2.9cm
\noindent{\bf Abstract:} We apply the methodology of our recent paper `The Dynamics of the Hubbard Model 
through Stochastic Calculus and Girsanov Transformation' [1] to thermodynamic correlation 
functions in the Fermi-Hubbard model. They can be obtained from a stochastic differential 
equation (SDE) system. To this SDE system, a Girsanov transformation can be applied. This 
has the effect that the usual determinant or pfaffian which shows up in a pfaffian quantum 
Monte Carlo (PfQMC) representation [2] basically gets absorbed into the new integration 
variables and information from that pfaffian moves into the drift part of the transformed 
SDE system.  While the PfQMC representation depends heavily on the choice of how the quartic 
interaction has been factorized into quadratic quantities in the beginning, the Girsanov transformed formula 
has the very remarkable property that it is nearly independent of that choice, the drift part of the transformed 
SDE system as well as a remaining exponential which has the obvious meaning of energy are always the same and do 
not depend on the Hubbard-Stratonovich details. The resulting formula may serve as a starting point for further 
theoretical or numerical investigations. Here we consider the spin-spin correlation at half-filling 
on a bipartite lattice and obtain an analytical proof that the signs of these correlations 
have to be of antiferromagnetic type, at arbitrary temperatures. 
Also, by checking against available benchmark data [3], we find that approximate ground state energies 
can be obtained from an ODE system. This may even hold for exact ground state energies, 
but future work would be required to prove or disprove this. As in [1], the methodology is generic and 
can be applied to arbitrary quantum many body or quantum field theoretical models.

\newpage

$$\phantom{.}$$
\noindent{\bf Content and Outline: } 

{\footnotesize

\medskip
\bigskip
\begin{itemize}
\item[{ 1.}] Why Girsanov Transformation and Stochastic Calculus? A Motivating Example  \hfill 1

\medskip
\item[{ 2.}] Setup and Main Result  \hfill 6
\begin{itemize}
\item[{ 2.1}] The Hubbard Hamiltonian in terms of Majorana Fermion Operators \hfill 6 
\item[{ 2.2}] Hubbard-Stratonovich Transformation  \hfill 8 
\item[{ 2.3}] Main Theorem \hfill 10 
\end{itemize}

\smallskip
\item[{ 3.}] Symmetry Relations  \hfill 12  \\
$\phantom{w}${\tiny Theorem 3.1 \;(writes the complex $4\G\times 4\G$ SDE system of the Main Theorem in terms of real $\,2\G\times 2\G$ matrices) }\hfill {\tiny 13}   \\ 
$\phantom{w}${\tiny Theorem 3.2 \;(symmetry relations for arbitrary chemical potential $\mu$, further reduction to $\G\times \G$ system) } \hfill {\tiny 14} 

\medskip
\item[{ 4.}] The Model at Half Filling \hfill 17 
\begin{itemize}
\item[{ 4.1}] Symmetry Relations at Half Filling and Constant Density Theorem  \hfill 17 \\
  {\tiny Theorem 4.1 \;(symmetry relations at half filling $\,\mu=0\,$ on a bipartite lattice)} \hfill {\tiny 18\,}\\
  {\tiny Theorem 4.2 \;(SDEs at half filling for zero external pairing and exchange terms $r=s=0\,$) } \hfill {\tiny 19\!\!}
\smallskip
\item[{ 4.2}] Numerical Test: Main Theorem vs.~Exact Diagonalization  \hfill 20\!\! 
\smallskip
\item[{ 4.3}] Spin-Spin and Pair-Pair Correlation  \hfill 23 \\
  {\tiny Theorem 4.3 \;(spin-spin correlations have to have antiferromagnetic signs at arbitrary temperatures)}  \hfill {\tiny 23\!\! } \\
  {\tiny Theorem 4.4 \;(pair-pair correlations are nonnegative for attractive couplings)} \hfill {\tiny 24} \\ 
  {\tiny Numerical Check: $w_1=1$ representation vs.~$w_2=1$ representation} \hfill {\tiny 25} 
\item[{ 4.4}] Temperature Zero Limit  \hfill 27 \\
  {\tiny Numerical Test: Energies vs.~Benchmark Data}  \hfill {\tiny 28} 
\end{itemize}

\smallskip
\item[{ 5.}] Proof of Main Theorem \hfill 32 
\begin{itemize}
\item[{ 5.1}] Pfaffian Quantum Monte Carlo Representation \hfill 32 \\ 
  {\tiny Theorem 5.1 \;(traces of products of exponentials of arbitrary quadratic operators as pfaffians)} \hfill {\tiny 32}\\ 
  {\tiny Theorem 5.2 \;(pfaffian quantum Monte Carlo representation for the Hubbard model) \hfill 33}
\item[{ 5.2}] Stochastic Differential Equation Representation \hfill 35 \\ 
  {\tiny Theorem 5.3 \;(SDEs for thermodynamic evolution matrix $U$, density matrix $G$ and pfaffian $Z$) \hfill 36}
\item[{ 5.3}] Girsanov Transformation \hfill 43 \\ 
  {\tiny Proposition 5.1 (introduces Girsanov transformation)  \hfill  45  \\ 
   Theorem 5.4 \;(calculates the Girsanov transformed SDE for the density matrix $G$) \hfill 47  \\ 
   Theorem 5.5 \;(identifies the remaining exponential as the energy) \hfill 51 }
\item[{ 5.4}] Correlation Functions \hfill 52 \\ 
  {\tiny finalizes proof of Main Theorem  \hfill  }
\end{itemize}

\bigskip

\item[{ A.}] Appendix: Proofs of Theorems of Section 4 \hfill 55 
{\tiny 
\begin{itemize}
\item[A.1] Proof of Theorem 4.1  \hfill 55  
\item[A.2] Proof of Theorem 4.2  \hfill 58  
\item[A.3] Proof of Theorem 4.3  \hfill 63  
\item[A.4] Proof of Theorem 4.4  \hfill 66 
\end{itemize}
}

\end{itemize}

\bigskip
\vfill
\noindent{References}  \hfill 71

}

\newpage

\pagenumbering{arabic}

\bigskip
\bigskip
\bigskip
{\large\bf 1. Why Girsanov Transformation and Stochastic Calculus? } 

\medskip
{\large\bf \phantom{1.} A Motivating Example}

\numberwithin{equation}{section}
\renewcommand\thesection{1}
\setcounter{equation}{0}

\medskip
\bigskip
\bigskip
Before we start with the actual topic of this paper, correlation functions in the Fermi-Hubbard model, let us first consider a very simple example 
which allows the illustration of Girsanov transformation and stochastic calculus in this context in a very transparent way. Let $G=G(x)$ be 
some function of one variable. Suppose we want to calculate the quantity 
\beqn
\la G\ra &:=& {\phantom{\Bigl|} \int_\R \, G(x)\;Z_\beta(x)\; e^{-{x^2\over 2}}\; {dx\over \sqrt{2\pi}} \phantom{\Bigl|}\over  \phantom{\Bigl|}
	\int_\R \,Z_\beta(x)\; e^{-{x^2\over 2}}\; {dx\over \sqrt{2\pi}} \phantom{\Bigl|} }  \pI \lbeq{1.1}
\eeqn
with
\beqn
Z_\beta(x)&:=&\cosh(\sqrt{\beta}\,x)\;\;=\;\;\ts {1\over 2}\,(\, e^{\,+\,\sqrt{\beta}\,x} \,+\, 
  e^{\,-\,\sqrt{\beta}\,x} \,)  \pI \lbeq{1.2}
\eeqn
with a Monte Carlo simulation. It is obvious that for larger values of $\beta$ a simulation with standard normal random numbers for $x$ has to 
fail since the main contributions come from $\,x\,\approx\, \pm\sqrt{\beta}\,$. The problem can be fixed by moving 
the $Z_\beta$ into a new integration measure as follows: We write 
\beq
\beta&=& k\,dt \;\;=:\;\;t_k  \pI
\eeq
and substitute 
\beqn
\sqrt{\beta}\,x&=&\sqrt{dt}\,\summ_{\ell=1}^k \phi_\ell \;\;=:\;\;x_{t_k} \pI \lbeq{1.3}
\eeqn
to obtain the exact identity
\beqn
\la G\ra &=& {\phantom{\Bigl|} \int_{\R^k} \, G(x_{t_k}/\sqrt{\beta})\;\cosh(x_{t_k})\; dP_k(\phi) \phantom{\Bigl|}\over  \phantom{\Bigl|}
	\int_{\R^k} \,\cosh(x_{t_k})\; dP_k(\phi) \phantom{\Bigl|} }  \pI  \lbeq{1.4}
\eeqn
with
\beqn
dP_k&:=&\pro_{\ell=1}^k \, e^{\,-\,{\phi_\ell^2\over 2}}\;{\ts {d\phi_\ell\over \sqrt{2\pi}} } \pI 
\eeqn
being a discretized Wiener measure and $x_{t_k}$ on the right hand side of (\req{1.3}) being a discretized Brownian motion. 
Now we write (with $x_{t_0}=x_0:=0$\,)
\beqn
\cosh(x_{t_k})&=&\exp\bigl\{\; \log\cosh(x_{t_k})\;\bigr\} \;\;=\;\;\exp\bigl\{\; \log\cosh(x_{t_k})\,-\,\log\cosh(0)\;\bigr\}  \pI \nn  \\ 
&=&\exp\Bigl\{\; \summ_{\ell=1}^k\,\bigl[\,\log\cosh(x_{t_\ell})\,-\,\log\cosh(x_{t_{\ell-1}})\,\bigr]\;\Bigr\}  \pI \lbeq{1.6}
\eeqn
and Taylor expand, with $\,dx_{t_\ell}\;:=\;x_{t_\ell}-x_{t_{\ell-1}}\;=\;\sqrt{dt}\,\phi_\ell\,$, 
\beqn
\log\cosh(x_{t_\ell})&=&\log\cosh(x_{t_{\ell-1}}+dx_{t_\ell})  \pI  \lbeq{1.7} \\ 
&=&\log\cosh(x_{t_{\ell-1}})\;+\;\tanh(x_{t_{\ell-1}})\, dx_{t_\ell} \;+\;\ts {1\over 2}\,{1\over \cosh^2(x_{t_{\ell-1}})}\; (dx_{t_\ell})^2 \;+\; O(\,dt^{3/2}\,) \pI \nn
\eeqn
We use the Brownian motion calculation rule (see the appendix of [1] for more background)
\beqn
(dx_t)^2&=&dt \pI \lbeq{1.8}
\eeqn
and ignore the higher order terms to obtain the approximate identity
\beqn
\cosh(x_{t_k})&\approx&\exp\Bigl\{\; \sqrt{dt}\,\summ_{\ell=1}^k\,\tanh(x_{t_{\ell-1}})\,\phi_\ell 
  \;+\;{\ts {1\over 2}}\, \summ_{\ell=1}^k\, \ts {dt \over \cosh^2(x_{t_{\ell-1}})}  \;\Bigr\}  \pI \lbeq{1.9}
\eeqn
which becomes exact in the limit $dt\to 0$ which is implicitely understood in all what follows, so we proceed with an exact equal sign. 
Then we make the substitution of variables or Girsanov transformation 
\beqn
\ti\phi_\ell&:=&\phi_\ell\,-\sqrt{dt}\,\tanh(x_{t_{\ell-1}})   \pS \nn\\ 
\Leftrightarrow\;\;\;\;\;\;d\ti x_{t_\ell}&=&dx_{t_\ell}\,-\,dt\,\tanh(x_{t_{\ell-1}}) \phantom{\Leftrightarrow\;\;\;\;\;\;}  \pS
\eeqn
to obtain (observe that $\,\det[\,\pt\ti\phi/ \pt\phi \,]\,=\,1\,$)
\beqn
\lefteqn{
\cosh(x_{t_k})\; dP_k(\phi)  } \pS  \nn\\ 
&=&\exp\Bigl\{\; -\,{\ts {1\over 2}}\,\summ_{\ell=1}^k\,\Bigl[\,\phi_\ell^2\,-\,2\sqrt{dt}\,\tanh(x_{t_{\ell-1}})\,\phi_\ell \,\Bigr] 
  \;+\;{\ts {1\over 2}}\, \summ_{\ell=1}^k\, {\ts {dt \over \cosh^2(x_{t_{\ell-1}})} } \;\Bigr\} \;\pro_{\ell=1}^k \,\ts {d\phi_\ell \over \sqrt{2\pi}}  \pI \nn \\ 
&& \nn\\
&=& \exp\Bigl\{\; -\,{\ts {1\over 2}}\,\summ_{\ell=1}^k\,\ti\phi_\ell^2\,+\,{\ts{dt\over 2}}\,\summ_{\ell=1}^k\,\tanh^2(x_{t_{\ell-1}}) 
  \;+\;{\ts {1\over 2}}\, \summ_{\ell=1}^k\, {\ts {dt \over \cosh^2(x_{t_{\ell-1}})} } \;\Bigr\} \;\pro_{\ell=1}^k \,\ts {d\ti\phi_\ell \over \sqrt{2\pi}}  \pI  \nn \\ 
&& \nn\\
&=& \exp\Bigl\{\;{\ts {dt\over 2}}\, \summ_{\ell=1}^k\, {\ts {\sinh^2(x_{t_{\ell-1}})\,+\,1 \over \cosh^2(x_{t_{\ell-1}})} } \;\Bigr\} \;dP_k(\ti\phi)
  \;\;=\;\; e^{\,+\,{t_k\over 2}} \; dP_k(\ti\phi)  \pI  
\eeqn
Thus we arrive at 
\beqn
\la G\ra &=& {\phantom{\Bigl|} \int_{\R^k} \, G(x_{t_k}/\sqrt{\beta})\;\cosh(x_{t_k})\; dP_k(\phi) \phantom{\Bigl|}\over  \phantom{\Bigl|}
	\int_{\R^k} \,\cosh(x_{t_k})\; dP_k(\phi) \phantom{\Bigl|} } 
\;\;=\;\; {\phantom{\Bigl|} \int_{\R^k} \, G(x_{t_k}/\sqrt{\beta})\; dP_k(\ti\phi) \phantom{\Bigl|}\over  \phantom{\Bigl|}
	\int_{\R^k} \, dP_k(\ti\phi) \phantom{\Bigl|} }  \pI \nn\\ 
&=& \ts\int_{\R^k} \, G(x_{t_k}/\sqrt{\beta})\; dP_k(\ti\phi)   \pI \lbeq{1.12}
\eeqn
with $\,x_{t_k}=x_{t_k}\bigl(\,\{\ti\phi_\ell\}_{\ell=1}^k\,\bigr)\,$ given by the recursion or stochastic differential equation 
\beq
dx_{t_\ell}&=&d\ti x_{t_\ell}\,+\,dt\,\tanh(x_{t_{\ell-1}})   \pS
\eeq
or 
\beqn
x_{t_\ell}&=&x_{t_{\ell-1}} \;+\; \tanh(x_{t_{\ell-1}}) \, dt \;+\; \sqrt{dt}\;\ti\phi_\ell \pS \lbeq{1.13}
\eeqn
with $x_{t_0}=0\,$. Let's choose some concrete $G$, say, 
\beqn
G(x)&=&\cos(\mu x) \pS 
\eeqn
Then the exact result is found to be 
\beqn
\la G\ra &=& e^{\,-\,{\mu^2\over 2}}\,\cos\bigl[\,\mu\sqrt{\beta}\,\bigr] \pI
\eeqn
Plain Monte Carlo produces the following typical picture, with $N=$ 100'000 random numbers for the untransformed representation given by (\req{1.1}) (using the 
same random numbers for all $\beta\in[0,200]\,$) and also $N=$ 100'000 MC paths (but more random numbers of course, 100K per time step) for the Girsanov transformed 
representation given by (\req{1.12},\req{1.13}). Shown is the quantity $\,e^{+\mu^2/2}\,\la G\ra\,$ for $\mu=2\,$, as a function of $\beta\,$: 

\bigskip
\centerline{\includegraphics[width=13cm]{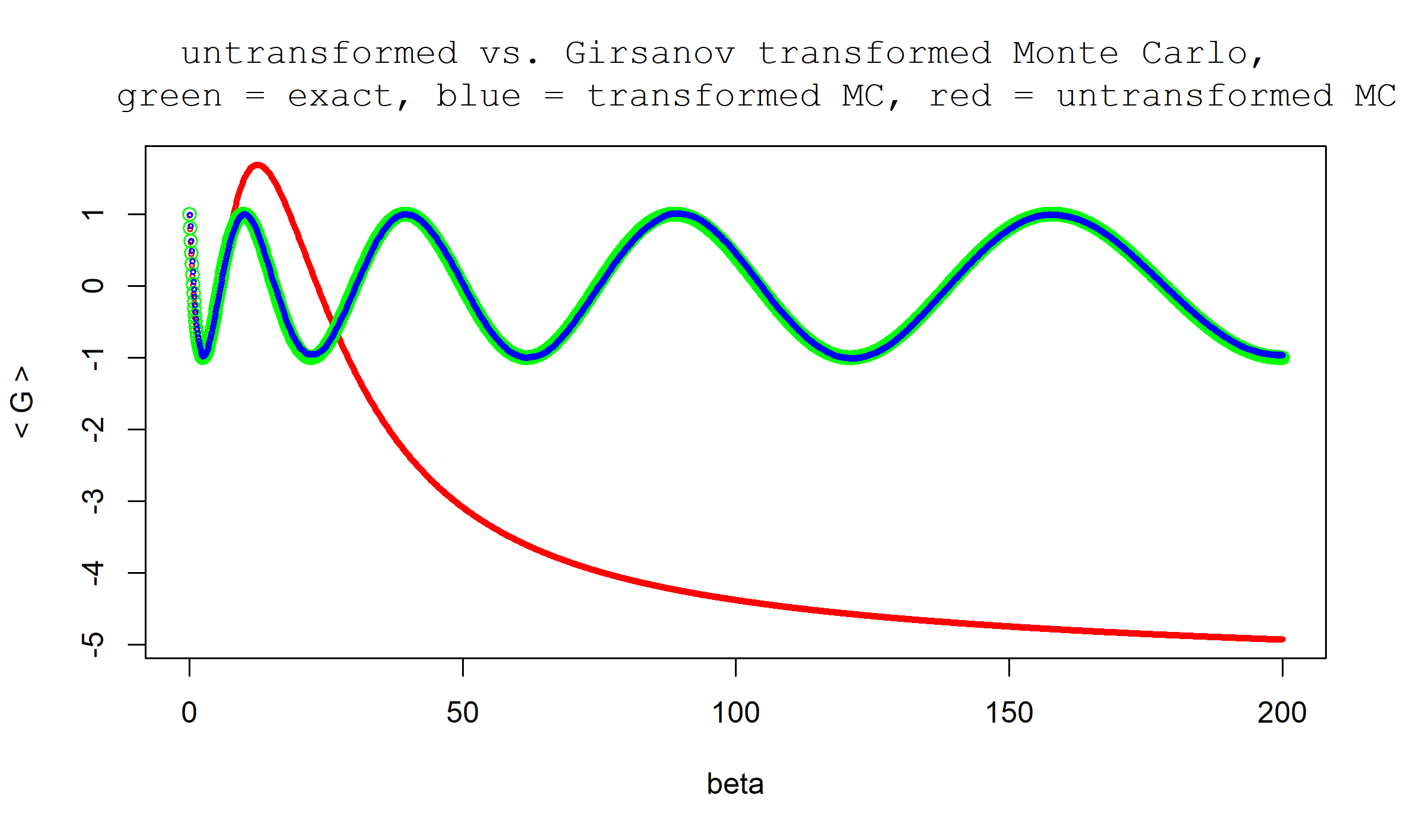}}

\bigskip
An obvious generalization of (\req{1.1},\req{1.2}) is 
\beqn
\la G\ra &:=& {\phantom{\Bigl|} \int_\R \, G(x)\;{\ds  Z(\sqrt{\beta}\,x)}\; e^{-{x^2\over 2}}\; {dx\over \sqrt{2\pi}} \phantom{\Bigl|}\over  \phantom{\Bigl|}
	\int_\R \,{\ds Z(\sqrt{\beta}\,x)}\; e^{-{x^2\over 2}}\; {dx\over \sqrt{2\pi}} \phantom{\Bigl|} }  \pI 
\eeqn
with, say,
\beqn
Z(x)&:=&\summ_{i=1}^n\, e^{\,\lam_i\, x}  \pI \lbeq{1.17}
\eeqn
As in (\req{1.4}), we have the exact identity
\beqn
\la G\ra &=& {\phantom{\Bigl|} \int_{\R^k} \, G(x_{t_k}/\sqrt{\beta})\;Z(x_{t_k})\; dP_k(\phi) \phantom{\Bigl|}\over  \phantom{\Bigl|}
	\int_{\R^k} \,Z(x_{t_k})\; dP_k(\phi) \phantom{\Bigl|} }  \pI  \lbeq{1.18}
\eeqn
with $x_{t_k}$ being a Brownian motion. We write $\,Z\,=\,\exp\{\,\log Z\,\}\,$ (which we only can do if there is no fermionic sign problem) and calculate $d\log Z\,$. 
Using continuous time notation now, the analog of (\req{1.7}) is
\beqn
d\log Z&=&\ts {dZ \over Z} \;-\; {1\over 2}\,\bigl( {dZ\over Z}\bigr)^2 \pS \lbeq{1.19}
\eeqn
with
\beqn
dZ &=&\ts Z'\,dx \;+\;{1\over 2}\,Z''\,(dx)^2 \;\;\buildrel (\req{1.8})\over =\;\;Z'\,dx \;+\;{1\over 2}\,Z''\,dt \pS 
\eeqn
Using $\,dx\cdot dt\,=\,0\,$, we have 
\beqn
\ts \bigl( {dZ\over Z}\bigr)^2&=&\ts \bigl( {Z'\over Z}\bigr)^2\,dt \pS 
\eeqn
such that
\beqn
d\log Z&=&\ts {Z' \over Z}\,dx \;+\;{1\over 2}\,\bigl[\,{Z''\over Z}- \bigl( {Z'\over Z}\bigr)^2\,\bigr]\,dt   \pI 
\eeqn
If we continue to use continuous time notation, the analog of (\req{1.9}) would read, with $\beta=t_k$ and $Z_\beta=Z(x_{t_k})\,$,
\beqn
Z_\beta&=&\exp\Bigl\{\;\ts \int_0^\beta \,d\log Z \;\Bigr\}\; \cdot\; Z_0  \pI \nn\\ 
&=&\exp\Bigl\{\;\ts \int_0^\beta \,{Z' \over Z}\,dx_t \,+\,{1\over 2}\int_0^\beta \,\bigl[\,{Z''\over Z}- \bigl( {Z'\over Z}\bigr)^2\,\bigr]\,dt \;\Bigr\}\; \cdot\; Z_0
  \lbeq{1.23}  \pI
\eeqn
Here it is crucial to keep in mind that the correct discretized version for the $dx_{t\phantom{.}}$-integral, for the stochastic integral, is 
\beqn
\ts \int_0^\beta \,{Z' \over Z}\,dx_t &:=&\summ_{\ell=1}^k \,{\ts {Z'\over Z}}(x_{t_{\ell-1}})\,dx_{t_\ell} 
  \;\;=\;\;\summ_{\ell=1}^k \,\ts   {Z'\over Z}(x_{t_{\ell-1}})\,\sqrt{dt}\;\phi_\ell   \pI \lbeq{1.24}
\eeqn
but not $\,\summ_{\ell=1}^k {Z'\over Z}(x_{t_{\ell}})\,dx_{t_\ell}\,$. Exactly this is meant by saying that the stochastic integral is an integral in the Ito-sense, 
and this holds for all stochastic integrals showing up in this paper. Then the Girsanov transformation in this case is obtained through 
\beqn
\lefteqn{
\ts \exp\Bigl\{\,  {Z'\over Z}(x_{t_{\ell-1}})\,\sqrt{dt}\,\phi_\ell\,\Bigr\}\; e^{\,-\,{\phi_\ell^2\over 2}} \;\;=\;\;  } \pI \nn\\ 
&&\ts\exp\Bigl\{\;-\,{1\over 2}\,\Bigl[\,\phi_\ell\,-\, \sqrt{dt}\, {Z'\over Z}(x_{t_{\ell-1}})\,\Bigr]^2\;\Bigr\} \;\cdot\; 
\exp\Bigl\{\;+\,{dt\over 2}\,\bigl( {Z'\over Z}\bigr)^2(x_{t_{\ell-1}})\;\Bigr\}  \pI\pI \lbeq{1.25}
\eeqn
That is, 
\beqn
\ti\phi_\ell &:=& \ts \phi_\ell\;-\; \sqrt{dt}\; {Z'\over Z}(x_{t_{\ell-1}})  \pS \nn \\ 
\Leftrightarrow\;\;\;\;\;d\ti x_{t_\ell}&=&\ts dx_{t_\ell}\;-\;dt\,{Z'\over Z}(x_{t_{\ell-1}})  \pI
\eeqn
Since the $(Z'/Z)^2$ terms in (\req{1.23}) and (\req{1.25}) cancel out, we obtain 
\medskip
\beqn
Z(x_{t_k})\,dP_k(\phi)&=&\exp\Bigl\{\;{\ts {1\over 2}}\,\summ_{\ell=1}^k\,\ts {Z''\over Z}(x_{t_{\ell-1}})\, dt \;\Bigr\}\;dP_k(\ti\phi) \; \cdot\;  Z(0)  \pI 
\eeqn

\medskip
and arrive at (for $dt$-integrals actually it does not matter whether we have an $x_{t_{\ell-1}}$ or an $x_{t_\ell}\,$)
\beqn
\la G\ra &=& {\phantom{\Bigl|} \int_{\R^k} \, G(x_{t_k}/\sqrt{\beta})\;\exp\Bigl\{\;{\ts {1\over 2}}\,\ds\summ_{\ell=1}^k\,\ts {Z''\over Z}(x_{t_{\ell-1}})\, dt \;\Bigr\}\;
   dP_k(\ti\phi) \phantom{\Bigl|}\over  \phantom{\Bigl|}
	\int_{\R^k} \,\exp\Bigl\{\;{\ts {1\over 2}}\,\ds\summ_{\ell=1}^k\,\ts {Z''\over Z}(x_{t_{\ell-1}})\, dt \;\Bigr\}\;dP_k(\ti\phi) \phantom{\Bigl|} }  \pI  \lbeq{1.28}
\eeqn
with $\,x_{t_k}=x_{t_k}\bigl(\,\{\ti\phi_\ell\}_{\ell=1}^k\,\bigr)\,$ given by the recursion or stochastic differential equation 
\beqn
x_{t_\ell}&=&\ts x_{t_{\ell-1}} \;+\;  {Z'\over Z}(x_{t_{\ell-1}}) \; dt \;+\; \sqrt{dt}\;\ti\phi_\ell \pI \lbeq{1.29} 
\eeqn
Let's check again with a concrete simulation. For $\,G(x)=\cos(\mu x)\,$ and $Z(x)$ given by (\req{1.17}) above, the exact result is found to be 
\beqn
\la G\ra&=&e^{\,-\,{\mu^2\over 2}}\; { \phantom{\Bigl|} \summ_{i=1}^n\, e^{\,+\,\beta\,{\lam_i^2\over 2}} \;\ds\cos(\,\mu\lam_i\sqrt{\beta}\,) \phantom{\Bigl|} \over 
  \phantom{\Bigl|} \summ_{i=1}^n\, e^{\,+\,\beta\,{\lam_i^2\over 2}}\phantom{\Bigl|} }  \pI 
\eeqn
We choose $\mu=2$, $n=3$ and
\beq
\lam_1\;=\;-1\,,\;\;\;\;\lam_2\;=\;+{1\over 2}\,,\;\;\;\;\lam_3\;=\;+{3\over 2}  \pS
\eeq
We plot the quantity $\,e^{\,+\,{\mu^2\over 2}}\,\la G\ra\,$ as a function of $\beta\in[0,200]\,$, again with 100K Monte Carlo simulations:

\bigskip
\centerline{\includegraphics[width=13cm]{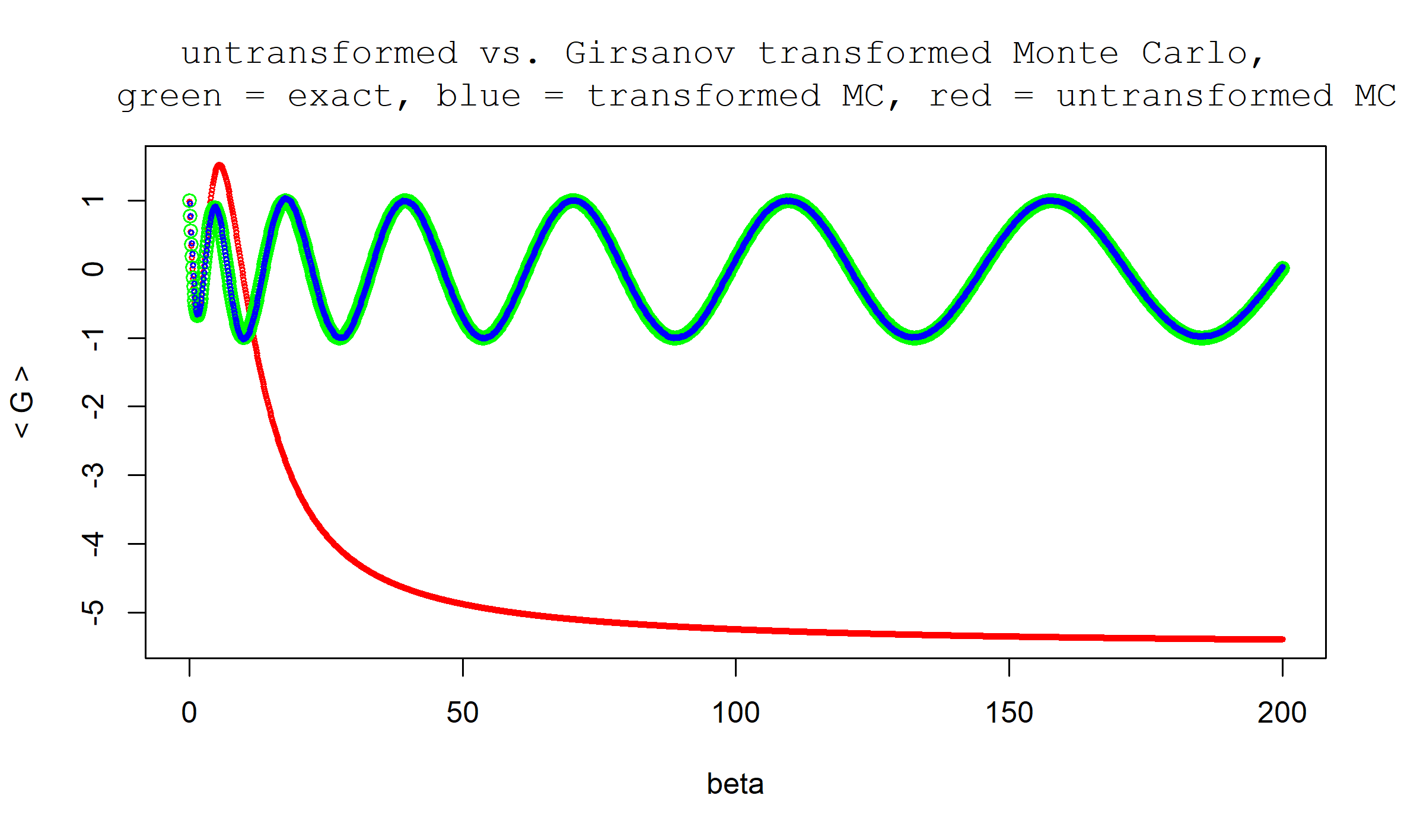}}

\bigskip

Roughly speaking, one may say that the rest of the paper writes down the passage from the untransformed representation (\req{1.18}) to the Girsanov transformed 
representation (\req{1.28}) with its SDE (\req{1.29}) for thermodynamic correlation functions in the Fermi-Hubbard model. Thereby the untransformed representation is 
given by a determinant or pfaffian quantum Monte Carlo representation which would be very standard if there are no symmetry breaking pairing or exchange terms in the 
Hamiltonian. However, if these terms are present, the PfQMC representation is less obvious and we were in the very lucky position to be able to use a very 
recent result of Han, Wan and Yao [2] which is summarized in Theorems 5.1 and 5.2\,. The main result of this paper then are the analogs 
of (\req{1.28}) and (\req{1.29}) for the Fermi-Hubbard model, they are given by (\req{2.26}) and (\req{2.29}) of the Main Theorem in section 2. Several analytical and numerical 
consistency checks have been performed (atomistic limit, exact dia\-gonalization vs.~Monte Carlo for small lattice size, constant density theorem at half filling, 
antiferromagnetic spin-spin correlations for repulsive couplings, nonnegative pair-pair correlations for attractive couplings, energies from SDE/ODE system 
vs.~2D benchmark data). The numerical computations have been performed on a standard stand-alone desktop pc with the exception for the calculations in section 4.3 
for the spin-spin and pair-pair correlations where an AWS EC2 instance with 192 cores has been used to speed up the calculations. All code, which is R code, is available at 
\,{\small \url{https://www.hsrm-mathematik.de/lehmann/code-fermi-hubbard}}\,.

\newpage
{\large\bf 2. Setup and Main Result}
\setcounter{equation}{0}
\numberwithin{equation}{section}
\renewcommand\thesection{2}

\medskip
\bigskip
\bigskip
{\bf 2.1 \;The Hubbard Hamiltonian in terms of Majorana Fermion Operators}

\bigskip
\bigskip
We consider the $d$-dimensional Fermi-Hubbard model with grand canonical Hamiltonian (we define a final hamiltonian without subscript in (\req{Hmain}) below)
\beqn
H_{\rm tot}&=&H_{\rm kin}\;+\;H_\rint \pI  \lbeq{2.1}
\eeqn
with quadratic part given by 
\beqn
H_{\rm kin}&=&\summ_{i,j}\;(\vep_{ij}-\mu\,\delta_{ij})\,(\,c_{i\su}^+\,c_{j\su} \,+\, c_{i\sd}^+\,c_{j\sd} \,) \pI \lbeq{2.2} \\
&&\phantom{mm}   \;+\;s\,\summ_j\,(\,c_{j\su}^+\,c_{j\sd} \,+\, c_{j\sd}^+\,c_{j\su}\, ) 
\;+\;r\,\summ_j\,(\,c_{j\su}^+\,c_{j\sd}^+ \,+\, c_{j\sd}\,c_{j\su}\, )   \pS \nn\\ 
&=:&c^+_\su(\vep-\mu)\,c_\su \,+\,c^+_\sd(\vep-\mu)\,c_\sd\,+\,s\,(\,c^+_\su c_\sd\,+\,c^+_\sd c_\su\,)\,+\,r\,(\,c^+_\su c^+_\sd\,+\,c_\sd c_\su\,)  \nn \pI
\eeqn
and a quartic part 
\beqn 
H_\rint&:=&  u\,\summ_j \,(\,c_{j\su}^+\,c_{j\su}-1/2\,)\,(\,c_{j\sd}^+\,c_{j\sd}-1/2\,) \pI  \lbeq{2.3}
\eeqn
We added symmetry breaking terms in the quadratic part, the $r$- and $s$-terms, to probe for potentially nonzero anomalous exchange and pairing 
correlations $\la c_{j\su}^+\,c_{j\sd}\ra,\,\la c_{j\sd}^+\,c_{j\su}\ra$ and $\la c_{j\su}^+\,c_{j\sd}^+\ra,\,\la c_{j\sd}\,c_{j\su}\ra\,$. 
Since we subtracted off the one halfs in the quartic part, half filling is at $\mu=0\,$. To be specific, we consider a cubic lattice 
\beqn
\Gamma&:=&\Bigl\{\;j\;\bigr|\;\;j\;=\;(j_1,\cdots,j_d)\;\;\in\;\;\{1,2,\cdots,L\}^d\;\;\Bigr\}  \pI \lbeq{2.4}
\eeqn
and for sums over lattice sites $i$ and $j$ we abbreviate $\Sigma_j:=\Sigma_{j\in\Gamma}$, $\;\Sigma_{i,j}:=\Sigma_{i,j\in\Gamma}\,$. The hopping matrix 
can be an arbitrary real symmetric $\G\times\G$ matrix, 
\beqn
\vep&=&(\,\vep_{ij}\,)_{i,j\in\Gamma}\;\;=\;\;(\,\vep_{ji}\,)_{i,j\in\Gamma}\;\;\in\;\;\R^{\G\times\G}  \pI 
\eeqn
and the $c_{j\sigma},c_{j\sigma}^+$ are standard fermionic annihilation and creation operators with anticommutation relations
\beqn
\{c_{i\sigma},c_{j\tau}^+\}\;\;:=\;\; c_{i\sigma}\,c_{j\tau}^+\,+\,c_{j\tau}^+\,c_{i\sigma}&=&\delta_{i,j}\,\delta_{\sigma,\tau}  \pS \nn\\
\{c_{i\sigma},c_{j\tau}\}\;\;=\;\; \{c_{i\sigma}^+,c_{j\tau}^+\}&=& 0  \pS
\eeqn

\medskip
As in [2], we find it convenient to work with Majorana fermions operators which are given by  
\beqn
a_{j\sigma}\;\;:=\;\;c_{j\sigma}\,+\,c^+_{j\sigma}  \;,&&\;\;\;\;\;\;  b_{j\sigma}\;\;:=\;\;{\ts {1\over i}}\,(\,c_{j\sigma}\,-\,c^+_{j\sigma}\,)  \pS  \nn \\
c_{j\sigma}\;\;=\;\;{\ts{1\over 2}}\,(\,a_{j\sigma}\,+\,i\,b_{j\sigma}\,)\;,&&\;\;\;\;\;\;
c_{j\sigma}^+\;\;=\;\;{\ts{1\over 2}}\,(\,a_{j\sigma}\,-\,i\,b_{j\sigma}\,)   \pS  \lbeq{n2.7}
\eeqn

\medskip
Both $a$ and $b$ are self adjoint operators, for $(i,\sigma)\,\ne\,(j,\tau)$ all oparators anticommute, 
\beqn
\{\,a_{i\sigma}\,,\,a_{j\tau}\,\}\;\;=\;\;\{\,a_{i\sigma}\,,\,b_{j\tau}\,\}\;\;=\;\;\{\,b_{i\sigma}\,,\,b_{j\tau}\,\}\;\;=\;\;0   \pI  
\eeqn
and for  $\,(i,\sigma)\,=\,(j,\tau)\,$ we have
\beqn
(a_{i\sigma})^2\;\;=\;\;(b_{i\sigma})^2&=&1\,,\;\;\;\;\;\; \{ \,a_{i\sigma}\,,\,b_{i\sigma}\,\}\;\;=\;\; 0 \pI
\eeqn
The density operator becomes 
\beqn
c_{j\sigma}^+c_{j\sigma}&=&\ts {1\over 4}\,(\,a_{j\sigma}\,-\,i\,b_{j\sigma}\,) \,(\,a_{j\sigma}\,+\,i\,b_{j\sigma}\,)  \pI  \nn\\ 
&=&\ts {1\over 4}\,\bigl[\, a_{j\sigma}^2\,+\,b_{j\sigma}^2\,+\,i\,a_{j\sigma}b_{j\sigma}\,-\,i\,b_{j\sigma}a_{j\sigma} \,\bigr] \nn \pS\\ 
&=&\ts {i\over 2}\,a_{j\sigma}b_{j\sigma}\,+\,{1\over 2} \pI
\eeqn
More generally, there are the following identities (observe that for Majorana operators the combination $\,iab\,$ is self adjoint)
\beqn
i\;a_{i\sigma}\,b_{j\tau}&=&\ts c^+_{i\sigma}\,c_{j\tau} \, +\,c^+_{j\tau} \,c_{i\sigma} \;+\; c_{i\sigma}\,c_{j\tau}  \, + \,c^+_{j\tau}\, c^+_{i\sigma}
   \;-\;\delta_{i\sigma,j\tau} \pS \nn \\
a_{i\sigma}\,a_{j\tau}&=&\ts c^+_{i\sigma}\,c_{j\tau}\, - \,c^+_{j\tau}\,c_{i\sigma}\; + \;  c_{i\sigma}\,c_{j\tau} \,- \,c^+_{j\tau} \, c^+_{i\sigma}
  \;+\;\delta_{i\sigma,j\tau}  \pS \nn \\
b_{i\sigma}\,b_{j\tau}&=&\ts  c^+_{i\sigma}\,c_{j\tau}\, - \,c^+_{j\tau}\,c_{i\sigma}\; - \;c_{i\sigma}\,c_{j\tau}\,+\,c^+_{j\tau}\, c^+_{i\sigma}
   \;+\;\delta_{i\sigma,j\tau} \pS  \lbeq{n2.11} 
\eeqn
In particular, 
\beqn
\ts {i\over 2}\;(\,a_{i\sigma}\,b_{j\tau}\,+\,a_{j\tau}\,b_{i\sigma}\,)&=&\ts c^+_{i\sigma}\,c_{j\tau} \, + \,c^+_{j\tau}\,c_{i\sigma} \,-\,\delta_{i\sigma,j\tau} \pS \nn \\ 
\ts {i\over 2}\;(\,a_{i\sigma}\,b_{j\tau}\,-\,a_{j\tau}\,b_{i\sigma}\,)&=&\ts c_{i\sigma}\,c_{j\tau} \, + \, c^+_{j\tau}\,c^+_{i\sigma}    \pS \lbeq{n2.12}
\eeqn

\medskip
The Hamiltonian becomes 
\beqn
H_{\rm kin}&=& \ts {i\over 2}\, \Bigl\{\,a_\su(\vep-\mu)\,b_\su\,+\,a_\sd(\vep-\mu)\,b_\sd \,+\,(s-r)\,a_\su b_\sd \,+\, (s+r)\,a_\sd b_\su \,\Bigr\} \;+\;Tr[\vep-\mu] \pI\nn \\
&=:& H_0\;+\;Tr[\vep-\mu] \pI
\eeqn
and
\beqn
H_\rint&=& \; +\;{\ts {u\over 4}}\,\summ_j \,a_{j\su}\,a_{j\sd}\,b_{j\su}\,b_{j\sd}   \pI  \lbeq{2.11}
\eeqn
where we used the abbreviations
\beq
a_\sigma(\vep-\mu)\,b_\sigma&:=&\summ_{i,j}\,(\vep_{ij}-\mu\,\delta_{ij})\,a_{i\sigma}\,b_{j\sigma}\;, \;\;\;\;\;\;\;\; a_\su b_\sd\;\;:=\;\;\summ_j\,a_{j\su}\, b_{j\sd} \pI 
\eeq
and the trace in $H_{\rm kin}$ is a trace over $\mathbb C^\G$, 
\beq
Tr[\vep-\mu]\;\;=\;\;Tr_{\mathbb C^\G}[\vep-\mu]&=&\summ_j\,(\vep_{jj}-\mu\,\delta_{jj}) \;\;=\;\;\summ_j\,\vep_{jj}\;-\;\mu\,\G \pI 
\eeq
In the following, we drop the constant $\,Tr[\vep-\mu]\,$ and consider the Hamiltonian 
\beqn
H&:=&H_{\rm tot}\;-\;Tr[\vep-\mu] \;\;=\;\; H_0\;+\;H_\rint  \pI \lbeq{Hmain}\\ 
&=& {\ts {i\over 2}}\, \Bigl\{\,a_\su(\vep-\mu)\,b_\su\,+\,a_\sd(\vep-\mu)\,b_\sd \,+\,(s-r)\,a_\su b_\sd \,+\, (s+r)\,a_\sd b_\su \,\Bigr\} \;+\;
    {\ts {u\over 4}}\,\summ_j \,a_{j\su}\,a_{j\sd}\,b_{j\su}\,b_{j\sd} \nn  \pI 
\eeqn

\bigskip
\bigskip
{\bf 2.2 \;Hubbard-Stratonovich Transformation} 

\bigskip
\bigskip
To allow for some freedom in the Hubbard-Stratonovich transformation, we use the following identities: Let $\,\eps\,$ be a sign, $\;\eps\;\in\;\{-1,+1\}\;$. Then 
\beqn
a_{j\su}\,a_{j\sd}\,b_{j\su}\,b_{j\sd} &=& \eps\, {\ts {1\over 2}}\, (\,a_{j\su}\,b_{j\su}\,-\,\eps\,a_{j\sd}\,b_{j\sd} \,)^2\;+\;\eps \pI \pI \nn\\ 
&=&\eps\, {\ts {1\over 2}}\,(\,a_{j\su}\,b_{j\sd}\,+\,\eps\,a_{j\sd}\,b_{j\su} \,)^2 \;+\; \eps \nn \\ 
&=&\eps\, {\ts {1\over 2}}\,(\,a_{j\su}\,a_{j\sd}\,+\,\eps\,b_{j\su}\,b_{j\sd} \,)^2\;+\;\eps  \pI 
\eeqn
We choose some weights $\,w_1,w_2,w_3\,$ and some signs $\,\eps_1,\eps_2,\eps_3\,$ with 
\beqn
w_1+w_2+w_3&=&1 \pS  \nn\\ 
\eps_1,\eps_2,\eps_3&\in&\{-1,+1\} \pS\pS\pS
\eeqn
and write
\beqn
H_\rint&=& +\;{\ts {u\over 4}}\,\summ_j \,(w_1+w_2+w_3)\,a_{j\su}\,a_{j\sd}\,b_{j\su}\,b_{j\sd}   \pI  \nn\\
&=& +\, {\ts {u\over 4}}\,\summ_j \,\Bigl\{\;\phantom{+\,}\; w_1\eps_1\, {\ts {1\over 2}}\, (\,a_{j\su}\,b_{j\su}\,-\,\eps_1\,a_{j\sd}\,b_{j\sd} \,)^2\;+\;w_1\eps_1 \pI \nn\\ 
&&\phantom{+\, {\ts {u\over 4}}\,\summ_j \,\Bigl\{} +\; w_2\eps_2\, {\ts {1\over 2}}\,(\,a_{j\su}\,b_{j\sd}\,+\,\eps_2\,a_{j\sd}\,b_{j\su} \,)^2 \;+\; w_2\eps_2 \nn\\
&&\phantom{+\, {\ts {u\over 4}}\,\summ_j \,\Bigl\{} +\; w_3\eps_3\, {\ts {1\over 2}}\,(\,a_{j\su}\,a_{j\sd}\,+\,\eps_3\,b_{j\su}\,b_{j\sd} \,)^2\;+\;w_3\eps_3\, \;\;\Bigr\} 
  \pI\pI  \nn \\ 
&=:& H_1\;+\;H_2\;+\;H_3 \;\;+\;\, {\ts {u\over 4}}\,w\eps\,\G \pI \lbeq{2.14}
\eeqn
with
\beqn
H_1&:=&+\, {\ts {u\over 4}}\, w_1\eps_1\,\summ_j \, {\ts {1\over 2}}\, (\,a_{j\su}\,b_{j\su}\,-\,\eps_1\,a_{j\sd}\,b_{j\sd} \,)^2  \pI  \nn \\ 
H_2&:=&+\, {\ts {u\over 4}}\, w_2\eps_2\,\summ_j \, {\ts {1\over 2}}\, (\,a_{j\su}\,b_{j\sd}\,+\,\eps_2\,a_{j\sd}\,b_{j\su} \,)^2  \pI  \nn \\
H_3&:=&+\, {\ts {u\over 4}}\, w_3\eps_3\,\summ_j \, {\ts {1\over 2}}\,(\,a_{j\su}\,a_{j\sd}\,+\,\eps_3\,b_{j\su}\,b_{j\sd} \,)^2   \pI 
\eeqn
and
\beqn
w\eps&:=& w_1\eps_1+w_2\eps_2+w_3\eps_3 \pS
\eeqn

\medskip
To calculate thermodynamic quantities, we use the Trotter formula and write for some discretized inverse temperature $\beta=k dt$ 
\medskip
\beqn
e^{\,-\,\beta H}\;\;=\;\;e^{\,-\,kdt\, H}&=& e^{\,-\,kdt\,(H_0+H_1+H_2+H_3)}\;\times\; e^{\,-\,\beta{u\over 4}w\eps\G} \pI \nn \\ 
&\approx& \bigl(\, e^{\,-\,dt H_0}\;e^{\,-\,dt H_1}\;e^{\,-\,dt H_2}\;e^{\,-\,dt H_3}\,\bigr)^k \;\times\; e^{\,-\,\beta{u\over 4}w\eps\G} \phantom{mmm}\pI \lbeq{2.17}
\eeqn

\medskip
In all what follows, the limit $dt\to 0$ is always implicitely understood and we proceed with an exact equal 
sign in (\req{2.17}). We make the following Hubbard-Stratonovich transformations (the choice of the $-i$ instead of a $+i$ in the 
exponents is just convention):
\medskip
\beqn
e^{\,-\,dt H_1}&=&\prodd_j\, \int_\R \,\exp\Bigl\{\, -\,i\, {\ts {\sqrt{u dt}\over 2}}\, \sqrt{w_1\eps_1}\, (\,a_{j\su}\,b_{j\su}\,-\,\eps_1\,a_{j\sd}\,b_{j\sd} \,)\,\phi_j \,\Bigr\}
  \; e^{-{\phi_j^2\over 2}}\,{\ts {d\phi_j \over \sqrt{2\pi}}} \pI\pI \nn \\
e^{\,-\,dt H_2}&=&\prodd_j\, \int_\R \,\exp\Bigl\{\, -\,i\, {\ts {\sqrt{u dt}\over 2}}\, 
  \sqrt{w_2\eps_2}\, (\,a_{j\su}\,b_{j\sd}\,+\,\eps_2\,a_{j\sd}\,b_{j\su} \,)\,\xi_j \,\Bigr\} \; e^{-{\xi_j^2\over 2}}\,{\ts {d\xi_j \over \sqrt{2\pi}}} \pI \nn \\ 
e^{\,-\,dt H_3}&=&\prodd_j\, \int_\R \,\exp\Bigl\{\, -\,i\, {\ts {\sqrt{u dt}\over 2}}\, 
  \sqrt{w_3\eps_3}\, (\,a_{j\su}\,a_{j\sd}\,+\,\eps_3\,b_{j\su}\,b_{j\sd} \,)\,\theta_j \,\Bigr\} \; e^{-{\theta_j^2\over 2}}\,{\ts {d\theta_j \over \sqrt{2\pi}}} \pI \lbeq{n2.22}
\eeqn

\medskip
which leads to 
\medskip
\beqn
e^{\,-\,\beta H}&=&\int_{\R^{3k\G}} \,\prodd_{\ell=1}^k\,\Bigl[\, e^{\,-\,dt H_0}\,e^{\,-\,dQ_1(\phi_\ell)}\,e^{\,-\,dQ_2(\xi_\ell)}\,e^{\,-\,dQ_3(\theta_\ell)}\,\Bigr]\; 
   dP_k(\phi,\xi,\theta) \;\cdot\; e^{\,-\,\beta{u\over 4}w\eps\G}  \pI \pI\;\;  \lbeq{2.19}
\eeqn

\bigskip
with quadratic quantities
\beqn
dQ_1(\phi)&:=& +\,i\, {\ts {\sqrt{u dt}\over 2}}\, \sqrt{w_1\eps_1}\, \summ_j\, (\,a_{j\su}\,b_{j\su}\,-\,\eps_1\,a_{j\sd}\,b_{j\sd} \,)\,\phi_j  \pI \nn\\ 
dQ_2(\xi)&:=& +\,i\, {\ts {\sqrt{u dt}\over 2}}\, \sqrt{w_2\eps_2}\, \summ_j\, (\,a_{j\su}\,b_{j\sd}\,+\,\eps_2\,a_{j\sd}\,b_{j\su} \,)\,\xi_j  \pI \nn \\ 
dQ_3(\theta)&:=& +\,i\, {\ts {\sqrt{u dt}\over 2}}\, \sqrt{w_3\eps_3}\, \summ_j\, (\,a_{j\su}\,a_{j\sd}\,+\,\eps_3\,b_{j\su}\,b_{j\sd} \,)\,\theta_j \pI
\eeqn
and integration measure
\medskip
\beqn
dP_k(\phi,\xi,\theta) &:=& \prodd_{\ell=1}^k\,\prodd_{j\in\Gamma}\,  e^{ \,-\,{{1\over 2}}\,[\,\phi_{j,\ell}^2\,+\,\xi_{j,\ell}^2\,+\,\theta_{j,\ell}^2 \,] } \;\;
  {\ts {d\phi_{j,\ell} \,d\xi_{j,\ell} \,d\theta_{j,\ell}\over \phantom{\bigr|} (2\pi)^{3/2} \phantom{\bigr|} }}  \pI \lbeq{2.21}
\eeqn

\medskip
In matrix notation, 
\beq
dQ_1(\phi)&=& i\, {\ts {\sqrt{u}\over 4}}\, \sqrt{w_1\eps_1}\, \bmat a_\su & a_\sd & b_\su & b_\sd \emat  
\bmat 0 & 0           & +dx & 0           \\ 
      0 & 0           & 0   & -\eps_1\,dx \\ 
    -dx & 0           & 0   & 0           \\ 
      0 & +\eps_1\,dx & 0   & 0           \emat  \bmat a_\su \\ a_\sd \\ b_\su \\ b_\sd \emat  \pI
\eeq
\beq
dQ_2(\xi)&=& i\, {\ts {\sqrt{u}\over 4}}\, \sqrt{w_2\eps_2}\, \bmat a_\su & a_\sd & b_\su & b_\sd \emat  
\bmat 0 & 0           & 0           & +dy  \\ 
      0 & 0           & +\eps_2\,dy & 0    \\ 
      0 & -\eps_2\,dy & 0           & 0    \\ 
    -dy & 0           & 0           & 0    \emat  \bmat a_\su \\ a_\sd \\ b_\su \\ b_\sd \emat  \pI
\eeq
\beq
dQ_3(\theta)&=& i\, {\ts {\sqrt{u}\over 4}}\, \sqrt{w_3\eps_3}\, \bmat a_\su & a_\sd & b_\su & b_\sd \emat  
\bmat 0 & +dz          & 0           & 0           \\ 
    -dz & 0           & 0           & 0           \\ 
      0 & 0           & 0           & +\eps_3\,dz \\ 
      0 & 0           & -\eps_3\,dz & 0           \emat  \bmat a_\su \\ a_\sd \\ b_\su \\ b_\sd \emat  \pI
\eeq

\bigskip
where we introduced the diagonal matrices of Brownian motions 
\medskip
\beqn
dx&:=&\bigl(\,dx_j\;\delta_{i,j}\,\bigr)_{i,j\in\Gamma}\;\;:=\;\;\bigl(\,\sqrt{dt}\,\phi_j\;\delta_{i,j}\,\bigr)_{i,j\in\Gamma} \;\;\in\;\; \R^{\G\times\G }  \pI \nn\\
dy&:=&\bigl(\,dy_j\;\delta_{i,j}\,\bigr)_{i,j\in\Gamma}\;\;:=\;\;\bigl(\,\sqrt{dt}\;\xi_j\;\delta_{i,j}\,\bigr)_{i,j\in\Gamma} \;\;\in\;\; \R^{\G\times\G }   \nn \\
dz&:=&\bigl(\,dz_j\;\delta_{i,j}\,\bigr)_{i,j\in\Gamma}\;\;:=\;\;\bigl(\,\sqrt{dt}\;\theta_j\;\delta_{i,j}\,\bigr)_{i,j\in\Gamma} \;\;\in\;\; \R^{\G\times\G }  \pI\pI \lbeq{2.22}
\eeqn

\medskip
We also recall that (with the identifications $\,\mu\,\equiv\, \mu\, Id$,\;$\,r\,\equiv\, r\, Id$,\;$\,s\,\equiv\, s\, Id\,$\,)
\bigskip
\beqn
H_0&=&\ts {i\over 4}\, \bmat a_\su & a_\sd & b_\su & b_\sd \emat  
\bmat      0 & 0           & \vep-\mu    & s-r  \\ 
           0 & 0           & s+r     & \vep-\mu    \\ 
 -(\vep-\mu) & -(s+r)    & 0           & 0    \\ 
    -(s-r) & -(\vep-\mu) & 0           & 0    \emat  \bmat a_\su \\ a_\sd \\ b_\su \\ b_\sd \emat  \pI   \lbeq{2.23}
\eeqn

\bigskip
\bigskip
\bigskip
{\bf 2.3 \;Main Theorem} 

\medskip
\bigskip
We are now in a position to state the main result of this paper:

\bigskip
\bigskip
{\bf Main Theorem:} Let $H$ be the Hubbard-Hamiltonian (\req{Hmain}). Define the $\,4\G\times 4\G\,$ skew symmetric matrix 
of correlations $C_\beta$ in terms of Majorana fermion operators (\req{n2.7}) through {\footnotesize{(the very first $i$ is a $\sqrt{-1}$, not a lattice site)}}
\medskip
\beqn
C_\beta\;\;=\;\;\bmat C^{aa} & C^{ab} \\ C^{ba} & C^{bb} \emat&:=&
  i\,\bmat \la \,a_{i\sigma}\,a_{j\tau}\,\ra_\beta \,-\, \delta_{i\sigma,j\tau} & \la\, a_{i\sigma}\,b_{j\tau}\,\ra_\beta \\ 
  \la \,b_{i\sigma}\,a_{j\tau}\,\ra_\beta & \la\, b_{i\sigma}\,b_{j\tau}\,\ra_\beta \,-\, \delta_{i\sigma,j\tau}\emat \pI\pI  \lbeq{2.24} 
\eeqn

\medskip
with spin indices $\,\sigma,\tau\in\hbox{\footnotesize$\{\up,\down\}$}\,$ and lattice site indices $\,i,j\in\Gamma\,$ and 
\beqn
\la \,A\,\ra_\beta&:=& Tr_\cF\,A\,e^{-\beta H}\;\bigr/\;Tr_\cF \,e^{-\beta H}  \pI
\eeqn
with $\cF$ being the grand canonical Fock space. Then, with $\beta\,=\,k\,dt$,  
\medskip
\beqn
C_\beta&=&\la \,G_\beta\,\ra \;\;:=\;\;{  \phantom{\Bigr|} \int_{\R^{3k\G}}\,G_\beta(\phi,\xi,\theta)\; e^{\,-\,\int_0^\beta\, W(G_t)\;dt }\; dP_k(\phi,\xi,\theta) 
  \phantom{\Bigr|} \over \phantom{\Bigr|} \int_{\R^{3k\G}}\; e^{\,-\,\int_0^\beta\, W(G_t)\;dt }\; dP_k(\phi,\xi,\theta)  \phantom{\Bigr|} }  \pI \pI  \lbeq{2.26}
\eeqn

\medskip
with $dP_k$ given by the Gaussian measure (\req{2.21}) above,  
\beqn
{\ts \int_0^\beta\, W(G_t)\;dt }\;\;:=\;\;\summ_{\ell=0}^{k-1}\,W(G_{t_\ell})\;dt  
\eeqn
with
\beqn
W(G)&:=&{\ts {1\over 2}}\, Tr_{\mathbb C^\G}\Bigl[\; (\vep-\mu)\,(G^{ab}_{\su\su} + G^{ab}_{\sd\sd}) 
  \,+\,(s+r)\,G^{ab}_{\sd\su} + (s-r)\,G^{ab}_{\su\sd}  \;\Bigr] \pI \nn\\
&& \;+\; {\ts {u\over 4}}\,\summ_j\,\Bigl[\;G^{ab}_{\su\su,jj}\;G^{ab}_{\sd\sd,jj} 
   \,-\,G^{ab}_{\su\sd,jj}\;G^{ab}_{\sd\su,jj} \, -\,G^{aa}_{\su\sd,jj}\;G^{bb}_{\su\sd,jj} \;\Bigr]  \pI \lbeq{2.28}
\eeqn
and the skew symmetric matrix $G_t\,\in\, \mathbb C^{4\G\times 4\G} $, let's refer to it also as the density matrix, is given by the SDE system  
\medskip
\beqn
dG&=&\ts +\,{1\over 2}\,(\,G -i\,Id\,)\, \Bigl[\;-\,h_0\,dt \,+\,{u\over 2}\,DG\,dt\,-\,\sqrt{u}\,dB \; \Bigr] \, (\,G+i\,Id\,)  \pI  \lbeq{2.29}
\eeqn

\medskip
with initial value $G_{t=0}=0\,$. The quantities $h_0$, $DG$ and $dB$ are given by the following skew symmetric $4\G\times 4\G$ matrices 
(with the identifications $\,\mu\,\equiv\, \mu\, Id$,\;$\,r\,\equiv\, r\, Id$,\;$\,s\,\equiv\, s\, Id\,$\,):
\medskip
\beqn
h_0&:=&\bmat      0 & 0           & \vep-\mu    & s-r  \\ 
           0 & 0           & s+r     & \vep-\mu    \\ 
 -(\vep-\mu) & -(s+r)    & 0           & 0    \\ 
    -(s-r) & -(\vep-\mu) & 0           & 0    \emat   \pI \lbeq{2.30}
\eeqn
\beqn
DG&:=&\bmat 
       0                         & +{\rm diag}[\,G^{bb}_{\su\sd}\,] & -{\rm diag}[\,G^{ab}_{\sd\sd}\,] & +{\rm diag}[\,G^{ab}_{\sd\su}\,]  \\ 
-{\rm diag}[\,G^{bb}_{\su\sd}\,] &          0                & +{\rm diag}[\,G^{ab}_{\su\sd}\,] & -{\rm diag}[\,G^{ab}_{\su\su}\,]  \\ 
+{\rm diag}[\,G^{ab}_{\sd\sd}\,] & -{\rm diag}[\,G^{ab}_{\su\sd}\,]  &        0                         & +{\rm diag}[\,G^{aa}_{\su\sd}\,]  \\  
-{\rm diag}[\,G^{ab}_{\sd\su}\,] & +{\rm diag}[\,G^{ab}_{\su\su}\,]  & -{\rm diag}[\,G^{aa}_{\su\sd}\,] & 0                       \emat \pI  \lbeq{2.31}
\eeqn
\beqn
dB&:=& 
\bmat 0 & +\nu_3\,dz           & +\nu_1\,dx & +\nu_2\,dy           \\ 
      -\nu_3\,dz & 0           & +\nu_2\eps_2\,dy   & -\nu_1\eps_1\,dx \\ 
    -\nu_1\,dx & -\nu_2\eps_2\,dy     & 0   & +\nu_3\eps_3\,dz            \\ 
      -\nu_2\,dy & +\nu_1\eps_1\,dx & -\nu_3\eps_3\,dz    & 0           \emat  \pI \lbeq{2.32}
\eeqn

\medskip
with diagonal matrices of Brownian motions $dx,dy,dz\in\R^{\G\times\G}$ given by (\req{2.22}) above and  $\nu_i:=\sqrt{w_i\eps_i}\,$. 
Also $\;{\rm diag}[A]:=(\,A_{jj}\,\delta_{ij}\,)_{i,j\in\Gamma}\,$. In particular, we have the following expectations: 
\beqn
\la H\ra_\beta\;\;:=\;\; { Tr_\cF[\, H\,e^{\,-\,\beta H} \,] \over Tr_\cF[\, e^{\,-\,\beta H} \,] } &=&
  { \phantom{\Bigl|} \int_{\R^{3k\G}} \,  W(G_\beta) \; e^{\,-\,\int_0^\beta\, W(G_t)\;dt }\; dP_k(\phi,\xi,\theta) \phantom{\Bigl|}\over 
      \phantom{\Bigl|}\int_{\R^{3k\G}} \,   e^{\,-\,\int_0^\beta\, W(G_t)\;dt }\; dP_k(\phi,\xi,\theta) \phantom{\Bigl|} }  \pI  \lbeq{n2.37}\\
&& \nn \\ 
Tr_\cF[\, e^{\,-\,\beta H} \,]\,\bigr/\,Tr_\cF[\, Id \,]&=&
  \ts\int_{\R^{3k\G}} \,   e^{\,-\,\int_0^\beta\, W(G_t)\;dt }\; dP_k(\phi,\xi,\theta) \pI \lbeq{2.33}
\eeqn
and density matrix elements are given by
\medskip
\beqn
\ts \la \,c_{j\su}^+c_{j\su}\,\ra_\beta\;-\;{1\over 2} &=&\ts +\,{1\over 2}\,\bigl\la\; G^{ab}_{\beta,\su\su,jj}\; \bigr\ra  \lbeq{2.34}  \\
\la\,c_{j\su}^+\,c_{j\sd}\,+\,c_{j\sd}^+\,c_{j\su}\,\ra_\beta &=&+\,{\ts{1\over 2}}\,\bigl\la \;G^{ab}_{\beta,\su\sd,jj} \,+\,G^{ab}_{\beta,\sd\su,jj} \;\bigr\ra 
    \pI\pI  \lbeq{2.35} \\ 
\la \,c_{j\su}^+\,c^+_{j\sd}\,+\,c_{j\sd}\,c_{j\su}\,\ra_\beta&=&-\,{\ts{1\over 2}}\,\bigl\la\;G^{ab}_{\beta,\su\sd,jj} \,-\,G^{ab}_{\beta,\sd\su,jj}  \;\bigr\ra 
    \pS  \lbeq{2.36}
\eeqn

\medskip
The only quantity which depends explicitely on the details of the Hubbard-Stratonovich factorization (\req{2.19}) is the $dB$, all other quantities, 
in particular the $DG$ and the $W(G)$, do not depend on the $w_i$ or $\eps_i\,$. 

\bigskip
\bigskip
\bigskip
{\bf Remarks:} \;{\bf (i)} \,Recall the Hubbard-Stratonovich weights $w_1$, $w_2$ and $w_3$ and the signs $\eps_1$, $\eps_2$ and $\eps_3$ as they have been 
introduced in (\req{2.14}). In sections 3 and 4, we put $\,w_3=0\,$ and choose $\,\eps_1=\eps_2=\eps_u:=\sign\,u\,$. For that choice, the SDE system (\req{2.29}) 
for the skew symmetric complex $4\G\times 4\G$ generalized density matrix $G$ can be written in terms of real $2\G\times 2\G$ matrices $\rho$, $F^a$ and $F^b$, this is done 
in (\req{n3.4}) of Theorem 3.1 below.    

\medskip
{\bf (ii)} \;For the more specific choice $w_1=1$ and with zero external pairing and exchange terms $\,r=s=0\,$ in the Hamiltonian (\req{2.2}) or (\req{Hmain}), 
the system (\req{n3.4}) reduces further and can be written in terms of real $\,\G\times\G$ matrices $\rho_{\su\su},\rho_{\sd\sd}$ and $F_{\su\su},F_{\sd\sd}$, 
this is done in part (a) of Theorem 3.2 below in (\req{3.7n}).

\medskip
{\bf (iii)} \;For the choice $w_2=1$, but with arbitrary pairing and exchange terms $r,s\in\R$ in the Hamiltonian (\req{2.2}) or (\req{Hmain}), part (b) of Theorem 3.2 
rewrites the $2\G\times 2\G$ system (\req{n3.4}) of Theorem 3.1 in a more compact way, the result is (\req{3.10n}) which is still $2\G\times 2\G\,$,  
but it has less terms, (\req{n3.4}) has an $F^a$ and an $F^b$, but (\req{3.10n}) only has an $F$. 

\medskip
{\bf (iv)} \;At half filling $\,\mu=0\,$ there are further simplifications and they are summarized in Theorem 4.2 below. Part (a) covers the $w_1=1$ representation 
which results in (\req{4.16n}) with energy given by (\req{4.22n}) and part (b) covers the $w_2=1$ representation which results in (\req{4.20n}) with energy 
given by (\req{4.25nn}). 

\medskip
{\bf (v)} \;We also calculated the spin-spin and pair-pair correlations in both representations, in Theorems 4.3 
and 4.4. Here the fact that spin-spin correlations have to be of antiferromagnetic type, for repulsive couplings, comes out automatically in the $w_2=1$ representation, 
it holds pathwise for every Monte Carlo path, but in the $w_1=1$ representation this is obtained only after the MC average has been taken, individual MC paths can have a more 
irregular sign structure. On the other hand, the fact that pair-pair correlations are nonnegative for attractive couplings, comes out automatically in the $w_1=1$ 
representation, but this is less obvious in the $w_2=1$ representation. A similar remark also applies to the constant density theorem: for a perfectly spin symmetric hamiltonian, 
the density is pathwise constant in the $w_2=1$ representation (that is, to allow for symmetry breaking one should actually add a small spin dependent term to the 
hamiltonian), but in the $w_1=1$ representation, the density is not pathwise constant, but only after the Monte Carlo average has been taken. 

\medskip
{\bf (vi)} Finally just concerning the notation: In section 5 where we prove the Main Theorem, we have $G^{ab}$\,'s which correspond to expectations $\la a b\ra$. 
In section 5.4, we introduce a ${\ti G}^{ab}\,:=\,i\,G^{ab}$ which corresponds to expectations $i\la a b\ra$. The $G$ in the Main Theorem is actually a $\ti G$ corresponding 
to expectations $i\la a b\ra$, to keep the notation simple we omitted the tilde in the formulation of the Main Theorem. Observe that for Majorana operators 
$a,b$ the combination $iab$ is self adjoint.

\goodbreak

\bigskip
\bigskip
\bigskip
{\large\bf 3. Symmetry Relations}
\numberwithin{equation}{section}
\renewcommand\thesection{3}
\setcounter{equation}{0}

\bigskip
\bigskip
Recall the Hubbard-Stratonovich parameters $\,w_1,\,w_2,\,w_3\,$ and $\,\eps_1,\,\eps_2,\,\eps_3\,$ as they have been introduced in (\req{2.14}) and (\req{n2.22}).  
In this and the next section, we consider the following two choices:
\beqn
w_1\;\;=\;\;1,\;\;\;\;w_2\;=\;w_3\;=\;0   \lbeq{n3.1} \\ 
w_2\;\;=\;\;1,\;\;\;\;w_1\;=\;w_3\;=\;0   \lbeq{n3.2}
\eeqn
and the signs $\eps_i$ we always put equal to the sign of $u$, 
\beqn
\eps_1\;\;=\;\;\eps_2\;\;=\;\;\eps_u\;\;:=\;\;\sign\,u   \lbeq{n3.3}
\eeqn
For these choices, the $G^{aa}$ and $G^{bb}$ are pure imaginary and the $G^{ab}$ are real, there is the following

\bigskip
\bigskip
{\bf Theorem 3.1:} With the above choices of Hubbard-Stratonovich parameters and for arbitrary real $r,s$ and $\mu$ and $u$, we have
\beqn
{\rm Re}[\,G^{aa}\,]\;\;=\;\;{\rm Re}[\,G^{bb}\,]&=& 0 \nn \\ 
{\rm Im}[\,G^{ab}\,]\;\;=\;\;{\rm Im}[\,G^{ba}\,]&=& 0 
\eeqn
If we introduce the notation, with  $\,\rho\,,\,F^a\,,\,F^b\;\in\;\R^{2\G\times2\G} \,$, 
\beqn
\rho&:=&G^{ab} \nn \\
i\,F^a&:=&G^{aa} \nn \\ 
i\,F^b&:=&G^{bb}  \lbeq{3.5nn}
\eeqn
then the complex SDE system (\req{2.29}) for the skew symmetric $\,4\G\times 4\G\,$ matrix $G$ 
is equivalent to the following real $2\G\times 2\G$ SDE system 
\beqn
d\rho&=&\ts {1\over 2}\,\Bigl[\; (Id-F^a)\,dh\,(Id+F^b)\;-\; \rho\,dh^T \rho    \;\Bigr] \pI \nn \\ 
&&\;\;\;\; \;+\; \ts {udt\over 4}\,\Bigl[\; (Id-F^{a})\,DF^{b}\,\rho \;-\;\rho\,DF^{a}\,(Id+ F^{b})  \;\Bigr] \nn \\
dF^{a}&=&\ts {1\over 2}\,\Bigl[\; (Id-F^{a})\,dh\,\rho^T\,-\,\rho\,dh^T(Id+F^{a})  \; \Bigr]  \pI \nn \\ 
&&\;\;\;\; \;+\;\ts {udt\over 4}\,\Bigl[\; (Id-F^a)\,DF^{b}\,(Id+F^a)\,-\,\rho\,DF^{a}\, \rho^T  \;\Bigr]      \nn  \\
dF^{b}&=&\ts {1\over 2}\,\Bigl[\; (Id-F^{b})\,dh^T \rho \,-\,\rho^T\,dh\,(Id+F^{b})  \;\Bigr]  \pI\nn  \\ 
&&\;\;\;\; \;+\;\ts {udt\over 4}\,\Bigl[\;  (Id-F^b)\,DF^{a}\,(Id+F^b)  \,-\,\rho^T DF^{b}\,\rho  \; \Bigr]      \lbeq{n3.4}
\eeqn
with
\beqn
dh&:=&\ts -\,dt\,\bmat \vep-\mu & s-r \\  s+r  & \vep-\mu \emat  \;+\;{udt\over 2}\,D\rho 
  \;-\;\sqrt{\sabs u\sabs}\;\bmat  \sqrt{w_1}\;dx & \sqrt{w_2}\; dy \\ \eps_u\sqrt{w_2}\;dy & -\eps_u\sqrt{w_1}\;dx \emat  \pI \lbeq{n3.5}
\eeqn
\beqn
D\rho\;\;:=\;\;\bmat  -{\rm diag}[\,\rho_{\sd\sd}\,] & +{\rm diag}[\,\rho_{\sd\su}\,] \; \\ 
  +{\rm diag}[\,\rho_{\su\sd}\,] & -{\rm diag}[\,\rho_{\su\su}\,]  \; \emat,\;\;\;
DF^{a,b}\;\;:=\;\; \bmat  0  & +{\rm diag}[\,F_{\su\sd}^{a,b}\,] \; \\ -{\rm diag}[\,F_{\su\sd}^{a,b}\,] &  0 \emat  \pI  \lbeq{n3.6}
\eeqn

\medskip
and initial values $\,\rho_{t=0}\,=\,F^a_{t=0}\,=\,F^b_{t=0}\,=\,0\,$. 

\bigskip
\bigskip
{\bf Proof:} \;Follows directly from the Main Theorem by straightforward calculations.  \;$\blacksquare$

\goodbreak

\bigskip
\bigskip
\bigskip
{\bf Theorem 3.2 (symmetry relations, arbitrary $\bm{\mu}\,$):} As in Theorem 3.1, choose the Hubbard-Stratonovich parameters $w_i$ and $\eps_i$ 
as given by (\req{n3.1}-\req{n3.3}). Then, for arbitrary real $r,s$ and $\mu$ and $u$, the $4\G\times 4\G$ skew symmetric matrix 
\beq
G&=&\bmat G^{aa} & G^{ab} \\ G^{ba} & G^{bb} \emat\;\;\buildrel (\req{3.5nn})\over=\;\; \bmat \;i\,F^a & \rho\; \\ \;-\,\rho^T & i\,F^b\; \emat \pI 
\eeq 
from the Main Theorem fulfills the following symmetry relations: 

\medskip
\begin{itemize}
\item[{\bf a)}] Choose $w_1=1$ and, in addition, also put $\,r=0\,$. Then 
\beqn
F^b&=&F^a \nn \\ 
\rho^T&=& \rho \phantom{mmm}
\eeqn
If we also put $\,s=0\,$, we have the additional identities, with $F\equiv F^a$, 
\beqn
\rho_{\su\sd}\;\;=\;\;\rho_{\sd\su}&=& 0   \nn \\  
F_{\su\sd}\;\;=\;\;F_{\sd\su}&=& 0   \lbeq{3.7nn}
\eeqn
and the SDE system of Theorem 3.1 reduces to the following $\G\times\G$ system 
\beqn
d\rho_{\sigma\sigma}&=&\ts {1\over 2}\,\Bigl[\; (Id-F_{\sigma\sigma})\,dh_{\sigma\sigma}\,(Id+F_{\sigma\sigma})\;-\; 
   \rho_{\sigma\sigma}\,dh_{\sigma\sigma} \rho_{\sigma\sigma}    \;\Bigr] \pI \nn \\ 
dF_{\sigma\sigma}&=&\ts {1\over 2}\,\Bigl[\; (Id-F_{\sigma\sigma})\,dh_{\sigma\sigma}\,\rho_{\sigma\sigma} \;-\; 
  \rho_{\sigma\sigma}\,dh_{\sigma\sigma}\,(Id+F_{\sigma\sigma})  \; \Bigr]  \lbeq{3.7n}
\eeqn
with $\sigma\in\{\up,\down\}$ and
\beqn
dh_{\su\su}&=& -\,(\vep-\mu)\,dt \;-\;\ts {udt \over 2} \,{\rm diag}[\,\rho_{\sd\sd}\,] \;-\; \sqrt{\sabs u\sabs}\; dx \pI \nn \\ 
dh_{\sd\sd}&=& -\,(\vep-\mu)\,dt \;-\;\ts {udt \over 2} \,{\rm diag}[\,\rho_{\su\su}\,] \;+\; \eps_u \sqrt{\sabs u\sabs}\; dx  \lbeq{3.8n}
\eeqn

\medskip
The energy $W(G)$ of (\req{2.28}) simplifies to 
\beqn
W(G)&=&{\ts {1\over 2}}\, Tr_{\mathbb C^\G}\Bigl[\; (\vep-\mu)\,(\rho_{\su\su} + \rho_{\sd\sd})  \;\Bigr] 
  \;+\; {\ts {u\over 4}}\,\summ_j\, \rho_{\su\su,jj}\;\rho_{\sd\sd,jj}   \pI \lbeq{3.13n}
\eeqn
\item[{\bf b)}] Choose $w_2=1$. Then, with 
\beq
P&:=&\bmat 0&Id\\ Id& 0 \emat \;\;\in\;\;\R^{2\G\times 2\G} \pI 
\eeq
we have for arbitrary $r,s\in\R$ 
\beqn
F^b &=& P F^a P \\ 
\rho^T &=& P \rho \,P \phantom{mmm}
\eeqn
or more explicitely, in terms of $\G\times \G$ block matrices,
\medskip
\beq
\bmat F^{b}_{\su\su} &F^{b}_{\su\sd} \\ F^{b}_{\sd\su} & F^{b}_{\sd\sd} \emat  \;\;=\;\;  
  \bmat F^{a}_{\sd\sd} & F^{a}_{\sd\su}  \\  F^{a}_{\su\sd} & F^{a}_{\su\su} \emat, \;\;\;\;\;
\bmat \phantom{\bigr|} \rho_{\su\su}^T & \rho_{\sd\su}^T \phantom{\Bigr|} \\ \phantom{\bigr|} \rho_{\su\sd}^T & \rho_{\sd\sd}^T \phantom{\bigr|} \emat\;\;=\;\;
  \bmat \rho_{\sd\sd} & \rho_{\sd\su}  \\  \rho_{\su\sd} & \rho_{\su\su} \emat  \pI 
\eeq

\medskip
The SDE system of Theorem 3.1 becomes, with $F\equiv F^a$ and $\ti F:=PFP=F^b$ and $DF^b=-DF^a$,  
\beqn
d\rho&=&\ts {1\over 2}\,\Bigl[\; (Id-F)\,dh\,(Id+\ti F)\;-\; \rho\,dh^T \rho    \;\Bigr] \pI \nn \\ 
&&\;\;\;\; \;-\; \ts {udt\over 4}\,\Bigl[\; (Id-F)\,DF\,\rho \;+\;\rho\,DF\,(Id+ \ti F)  \;\Bigr] \nn \\
dF&=&\ts {1\over 2}\,\Bigl[\; (Id-F)\,dh\,\rho^T\,-\,\rho\,dh^T(Id+F)  \; \Bigr]  \pI \nn \\ 
&&\;\;\;\; \;-\;\ts {udt\over 4}\,\Bigl[\; (Id-F)\,DF\,(Id+F)\,+\,\rho\,DF\, \rho^T  \;\Bigr]  \pM   \lbeq{3.10n}   
\eeqn
with
\beqn
dh&=&\ts -\,dt\,\bmat \vep-\mu & s-r \\  s+r  & \vep-\mu \emat  \;+\;{udt\over 2}\,D\rho 
  \;-\;\sqrt{\sabs u\sabs}\;\bmat  0 &  dy \\ \eps_u\,dy & 0 \emat  \pI 
\eeqn
\beqn
D\rho\;\;=\;\;\bmat  -{\rm diag}[\,\rho_{\sd\sd}\,] & +{\rm diag}[\,\rho_{\sd\su}\,] \; \\ 
  +{\rm diag}[\,\rho_{\su\sd}\,] & -{\rm diag}[\,\rho_{\su\su}\,]  \; \emat,\;\;\;
DF\;\;=\;\; \bmat  0  & +{\rm diag}[\,F_{\su\sd}\,] \; \\ -{\rm diag}[\,F_{\su\sd}\,] &  0 \emat  \pI  
\eeqn
\end{itemize} 

\bigskip
{\bf Proof:} The general strategy for obtaining symmetry relations in our setup is as follows: First, to get an idea, simulate a couple of paths to check 
for a particular identity. Then, more mathematically, prove this identity by induction on time steps. For part (a), the induction looks as follows: 
Suppose at some time (or inverse temperature) step $t_k=kdt$ we have  
\beq
F^b_{t_k}&=&F^a_{t_k} \\ 
\rho^T_{t_k}&=& \rho_{t_k} \phantom{mmm}
\eeq
Then we have to prove
\beq
F^b_{t_{k+1}}\;\;=\;\;F^b_{t_k}\;+\;dF^b_{t_{k+1}}&=&F^a_{t_k}\;+\;dF^a_{t_{k+1}}\;\;=\;\;F^a_{t_{k+1}} \pS\\ 
\rho^T_{t_{k+1}}\;\;=\;\;\rho_{t_k}^T\;+\;d\rho_{t_{k+1}}^T&=&\rho_{t_k}\;+\;d\rho_{t_{k+1}}\;\;=\;\; \rho_{t_{k+1}} \pS
\eeq
which, using the induction hypothesis, is equivalent to  
\beqn
dF^b_{t_{k+1}}&=&dF^a_{t_{k+1}}  \lbeq{n3.8}\\ 
d\rho^T_{t_{k+1}}&=& d\rho_{t_{k+1}} \lbeq{n3.9}\pS
\eeqn
From Theorem 3.1, we have with $F^{a,b}\equiv F^{a,b}_{t_k}$ and $\rho\equiv \rho_{t_k}$, 
\beq
d\rho_{t_{k+1}}&=&\ts {1\over 2}\,\Bigl[\; (Id-F^a)\,dh_{t_{k+1}}\,(Id+F^b)\;-\; \rho\,dh^T_{t_{k+1}} \rho    \;\Bigr] \pI \nn \\ 
&&\;\; \;+\; \ts {udt\over 4}\,\Bigl[\; (Id-F^{a})\,DF^{b}\,\rho \;-\;\rho\,DF^{a}\,(Id+ F^{b})  \;\Bigr] \nn \\
dF^{a}_{t_{k+1}}&=&\ts {1\over 2}\,\Bigl[\; (Id-F^{a})\,dh_{t_{k+1}}\,\rho^T\,-\,\rho\,dh^T_{t_{k+1}}(Id+F^{a})  \; \Bigr] \pI \nn \\ 
&&\;\; \;+\; \ts {udt\over 4}\,\Bigl[\; (Id-F^a)\,DF^{b}\,(Id+F^a)\,-\,\rho\,DF^{a}\, \rho^T  \;\Bigr]      \nn  \\
dF^{b}_{t_{k+1}}&=&\ts {1\over 2}\,\Bigl[\; (Id-F^{b})\,dh^T_{t_{k+1}} \rho \,-\,\rho^T\,dh_{t_{k+1}}\,(Id+F^{b})  \;\Bigr]  \pI\nn  \\ 
&&\;\; \;+\; \ts {udt\over 4}\,\Bigl[\;  (Id-F^b)\,DF^{a}\,(Id+F^b)  \,-\,\rho^T DF^{b}\,\rho  \; \Bigr]     
\eeq
with, using $w_1=1$ and $r=0$, 
\beq
dh_{t_{k+1}}&=&\ts -\,dt\,\bmat \vep-\mu & s \\  s  & \vep-\mu \emat  \;\;+\;\;{udt\over 2}\,D\rho 
  \;\;-\;\;\sqrt{\sabs u\sabs}\;\bmat  dx_{t_{k+1}} & 0 \\ 0 & -\eps_u\,dx_{t_{k+1}} \emat  \pI 
\eeq
From the induction hypothesis, we get $[D\rho]^T=D\rho$ which implies 
\beq
dh_{t_{k+1}}^T&=& dh_{t_{k+1}} \pS 
\eeq
From this, the equations (\req{n3.8},\req{n3.9}) follow immediately, recall that $F^a$ and $F^b$ are skew symmetric matrices. 
Also, for $s=0$ and by making the induction hypothesis 
\beq
\rho_{\su\sd}\;\;=\;\;\rho_{\sd\su}\;\;=\;\;F_{\su\sd}\;\;=\;\;F_{\sd\su}&=& 0 \pS
\eeq
the only non-vanishing $(\sigma,\tau)$ components are the $(\up,\up)$ and $(\down,\down)$ components, and also (\req{3.7nn}) follows. 

\medskip
{Part (b):} In this case, we have to prove
\beqn
dF^b &=& P dF^a P \lbeq{3.14n} \\ 
d\rho^T &=& P d\rho \,P \pS  \lbeq{3.15n}
\eeqn
Since $P^2=Id\,$ and $P^T=P$, we can write 
\beq
P \,d\rho \,P&=&\ts {1\over 2}\,\Bigl[\; P(Id-F^a)P\,PdhP\,P(Id+F^b)P\;-\; P\rho P\,[PdhP]^T P\rho P    \;\Bigr] \pI \nn \\ 
&& \;+\; \ts {udt\over 4}\,\Bigl[\; P(Id-F^{a})P\,PDF^{b}P\,P\rho P \;-\;P\rho P\,PDF^{a}P\,P(Id+ F^{b})P  \;\Bigr] \nn 
\eeq
Now, 
\beq
PdhP&=&\ts -\,dt\,\bmat \vep-\mu & s+r \\  s-r  & \vep-\mu \emat  \;+\;{udt\over 2}\,PD\rho\, P 
  \;-\;\sqrt{\sabs u\sabs}\;\bmat  0 &  \eps_u\,dy \\ dy & 0 \emat  \pI 
\eeq
with 
\beq
PD\rho \,P &=&P\bmat  -{\rm diag}[\,\rho_{\sd\sd}\,] & +{\rm diag}[\,\rho_{\sd\su}\,] \; \\ 
  +{\rm diag}[\,\rho_{\su\sd}\,] & -{\rm diag}[\,\rho_{\su\su}\,]  \; \emat P 
\;\;=\;\; \bmat  -{\rm diag}[\,\rho_{\su\su}\,] & +{\rm diag}[\,\rho_{\su\sd}\,] \; \\ 
  +{\rm diag}[\,\rho_{\sd\su}\,] & -{\rm diag}[\,\rho_{\sd\sd}\,]  \; \emat  \pI \\ 
&&\phantom{I} \\ 
&\buildrel {\rm ind.hyp.} \over =&  \bmat  -{\rm diag}[\,\rho_{\sd\sd}^T\,] & +{\rm diag}[\,\rho_{\su\sd}\,] \; \\ 
  +{\rm diag}[\,\rho_{\sd\su}\,] & -{\rm diag}[\,\rho_{\su\su}^T\,]  \; \emat \;\;=\;\;
\bmat  -{\rm diag}[\,\rho_{\sd\sd}\,] & +{\rm diag}[\,\rho_{\su\sd}\,] \; \\ 
  +{\rm diag}[\,\rho_{\sd\su}\,] & -{\rm diag}[\,\rho_{\su\su}\,]  \; \emat \;\;=\;\; [D\rho]^T \pI
\eeq
Thus, 
\beq
P\,dh\,P&=& dh^T \pS 
\eeq
and we get, using again the induction hypothesis, 
\beq
P \,d\rho \,P&=&\ts {1\over 2}\,\Bigl[\; (Id-F^b)\,dh^T\,(Id+F^a)\;-\; \rho^T\,dh\, \rho^T    \;\Bigr] \pI \nn \\ 
&& \;+\; \ts {udt\over 4}\,\Bigl[\; (Id-F^{b})\,PDF^{b}P\,\rho^T \;-\;\rho^T\,PDF^{a}P\,(Id+ F^{a})  \;\Bigr] \nn 
\eeq
Since 
\beq
PDF^{a,b}P&=& [\,DF^{a,b}\,]^T \pS
\eeq
and because of the skew symmetry of $F^{a}$ and $F^{b}$, we arrive at 
\beq
P \,d\rho \,P&=&\ts {1\over 2}\,\Bigl[\; (Id-F^a)\,dh\,(Id+F^b)\;-\; \rho\,dh^T \rho    \;\Bigr]^T \pI \nn \\ 
&& \;+\; \ts {udt\over 4}\,\Bigl[\; \rho\,DF^{b}\,(Id+F^b) \;-\;(Id-F^a)\,DF^{a}\,\rho  \;\Bigr]^T \;\;=\;\; d\rho^T \nn 
\eeq
where we used $DF^b=-DF^a$ in the last equal sign. This proves (\req{3.15n}). Equation (\req{3.14n}) follows in a similar way. $\;\blacksquare$

\goodbreak

\bigskip
\bigskip
\bigskip
\bigskip
{\large\bf 4. The Model at Half-Filling}
\numberwithin{equation}{section}
\renewcommand\thesection{4}
\setcounter{equation}{0}

\bigskip
\bigskip
\bigskip
{\bf 4.1 \;Symmetry Relations and Constant Density Theorem} 

\bigskip
\bigskip
We now restrict to the case of half-filling, so we put 
\beqn
\mu&=& 0 \pI 
\eeqn
We also assume the case of a bipartite lattice. That is, the lattice $\Gamma$ is a disjoint union 
\beqn
\Gamma&=& \GA\,\cup\,\GB 
\eeqn
such that the hopping matrix $\vep$ vanishes on $\,(\GA,\GA)\,$ or $\,(\GB,\GB)\,$ lattice sites, 
\beqn
\phantom{nmm} \vep_{ij}&=&0 \;\;\;\;\;\;{\rm if}\;\;\;\;(\,i\in \GA \,\land\, j\in \GA\,)\;\lor\; (\,i\in \GB \,\land\, j\in \GB\,)  \pS
\eeqn
which is the case in particular for nearest neighbor hopping on a cubic lattice. Before we state the next theorem, let us introduce some notation which will be used 
in the following. We define the $\G\times\G$ matrices 
\beqn
\chi^{\rm on}\;\;=\;\;(\,\chi^{\rm on}_{ij}\,)_{\atop i,j\in\Gamma}\;,\;\;\;\;\;\; \chi^{\rm off}\;\;=\;\;(\,\chi^{\rm off}_{ij}\,)_{\atop i,j\in\Gamma} \pI 
\eeqn
with matrix elements
\beqn
\chi^{\rm on}_{ij}&:=&\begin{cases} \,1 &  \;\;\;\;\;\;{\rm if}\;\;\;\;(\,i\in \GA \,\land\, j\in \GA\,)\;\lor\; (\,i\in \GB \,\land\, j\in \GB\,) \\ 
                                    \,0 &  \;\;\;\;\;\;{\rm if}\;\;\;\;(\,i\in \GA \,\land\, j\in \GB\,)\;\lor\; (\,i\in \GB \,\land\, j\in \GA\,) \end{cases}  \pI \nn \\ 
&&\phantom{I} \nn \\ 
\chi^{\rm off}_{ij}&:=&\begin{cases} \,0 &  \;\;\;\;\;\;{\rm if}\;\;\;\;(\,i\in \GA \,\land\, j\in \GA\,)\;\lor\; (\,i\in \GB \,\land\, j\in \GB\,) \\ 
                                     \,1 &  \;\;\;\;\;\;{\rm if}\;\;\;\;(\,i\in \GA \,\land\, j\in \GB\,)\;\lor\; (\,i\in \GB \,\land\, j\in \GA\,) \end{cases}  \pI 
\eeqn
such that 
\beqn
\chi^{\rm on}_{ij} \;+\;\chi^{\rm off}_{ij} &=&1 \pS 
\eeqn
for arbitrary lattice sites $i$ and $j$. Also, for arbitrary matrices $\chi=(\chi_{ij})$ and $M=(M_{ij})$, let us define an element-wise matrix multiplication,  
denoted with the $\ast$ operator, through
\beqn
\chi\ast M&:=&\bigl(\, \chi_{ij}\, M_{ij}\,\bigr)_{i,j\in\Gamma} \pI 
\eeqn
Then, for some arbitrary $\G\times \G$ matrix $M=(M_{ij})$, we define an on-part and an off-part through
\beqn
M^{\rm on}&:=&\chi^{\rm on} \;\ast\; M  \pS \nn \\ 
M^{\rm off}&:=&\chi^{\rm off} \;\ast\; M \lbeq{4.4n}
\eeqn
such that $\,M=M^{\rm on}+M^{\rm off}\,$. Then there is the following

\medskip
\bigskip
\bigskip
{\bf Theorem 4.1 (symmetry relations at half-filling $\bm{\mu=0}\,$ and $\bm{w_2=1}\,$):} Choose the Hubbard-Stratonovich parameters $\,w_2=1\,$, $\,w_1=w_3=0\,$ and 
$\,\eps_2=\eps_u=\sign\,u\,$. As in Theorem 3.2, we use the abbreviations 
\beqn
\rho\;\;:=\;\;G^{ab}&\in&\R^{2\G\times 2\G}  \pS \nn \\
F\;\;:=\;\;-\,i\,G^{aa} &\in&\R^{2\G\times 2\G}
\eeqn
Then, at half-filling $\,\mu=0\,$ and on a bipartite lattice $\Gamma=\GA\cup \GB$\,, we have the following pathwise properties, for arbitrary real $r,s$ and $u\,$:  
\beqn
\rho_{\su\su}^{\rm on}\;\;=\;\;\rho_{\sd\sd}^{\rm on}&=&0  \lbeq{n4.6} \\ 
\rho_{\su\sd}^{\rm off}\;\;=\;\;\rho_{\sd\su}^{\rm off}&=&0 \pS\\
F_{\su\su}^{\rm off}\;\;=\;\;F_{\sd\sd}^{\rm off}&=&0  \lbeq{n4.8}\\ 
F_{\su\sd}^{\rm on}\;\;=\;\;F_{\sd\su}^{\rm on}&=&0  \pS  \lbeq{n4.9}
\eeqn
In particular, for arbitrary lattice sites $j\in\Gamma$, 
\beqn
\rho_{\su\su}(j,j)\;\;=\;\;\rho_{\sd\sd}(j,j)\;\;=\;\; 0 \pS
\eeqn
and the density is constant [4], we have using (\req{2.34}) 
\beqn
\phantom{mm}\ts \la c_{j\sigma}^+c_{j\sigma}\ra \;\;=\;\;{1\over 2}\;+\;{1\over 2}\,\la \rho_{\sigma\sigma}(j,j)\,\ra 
  \;\;=\;\; {1\over 2} \;\;\;\;\;\;\;\;\forall \,j\in\Gamma  \pS 
\eeqn
Since $DF=0$ because of (\req{n4.9}), the SDE system (\req{3.10n}) of part (b) of Theorem 3.2 simplifies to 
\beqn
d\rho&=&\ts {1\over 2}\,\Bigl[\;(Id-F)\,dh\,(Id+\ti F) \;-\; \rho\,dh^T \rho  \;\Bigr] \pS   \nn \\
dF&=&\ts {1\over 2}\,\Bigl[\; [\,\rho\,dh^T(Id+ F)\,]^T \,-\,\rho\,dh^T(Id+ F)  \; \Bigr]  \lbeq{n4.12} \pI 
\eeqn
with $dh$ given by 
\beqn
dh&=&\ts -\,dt\,\bmat \vep & s-r \\  s+r  & \vep \emat  \;+\;{udt\over 2}\,\bmat  0 & {\rm diag}[\,\rho_{\sd\su}\,] \; \\ {\rm diag}[\,\rho_{\su\sd}\,] & 0  \; \emat
  \;-\;\sqrt{\sabs u\sabs}\;\bmat  0 & dy \\ \eps_u\,dy & 0 \emat  \pI\pI\pI \lbeq{4.17nn}
\eeqn

\medskip
\bigskip
{\bf Proof:} Follows by induction on time steps through straightforward calculations, see the appendix for the details. \;$\blacksquare$

\goodbreak
\medskip
\bigskip
\bigskip
{\bf Theorem 4.2 (SDEs at half-filling for $\bm{r=s=0}\,$):} Choose $\mu=0$ and zero external pairing and exchange terms $\,r=s=0\,$ in the Hamiltonian (\req{2.2}) 
or (\req{Hmain}).  
\begin{itemize}
\item[{\bf a)}] Choose $w_1=1$ and recall the notation (\req{4.4n}). Then:
\beqn
\rho_{\su\su}^{\rm on}\;\;=\;\;-\;\eps_u\,\rho_{\sd\sd}^{\rm on}\;,\;\;\;\;\;&& \rho_{\su\su}^{\rm off}\;\;=\;\;\rho_{\sd\sd}^{\rm off}\; \pS  \nn \\
 F_{\su\su}^{\rm off}\;\;=\;\;-\;\eps_u\,F_{\sd\sd}^{\rm off}\;,\;\;\;\;\;&& F_{\su\su}^{\rm on}\;\;=\;\;F_{\sd\sd}^{\rm on}\; \pS  \lbeq{4.18nn}
\eeqn
The only independent quantities are the real $\,\G\times \G\,$ matrices $\,\rho_{\su\su}\,$ and $\,F_{\su\su}\,$ with $\,\rho_{\su\su}^T\,=\,\rho_{\su\su}\,$ and $\,F_{\su\su}^T\,=\,-F_{\su\su}\,$. They are given by the SDE system
\beqn
d\rho_{\su\su}&=&\ts {1\over 2}\,\Bigl[\; (Id-F_{\su\su})\,dh_{\su\su}\,(Id+F_{\su\su})\;-\; 
   \rho_{\su\su}\,dh_{\su\su} \rho_{\su\su}    \;\Bigr] \pI \nn \\ 
dF_{\su\su}&=&\ts {1\over 2}\,\Bigl[\; (Id-F_{\su\su})\,dh_{\su\su}\,\rho_{\su\su} \;-\; \bigl[\,(Id-F_{\su\su})\,dh_{\su\su}\,\rho_{\su\su}\,\bigr]^T \; \Bigr]  \lbeq{4.16n}
\eeqn
with
\beqn
dh_{\su\su}&=& -\,\vep\,dt \;+\;\ts  {|u|dt \over 2} \,{\rm diag}[\,\rho_{\su\su}\,] \;-\; \sqrt{\sabs u\sabs}\; dx \lbeq{4.21n} 
\eeqn
The energy (\req{2.28}) simplifies to
\beqn
W(\rho)&=& Tr_{\mathbb C^\G}\bigl[\; \vep\,\rho_{\su\su}  \;\bigr]  \;-\; {\ts {|u|\over 4}}\,\summ_j\, \rho_{\su\su,jj}^2  \pS \lbeq{4.22n}
\eeqn

\item[{\bf b)}] Choose $w_2=1$. Then:
\beqn
\rho_{\sd\sd}\;\;=\;\;\rho_{\su\su}\;,&&\;\;\;\;F_{\sd\sd}\;\;=\;\;F_{\su\su}  \nn \\ 
\rho_{\sd\su}\;\;=\;\;\eps_u\,\rho_{\su\sd}\;,&&\;\;\;\;F_{\sd\su}\;\;=\;\;\eps_u\,F_{\su\sd}  \pS \lbeq{4.23n}
\eeqn
and with 
\beq
\rho\;\;=\;\;\bmat \rho_{\su\su} & \rho_{\su\sd} \\ \eps_u\,\rho_{\su\sd} & \rho_{\su\su} \emat,\;\;\;\; 
F\;\;=\;\;\bmat F_{\su\su} & F_{\su\sd} \\ \eps_u\,F_{\su\sd} & F_{\su\su} \emat,\;\;\;\; 
\ti F\;\;=\;\;\bmat F_{\su\su} & \eps_u\,F_{\su\sd} \\ F_{\su\sd} & F_{\su\su} \emat  \pI
\eeq
we have 
\beqn
d\rho&=&\ts {1\over 2}\,\Bigl[\;(Id-F)\,dh\,(Id+\ti F) \;-\; \rho\,dh^T \rho  \;\Bigr] \pS \nn \\
dF&=&\ts {1\over 2}\,\Bigl[\; (Id- F)\,dh\,\rho^T \,-\,\bigl[\,(Id- F)\,dh\,\rho^T\,\bigr]^T \; \Bigr]  \lbeq{4.20n} \pI 
\eeqn
with 
\beqn
dh&=&\ts -\,dt\,\bmat \vep & 0 \\  0  & \vep \emat  \;+\;{|u|dt\over 2}\,\bmat  0 & {\rm diag}[\,\rho_{\su\sd}\,] \; \\ \eps_u\,{\rm diag}[\,\rho_{\su\sd}\,] & 0  \; \emat
  \;-\;\sqrt{\sabs u\sabs}\;\bmat  0 & dy \\ \eps_u\,dy & 0 \emat  \pI\pI \lbeq{4.24nn}
\eeqn

\medskip
The energy (\req{2.28}) becomes (recall $\,\rho_{\su\su,jj}=0\,$ for $w_2=1\,$)
\beqn
W(\rho)&=& Tr_{\mathbb C^\G}\bigl[\, \vep\,\rho_{\su\su} \,\bigr]  \;-\; {\ts { |u|\over 4}}\,\summ_j\, \rho_{\su\sd,jj}^2 \pI \lbeq{4.25nn}
\eeqn
Thus, in this case we have $\rho_{\su\su},F_{\su\su}$ and $\rho_{\su\sd},F_{\su\sd}$ as independent quantities, but, because of Theorem 4.1, half of the 
matrix elements are actually zero. 
\end{itemize}

\medskip
\bigskip
{\bf Proof:} Follows by induction on time steps through straightforward calculations, see the appendix for the details. \;$\blacksquare$ 

\medskip
\bigskip
\bigskip
{\bf Remark:} It is well known (see for example appendix C of [5] or section 9.3.3 of [6]\,) that if the interacting part of the Hamiltonian is chosen to be written 
in the `symmetrized' form (\req{2.3}) with the one half's subtracted off in the interacting part, then at half filling on a bipartite lattice the expectation value 
of the energy depends only on the absolute value of $u$. 
In the $w_1=1$ representation, this is immediately obvious since we have $\,W(\rho(u))\,=\,W(\rho(-u))\,$ pathwise for every Monte Carlo path, the expression 
(\req{4.22n}) for the energy and also the SDE for $\rho_{\su\su}$ with the $dh$ given by (\req{4.21n}) depends only on the absolute value of $u$. 
In the $w_2=1$ representation, the SDE system is $\,2\G\times 2\G$, the sign $\eps_u$ enters through (\req{4.24nn}), and it is not obvious whether the energy is pathwise 
independent of $\eps_u$. Actually it is not, only after taking the Monte Carlo average we recover the result $\,\la H(u)\ra_\beta\,=\,\la H(-u)\ra_\beta\,$.

\goodbreak
\bigskip
\bigskip
\bigskip
{\bf 4.2 \;Numerical Test: Main Theorem vs.~Exact Diagonalization} 

\bigskip
\bigskip
Let's make a numerical check for the formulae obtained in Theorem 4.2. Since the purpose here is validation only, we stick to small system size. 
We calculate the grand canonical expectation value of the energy on a $3\times 2$ lattice (section 4.4 below has calculations on a $12\times 12$ lattice).  
We put the external pairing and exchange terms to zero, $\,r=s=0\,$, 
and calculate at half filling $\mu=0$ such that Theorem 4.2 applies. We use no boundary conditions, that is, a particle on 
lattice site $(1,1)$ can only hop to $(1,2)$ or $(2,1)$, but nowhere else. The Hamiltonian is, with the notations of section 2, 
\beq
H&=&H_{\rm tot}\;-\;Tr[\vep-\mu] \;\;=\;\; {\ts {i\over 2}}\, \bigl\{\,a_\su(\vep-\mu)\,b_\su\,+\,a_\sd(\vep-\mu)\,b_\sd \,\bigr\} \;+\;
    {\ts {u\over 4}}\,\summ_j \,a_{j\su}\,a_{j\sd}\,b_{j\su}\,b_{j\sd} \nn  \pS \\ 
&=&c^+_\su(\vep-\mu)\,c_\su \,+\,c^+_\sd(\vep-\mu)\,c_\sd \;+\;  u\,\summ_j \,(\,c_{j\su}^+\,c_{j\su}-1/2\,)\,(\,c_{j\sd}^+\,c_{j\sd}-1/2\,) 
  \;-\;Tr[\vep-\mu] \pS 
\eeq
At half filling $\mu=0$, this reduces to 
\beqn
H&=& {\ts {i\over 2}}\, \summ_{i,j}\,\vep_{ij}\,(\,a_{i\su} \,b_{j\su}\,+\,a_{i\sd}\,b_{j\sd} \,) \;+\;
    {\ts {u\over 4}}\,\summ_j \,a_{j\su}\,a_{j\sd}\,b_{j\su}\,b_{j\sd} \nn  \pS \\ 
&=&\summ_{i,j}\,\vep_{ij}\,(\,c^+_{i\su}\,c_{j\su} \,+\,c^+_{i\sd}\,c_{j\sd}\,) \;+\;  u\,\summ_j \,(\,c_{j\su}^+\,c_{j\su}-1/2\,)\,(\,c_{j\sd}^+\,c_{j\sd}-1/2\,)  \pS \nn \\ 
&=&\summ_{i,j}\,\vep_{ij}\,(\,c^+_{i\su}\,c_{j\su} \,+\,c^+_{i\sd}\,c_{j\sd}\,) 
  \;+\;  u\,\summ_j \,c_{j\su}^+\,c_{j\su}\,c_{j\sd}^+\,c_{j\sd} \;-\; \hat N\,u/2 \;+\; \G\,u/4  \pS  \lbeq{4.27nn}
\eeqn
with $\hat N$ being the number operator. If we write the decomposition of the grand canonical hamiltonian $H$ in canonical $H_N$'s as  
\beqn
H&=&\mathop{\oplus}_{\phantom{\ds i}N=0\phantom{\ds i}}^{\phantom{\ds|}2\G\phantom{\ds|}} H_N \pI 
\eeqn
and if we let $\lambda_{i,N}$ denote the eigenvalues of the canonical $H_N$'s, then the grand canonical expectation from exact diagonalization is 
simply 
\medskip
\beqn
\la H\ra_\beta\;\;=\;\; { Tr_\cF[\, H\,e^{\,-\,\beta H} \,] \over Tr_\cF[\, e^{\,-\,\beta H} \,] }  &=& 
  { \phantom{\Bigl|}\ds\summ_{N=0}^{2\G} \ds\summ_i \; \lambda_{i,N}\; e^{\,-\,\beta\,\lambda_{i,N}} \phantom{\Bigl|} \over  
   \phantom{\Bigl|} \ds\summ_{N=0}^{2\G} \ds\summ_i \; e^{\,-\,\beta\,\lambda_{i,N}} \phantom{\Bigl|} } \;\;=:\;\; \la H\ra_\beta^{\rm ED} \pI 
\eeqn
On the other hand, according to (\req{n2.37}) from the Main Theorem, we have with $\beta=k\,dt\,$:
\beq
\la H\ra_\beta\;\;=\;\; { Tr_\cF[\, H\,e^{\,-\,\beta H} \,] \over Tr_\cF[\, e^{\,-\,\beta H} \,] } &=&
  { \phantom{\Bigl|} \int_{\R^{3k\G}} \,  W(G_\beta) \; e^{\,-\,\int_0^\beta\, W(G_t)\;dt }\; dP_k(\phi,\xi,\theta) \phantom{\Bigl|}\over 
      \phantom{\Bigl|}\int_{\R^{3k\G}} \,   e^{\,-\,\int_0^\beta\, W(G_t)\;dt }\; dP_k(\phi,\xi,\theta) \phantom{\Bigl|} }  \;\;=:\;\; \la\, W\,\ra_\beta \pI 
\eeq
In the $w_1=1$ representation, this reduces to 
\beqn
\la \,W\,\ra_{\beta}^{\!\!{\atop{w_1=1\atop}}}&=& {\phantom{\Bigl|} \int_{\R^{k\G}} \,  W(\rho_{t_k}) \; e^{\,-\,\sum_{\ell=0}^{k-1}\, W(\rho_{t_\ell})\;dt }\; dP_k(\phi)
    \phantom{\Bigl|}\over  \phantom{\Bigl|}\int_{\R^{k\G}} \,   e^{\,-\,\sum_{\ell=0}^{k-1}\, W(\rho_{t_\ell})\;dt }\; dP_k(\phi) \phantom{\Bigl|} } \lbeq{4.29n}
\eeqn
with 
\beqn
W(\rho)&=& \summ_{i,j}\, \vep_{ij}\,\rho_{\su\su,ji}  \;-\; {\ts {|u|\over 4}}\,\summ_j\, \rho_{\su\su,jj}^2  \lbeq{4.30n}
\eeqn
and $\rho_{\su\su}$ given by the SDE system (\req{4.16n}). For $w_2=1$, we have to calculate 
\beqn
\la \,W\,\ra_{\beta}^{\!\!{\atop{w_2=1\atop}}}&=&
   { \phantom{\Bigl|} \int_{\R^{k\G}} \,  W(\rho_{t_k}) \; e^{\,-\,\sum_{\ell=0}^{k-1}\, W(\rho_{t_\ell})\;dt }\; dP_k(\xi) \phantom{\Bigl|}
  \over  \phantom{\Bigl|}\int_{\R^{k\G}} \,   e^{\,-\,\sum_{\ell=0}^{k-1}\, W(\rho_{t_\ell})\;dt }\; dP_k(\xi) \phantom{\Bigl|} }   \lbeq{4.31n}
\eeqn
with 
\beqn
W(\rho)&=& \summ_{i,j}\, \vep_{ij}\,\rho_{\su\su,ji}  \;-\; {\ts {|u|\over 4}}\,\summ_j\, \rho_{\su\sd,jj}^2  \lbeq{4.32n}
\eeqn
and $\rho_{\su\su}$ and $\rho_{\su\sd}$ given by the SDE system (\req{4.20n}). Recall that the standard normal random numbers $\,\phi_{j,\ell}\,$ 
and $\,\xi_{j,\ell}\,$ were introduced in (\req{n2.22}) which led to the integration measure $dP_k$ given by (\req{2.21}). 
Then plain Monte Carlo with 100'000 simulations produced the following 
picture as shown on top of the next page, displayed are the grand canonical energies per lattice site,
\beq
\la H\ra_\beta^{\rm ED}/\,\G \;&& {\rm in\;green\; color},  \pS \\
\la \,W\,\ra_{\beta}^{\!\!{\atop{w_1=1\atop}}}/\,\G\;,\;\;\;\la \,W\,\ra_{\beta}^{\!\!{\atop{w_2=1\atop}}} /\,\G \;&& {\rm in\;blue\; color}  \pS\pS\pS
\eeq
with nearest neighbor hopping parameter $\vep=-1$ and couplings $u=2$ and $u=4$. At the bottom, we also show these quantities in the atomistic limit $\vep=0$ for larger 
couplings $u=4$ and $u=8$ and for larger values of $\beta\in[0,20]$ and with just 100 (not 100'000 as in the first picture) Monte Carlo simulations:

\bigskip
\centerline{\includegraphics[width=12.5cm]{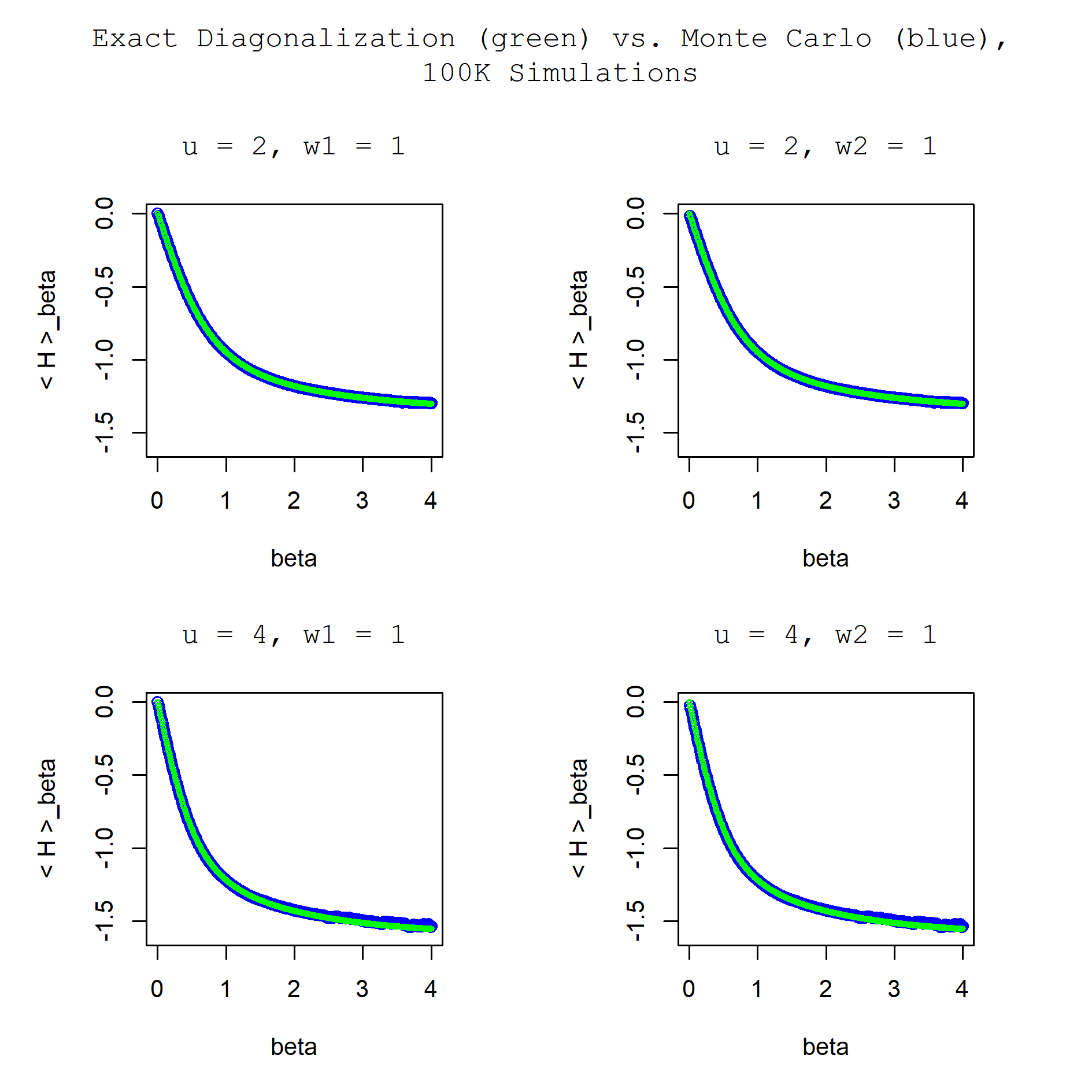}}

\bigskip

\bigskip
\centerline{\includegraphics[width=12.5cm]{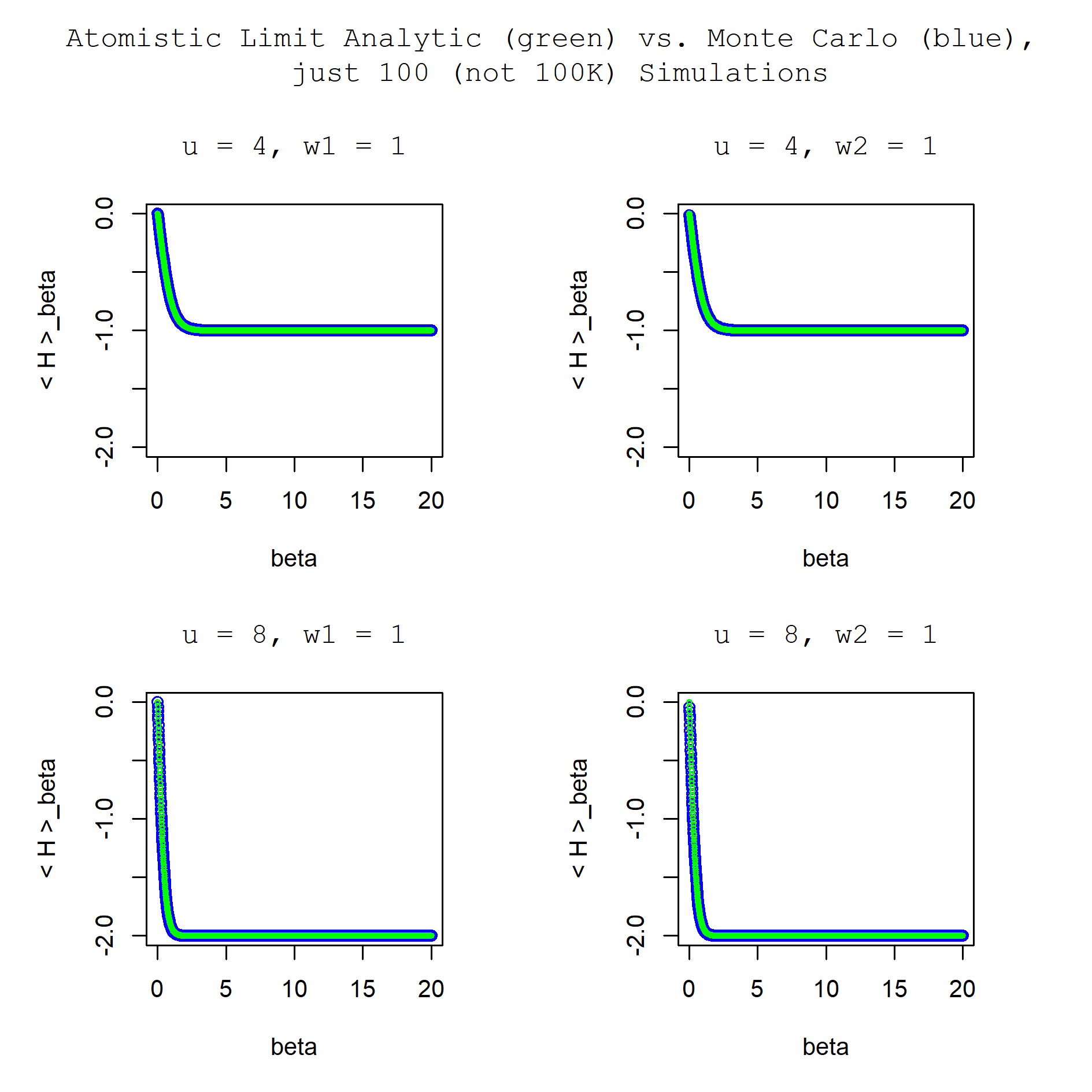}}

\bigskip

For nonzero hopping $\vep=-1$, the presence of the Boltzmann weight $\,e^{-\int_0^\beta W dt}\;$ makes relevant Monte Carlo paths sufficiently sparse such that 
already for values of $\,u \gtrsim 4\,$ and $\,\beta \gtrsim 2\,$ the plain Monte Carlo convergence starts to deteriorate. This is not due to the fact that 
the relevant integration region 
is proportional to $\sqrt{\beta}$ or $\beta$ and moves to infinity, exactly this problem has been taken care of by the Girsanov transformation. Rather, 
this is due to the fact that, say, inside the cube $\,[-5,+5]^{\,k\G}$, the relevant contributions $\,\{\,\phi_{j,\ell}\,\}$ or $\,\{\,\xi_{j,\ell}\,\}$ 
become thinner and thinner. Of course, this is a typical problem in the Monte Carlo area and some kind of accept-and-reject algorithm could be put in place. 
\,A natural candidate could probably be the so-called preconditioned Crank-Nicolson or pCN method, in [7] a very beautiful and compact motivation of this method 
has been given. Further in depth analysis is provided in [8]. Here, we limit ourselves to plain Monte Carlo and leave this as future work. 

\bigskip
As can be seen in the second picture at the bottom of the previous page, the atomistic limit $\vep=0$ has actually very fast Monte Carlo convergence, 
basically for arbitrary values of $u$ and $\beta$. Already with just 100 (not 100'000 as in the first picture) 
MC simulations, one gets very nice plots. The analytical expression is found to be 
\beqn
\la H\ra_\beta^{\,\vep=0}/\,\G&=&\ts -\,{u\over 4}\,\tanh\bigl[\,\beta{u\over 4}\,\bigr] 
  \;\;=\;\;-\,{|u|\over 4}\,\tanh\bigl[\,\beta{|u|\over 4}\,\bigr] \pS 
\eeqn
and is displayed in green color, Monte Carlo in blue.

\bigskip
\bigskip
\bigskip

{\bf 4.3 \;Spin-Spin and Pair-Pair Correlation} 

\bigskip
\bigskip
These correlation functions have been extensively studied in the determinant quantum Monte Carlo community [9-20] 
and also in the mathematical physics community [5,6], [21-28]. We use the definitions of Scalettar [17] and Tasaki [6] and consider the functions, 
for lattice sites $i,j\in\Gamma$, 
\beqn
C_{\rm spin}(i,j)&:=&\la\,(n_{i\su}-n_{i\sd})\,(n_{j\su}-n_{j\sd})\,\ra_\beta  \pS \\ 
C_{\rm pair}(i,j)&:=&\la\, c_{i\su}^+\,c_{i\sd}^+\,c_{j\sd}\,c_{j\su} \,\ra_\beta \pS
\eeqn
with $\,n_{j\sigma}=c^+_{j\sigma}c_{\sigma}\,$ and grand canonical expectation 
\beq
\la \,Q\,\ra_\beta &:=&  Tr_\cF[\,Q\,e^{-\beta H}\,] \;\bigr/\; Tr_\cF[\,e^{-\beta H}\,]  \pI
\eeq
Then we can formulate the following two theorems:

\bigskip
\bigskip
{\bf Theorem 4.3 (spin-spin correlation):} For the half filled Hubbard model on a bipartite lattice $\,\Gamma=\GA\cup \GB\,$ the following statements hold 
with $\eps_u={\rm sign}\,u$ being the sign of the coupling $u$\,:
\begin{itemize}
\item[{\bf a)}] Consider the $w_1=1$ representation with zero external pairing and exchange terms $\,r=s=0\,$ such that part (a) of Theorem 4.2 applies. 
Also recall the notations of Theorem 4.1. Then, with $\rho_{\sigma\tau,ij}^{2}:=[\rho_{\sigma\tau,ij}]^2$ and $F_{\sigma\tau,ij}^{2}:=[F_{\sigma\tau,ij}]^2$, 
\beqn
C_{\rm spin}(i,j)&\buildrel i\ne j\over =&\ts \,{1\over 2}\;\Bigl\la\, F_{\su\su,ij}^2 \,-\,\rho_{\su\su,ij}^2 
  \,+\,  {1\over 2}\,(1+\eps_u)^2\,\rho_{\su\su,ii}\,\rho_{\su\su,jj}  \,\Bigr\ra_\beta \pI   \lbeq{4.36nn} \\ 
C_{\rm spin}(j,j)&=& \ts {1\over 2}\,\bigl\la\, 1  \,+\,\eps_u\,\rho_{\su\su,jj}^2 \,\bigr\ra_\beta   \pS \nn
\eeqn
\item[{\bf b)}] Consider the $w_2=1$ representation with zero external pairing and exchange terms $\,r=s=0\,$ such that part (b) of Theorem 4.2 applies. Then  
\beqn
C_{\rm spin}(i,j)&\buildrel i\ne j\over =&\ts \,{1\over 2}\;\Bigl\la\, F_{\su\su,ij}^2\,-\,\rho_{\su\su,ij}^2
 \ts \,-\,\eps_u\,F_{\su\sd,ij}^2 \,+\,\eps_u\,\rho_{\su\sd,ij}^2 \,\Bigr\ra_\beta \pI   \lbeq{4.37nn}\\ 
C_{\rm spin}(j,j)&=&\ts {1\over 2}\;\bigl\la\, 1  \,+\,\eps_u\,\rho_{\su\sd,jj}^2 \,\bigr\ra_\beta \pS \nn
\eeqn
In particular, for repulsive coupling $u>0$, the spin-spin correlations have to have antiferromagnetic signs, with the notations of Theorem 4.1 we have
\beqn
C_{\rm spin}^{\,\rm on}(i,j)&=&\ts +\,{1\over 2}\;\bigl\la\, F_{\su\su,ij}^2 \,+\,\rho_{\su\sd,ij}^2 \,\bigr\ra_\beta \;\;\ge \;\; 0 \pI   \\
C_{\rm spin}^{\,\rm off}(i,j)&=&\ts -\,{1\over 2}\,\bigl\la\, \rho_{\su\su,ij}^2 \,+\,F_{\su\sd,ij}^2 \,\bigr\ra_\beta \;\;\le \;\;0  \pS \nn
\eeqn
for arbitrary lattice sites $i\ne j$\,. 
\smallskip
\end{itemize}
Here, all the expectations on the $F_{\sigma\tau,ij}$ and $\rho_{\sigma\tau,ij}$ on the right hand sides are given by 
\beqn
\bigl\la\;\cdot\;\bigr\ra_\beta&:=&\ts \int_{\R^{k\G}} \;\cdot\;\; e^{\,-\,\int_0^\beta  W(\rho_t)\,dt} \,dP_k \;\bigr/\; 
  \int_{\R^{k\G}} \, e^{\,-\,\int_0^\beta  W(\rho_t)\,dt} \,dP_k  \pS \lbeq{4.39nn}
\eeqn  

\bigskip
{\bf Proof:} Straightforward calculations, see the appendix for the collection of the individual terms. \;$\blacksquare$

\medskip
\bigskip
\bigskip
{\bf Theorem 4.4 (pair-pair correlation):} For the half filled Hubbard model on a bipartite lattice $\,\Gamma=\GA\cup \GB\,$ the following statements hold 
with $\eps_u={\rm sign}\,u$ being the sign of the coupling $u$\,:
\begin{itemize}
\item[{\bf a)}] Consider the $w_1=1$ representation with zero external pairing and exchange terms $r=s=0$ such that part (a) of Theorem 4.2 applies. 
Also recall the notations of Theorem 4.1. Then   
\beqn 
C_{\rm pair}(i,j)&\buildrel i\ne j\over=& \ts {1\over 4}\, (\chi_{ij}^{\rm on}-\eps_u \chi_{ij}^{\rm off})\;
  \Bigl\la\,(F_{\su\su,ij}+\rho_{\su\su,ij})(F_{\su\su,ij}-\,\eps_u \,\rho_{\su\su,ij}) \, \Bigr\ra_\beta \pI   \lbeq{4.40nn}\\   
C_{\rm pair}(j,j)&=& \ts {1\over 4}\,\bigl\la\;(\, 1\,+\,\rho_{\su\su,jj}\,)(\,1\,-\,\eps_u\, \rho_{\su\su,jj}\,)\;\bigr\ra_\beta  \nn  \pS 
\eeqn
In particular, for attractive coupling $u<0$, the pair-pair correlation is nonnegative, for $i\ne j$ we have 
\beqn
C_{\rm pair}(i,j)&=& \ts {1\over 4}\,\bigl\la\, (F_{\su\su,ij}+\rho_{\su\su,ij})^2 \, \bigr\ra_\beta \;\;\ge \;\; 0 \pS  
\eeqn
\item[{\bf b)}] Consider the $w_2=1$ representation with zero external pairing and exchange terms $r=s=0\,$ such that part (b) of Theorem 4.2 applies. Then  
\beqn
C_{\rm pair}(i,j)&\buildrel i\ne j\over =&\ts {1\over 4}\,\Bigl\la\,F_{\su\su,ij}^2\, + \,\rho_{\su\su,ij}^2 \,-\,{1+\eps_u\over 2}\,\bigl(\,F_{\su\sd,ij}^2
  \,+\,\rho_{\su\sd,ij}^2\,\bigr) \,+\,{1-\eps_u\over 2}\,\rho_{\su\sd,ii}\,\rho_{\su\sd,jj}  \;\Bigr\ra_\beta \pI\pI  \lbeq{4.42nn} \\ 
C_{\rm pair}(j,j)&=& \ts {1\over 4}\,\bigl\la \,1\,-\, \eps_u\,\rho_{\su\sd,jj}^2 \,\bigr\ra_\beta  \pS  \nn
\eeqn
\end{itemize} 
Again, all the expectations on the $F_{\sigma\tau,ij}$ and $\rho_{\sigma\tau,ij}$ on the right hand sides are given by (\req{4.39nn})\,.

\goodbreak

\medskip
\bigskip
{\bf Proof:} Straightforward calculations, see the appendix for the collection of the individual terms. \;$\blacksquare$

\bigskip
\bigskip

{\bf Numerical Test} 

\bigskip
As another consistency check, let us confirm that the $w_1=1$ and the $w_2=1$ representations do in fact produce the same expectations as long as we can 
ensure Monte Carlo convergence. Again, we stick to plain Monte Carlo, so we consider smaller values of $u$ and $\beta$. We calculate the spin-spin correlation for 
repulsive couplings $\,u=+2\,$ and $\,u=+4\,$ and the pair-pair correlation for attractive couplings $\,u=-2\,$ and $\,u=-4\,$ as a function of the inverse tempera\-ture $\beta$, 
on a $4\times 4$ lattice with periodic boundary conditions. These correlations are more sensitive than the energy 
with respect to Monte Carlo noise, so we used 10 million simulations or Monte Carlo paths or SDE solutions (with $\beta=t_k=k\,dt\,$)
\beq
\Bigl\{\;\, [\;\rho^{(i)}(t_\ell)\;]_{\ell=0}^k\;\, , \;\, [\;F^{(i)}(t_\ell)\;]_{\ell=0}^k\;\, \Bigr\}_{i=1}^{N_{\rm MC}=10^7} 
\eeq
and calculated $\,C_{\rm spin}\,$ and $\,C_{\rm pair}\,$ through the formulae \,(\req{4.36nn},\req{4.37nn})\, and \,(\req{4.40nn},\req{4.42nn})\, of Theorem 4.3 and 4.4 above. 
For the SDEs given by (\req{4.16n}) and (\req{4.20n}) of Theorem 4.2, a simple Euler discretization with $dt=0.01$ has been used, 
no higher order scheme had been put in place. 

\medskip
Then we plotted all 15 spin-spin and pair-pair correlations
\beq
C_{\rm spin}\bigl(\;(1,1)\,,\,(j_x,j_y)\;\bigr)\,(t_\ell)\;,\;\;\;\;\; C_{\rm pair}\bigl(\;(1,1)\,,\,(j_x,j_y)\;\bigr)\,(t_\ell)  \pM 
\eeq
for
\beq
j\;\;=\;\;(j_x,j_y)&\in& \bigl\{\,1\,,\,2\,,\,3\,,\,4\,\bigr\}\;\times\; \bigl\{\,1\,,\,2\,,\,3\,,\,4\,\bigr\}  \pM
\eeq
with $j\ne(1,1)$ which has been used as the reference site, with periodic boundary conditions, in the same diagram 
as a function of the inverse temperature $\beta$. That is, per diagram we have 15 curves. Thereby correlations belonging to the same bipartite sublattice 
were displayed  in similar colors, either all bluish or all reddish, we used the following coloring scheme:
\beq
C_{\rm spin/pair}\bigl(\;(1,1)\,,\,(j_x,j_y)\;\bigr)\;\;\;{\rm with}\;\;\; j_x+j_y\;\;{\rm even}\; &\Rightarrow&\;\;\;{\rm colors\; red\; to\; magenta} \pS \\
C_{\rm spin/pair}\bigl(\;(1,1)\,,\,(j_x,j_y)\;\bigr)\;\;\;{\rm with}\;\;\; j_x+j_y\;\;{\rm odd}\phantom{i}\, &\Rightarrow&\;\;\;{\rm colors\; blue\; to\; cyan} \pS 
\eeq
On the next page, those blue curves which are largest in absolute value belong to the four nearest neighbor correlations at sites (1,2), (1,4) and (2,1) and (4,1)\,:

\vfill
\newpage

\bigskip
\centerline{\includegraphics[width=12.5cm]{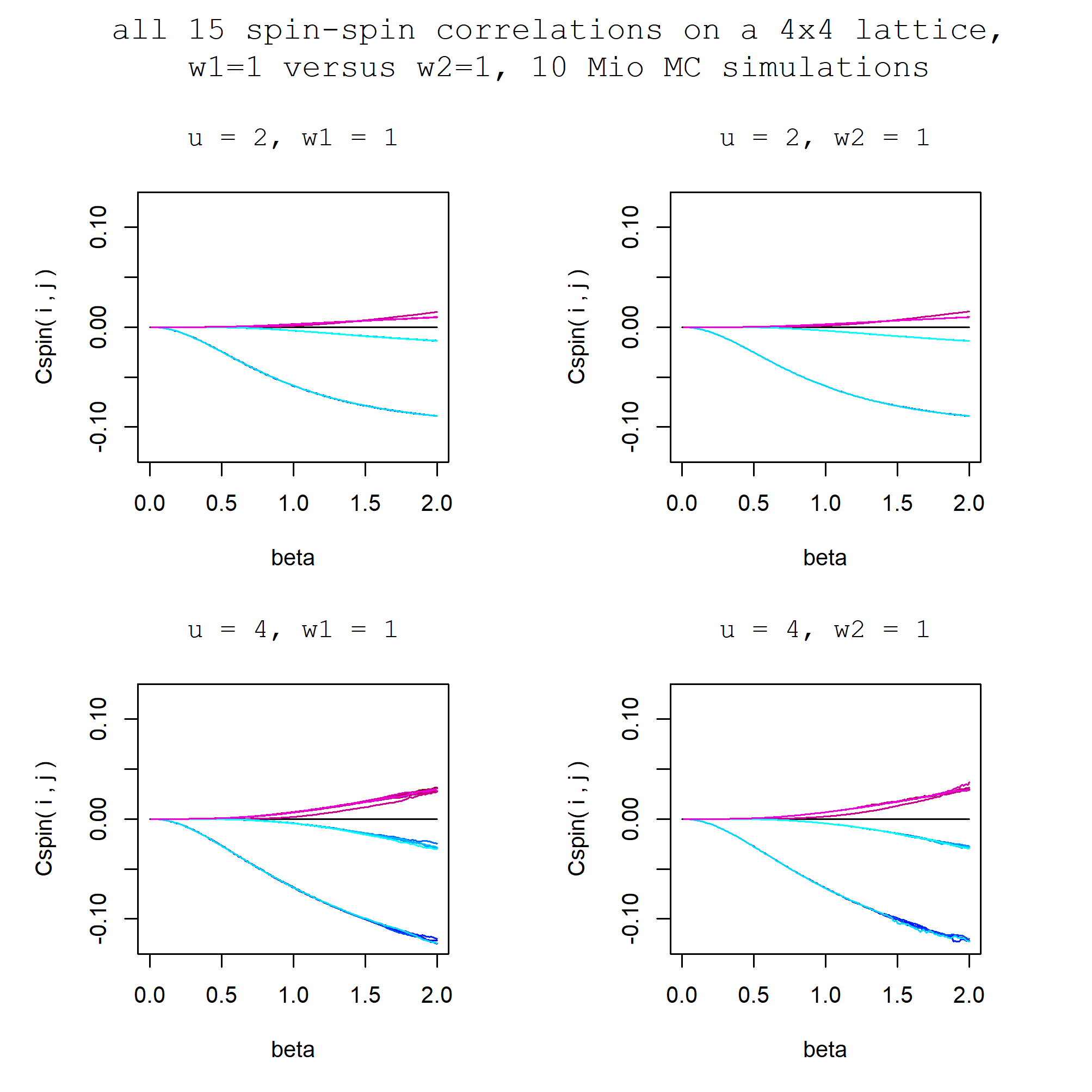}}

\bigskip

\bigskip
\centerline{\includegraphics[width=12.5cm]{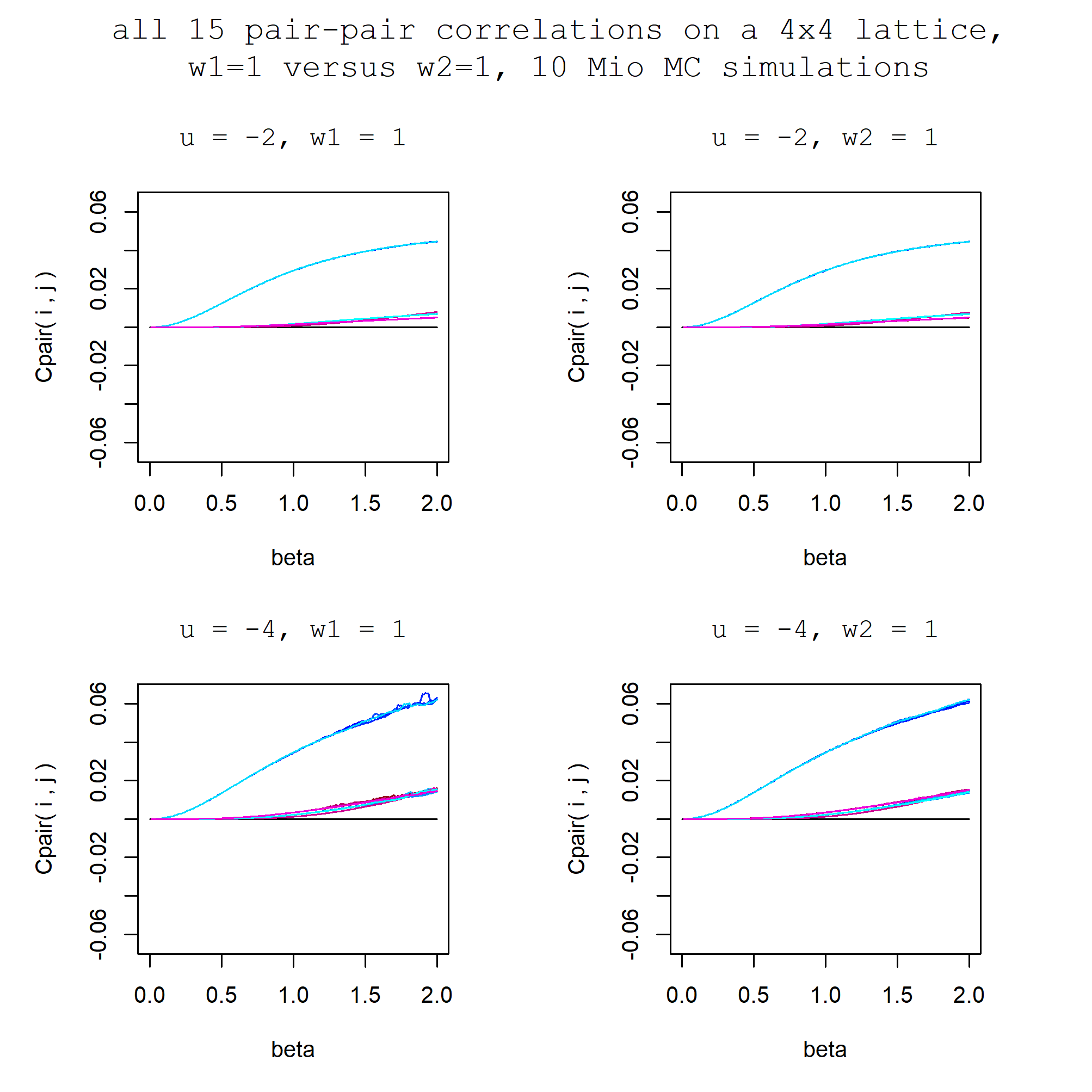}}

\bigskip

\newpage

\bigskip
\bigskip
\bigskip
{\bf 4.4 \;Temperature Zero Limit}

\bigskip
\bigskip
Recall the formulae (\req{4.29n}-\req{4.32n}) from section 4.2. In the $w_1=1$ representation, the energy is given by 
\beq
\la \,H\,\ra_\beta&=& {\phantom{\Bigl|} \int_{\R^{k\G}} \,  W(\rho_{t_k}) \; e^{\,-\,\sum_{\ell=0}^{k-1}\, W(\rho_{t_\ell})\;dt }\;
  {\ds \pro_{\ell=1}^k\pro_{j\in\Gamma}}\; e^{\,-\,{\phi_{j,\ell}^2\over 2}}\; d\phi_{j,\ell} 
    \phantom{\Bigl|}\over  \phantom{\Bigl|}\int_{\R^{k\G}} \,   e^{\,-\,\sum_{\ell=0}^{k-1}\, W(\rho_{t_\ell})\;dt }\; 
   {\ds \pro_{\ell=1}^k\pro_{j\in\Gamma}}\; e^{\,-\,{\phi_{j,\ell}^2\over 2}}\; d\phi_{j,\ell}  \phantom{\Bigl|} }  \pI
\eeq
with $W(\rho_t)$ given by (\req{4.30n}) and $\rho_{\su\su}$ given by the SDE system (\req{4.16n}), 
\beq
d\rho_{\su\su}&=&\ts {1\over 2}\,\Bigl[\; (Id-F_{\su\su})\,dh_{\su\su}\,(Id+F_{\su\su})\;-\; 
   \rho_{\su\su}\,dh_{\su\su} \,\rho_{\su\su}    \;\Bigr] \pI \nn \\ 
dF_{\su\su}&=&\ts {1\over 2}\,\Bigl[\; (Id-F_{\su\su})\,dh_{\su\su}\,\rho_{\su\su} \;-\; \bigl[\,(Id-F_{\su\su})\,dh_{\su\su}\,\rho_{\su\su}\,\bigr]^T \; \Bigr] 
\eeq
with
\beq
dh_{\su\su,\,t_\ell}&=& -\,\vep\,dt \;+\;\ts  {|u|dt \over 2} \,{\rm diag}[\,\rho_{\su\su,\,t_{\ell-1}}] 
  \;-\; \sqrt{\sabs u\sabs\,dt}\; (\,\delta_{ij}\,\phi_{j,\ell}\,)_{i,j\in\Gamma} \pS
\eeq
Let's make the substitution of variables
\beqn
\phi_{j,\ell}&=:& \sqrt{dt}\;\ti \phi_{j,\ell} \pS 
\eeqn
We obtain, renaming the $\ti\phi$ variables by $\phi$ again,  
\beqn
\la \,H\,\ra_\beta &=& {\phantom{\Bigl|} \int_{\R^{k\G}} \,  W(\rho_{t_k}) \; e^{\,-\,\beta\,V(\phi) }\; {\ds  \pro_{j,\ell}} \; d\phi_{j,\ell} 
    \phantom{\Bigl|}\over  \phantom{\Bigl|}\int_{\R^{k\G}} \,   e^{\,-\,\beta\,V(\phi) }\; {\ds  \pro_{j,\ell}} \; d\phi_{j,\ell}   \phantom{\Bigl|} } \pI
\eeqn
with
\beqn
V(\phi)&=& {\ts {1\over \beta}}\,\summ_{\ell=1}^k\;\Bigl[\; {\ts{1\over 2}}\summ_j\, \phi_{j,\ell}^2 \;+\; W(\rho_{t_{\ell-1}})\; \Bigr] \; dt  \pI
\eeqn
and $\rho_{t}$ given by the ODE system
\beqn
d\rho_{\su\su}&=&\ts {1\over 2}\,\Bigl[\; (Id-F_{\su\su})\,h_{\su\su}(\phi)\,(Id+F_{\su\su})\;-\; 
   \rho_{\su\su}\,h_{\su\su}(\phi)\, \rho_{\su\su}    \;\Bigr]\; dt \pI \nn \\ 
dF_{\su\su}&=&\ts {1\over 2}\,\Bigl[\; (Id-F_{\su\su})\,h_{\su\su}(\phi)\,\rho_{\su\su}
   \;-\; \bigl[\,(Id-F_{\su\su})\,h_{\su\su}(\phi)\,\rho_{\su\su}\,\bigr]^T \; \Bigr] \;dt \lbeq{4.37n}
\eeqn
with, omitting the time (or inverse temperature) indices, 
\beqn
h_{\su\su}(\phi)&:=& -\,\vep \;+\;\ts  {|u| \over 2} \,{\rm diag}[\,\rho_{\su\su}\,] \;-\;  \sqrt{\sabs u\sabs }\; (\,\delta_{ij}\,\phi_j\,)_{i,j\in\Gamma} \pI \lbeq{5.5n}
\eeqn
Thus, one may speculate that ground state properties could be characterized by the minimizers of
\smallskip
\beqn
V_0(\phi)&:=&\lim_{\beta\to\infty}\, {\ts {1\over \beta}}\,{\ts \int_0^\beta}\;\Bigl[\; {\ts{1\over 2}}\summ_j\, [\phi_j(t)]^2 \;+\; W(\rho_t)\; \Bigr] \; dt  \pS \lbeq{5.6n}
\eeqn
That is, 
\beqn
\lim_{\beta\to\infty} \,\la\,H\,\ra_\beta &\buildrel ?\over =&
  \lim_{\beta\to\infty} \,\bigl\la\, W\bigl(\rho_\beta(\phi_{\rm min})\,\bigr) \,\bigr\ra_{\phi_{\rm min}\;{\rm minimizer\; of\;} V_0(\phi)} \pS  \lbeq{4.40n}
\eeqn
with
\beqn
W(\rho_t)&=& \summ_{i,j}\, \vep_{ij}\,\rho_{\su\su,ji}(t)  \;-\; {\ts {|u|\over 4}}\,\summ_j\, [\,\rho_{\su\su,jj}(t)\,]^2  \pI
\eeqn
and $\,\rho_t=\rho_t(\phi)\,$ being the solution of the ODE system (\req{4.37n}). 

\bigskip
We can apply the same reasoning to the $w_2=1$ representation. In that case, we get 
\beqn
V_0(\xi)&:=&\lim_{\beta\to\infty}\, {\ts {1\over \beta}}\,{\ts \int_0^\beta}\;\Bigl[\; {\ts{1\over 2}}\summ_j\, [\xi_j(t)]^2 \;+\; W(\rho_t)\; \Bigr] \; dt  \pI \\ 
W(\rho_t)&=& \summ_{i,j}\, \vep_{ij}\,\rho_{\su\su,ji}(t)  \;-\; {\ts {|u|\over 4}}\,\summ_j\, [\,\rho_{\su\sd,jj}(t)\,]^2  \pI
\eeqn
with $\,\rho_t=\rho_t(\xi)\,$ given by the ODE system 
\beqn
{d\rho\over dt} &=&\ts {1\over 2}\,\Bigl[\;(Id-F)\,h(\xi)\,(Id+\ti F) \;-\; \rho\,h^T(\xi)\, \rho  \;\Bigr] \pI \nn \\
{dF\over dt}&=&\ts {1\over 2}\,\Bigl[\; (Id- F)\,h(\xi)\,\rho^T \,-\,\bigl[\,(Id- F)\,h(\xi)\,\rho^T\,\bigr]^T \; \Bigr]   \pI \lbeq{5.11n}
\eeqn
with 
\beqn
h(\xi)&:=&\ts -\,\bmat \vep & 0 \\  0  & \vep \emat  \;+\;{|u|\over 2}\,\bmat  0 & {\rm diag}[\,\rho_{\su\sd}\,] \; \\ \eps_u\,{\rm diag}[\,\rho_{\su\sd}\,] & 0  \; \emat
  \;-\;\sqrt{\sabs u\sabs}\;\bmat  0 & \delta_{ij}\,\xi_j \\ \eps_u\,\delta_{ij}\,\xi_j & 0 \emat  \pI\pI  \lbeq{5.12n}
\eeqn

\medskip
and one could speculate whether the following holds: 
\beqn
\lim_{\beta\to\infty} \,\la\,H\,\ra_\beta &\buildrel ?\over =&
  \lim_{\beta\to\infty} \,\bigl\la\, W\bigl(\rho_\beta(\xi_{\rm min})\,\bigr) \,\bigr\ra_{\xi_{\rm min}\;{\rm minimizer\; of\;} V_0(\xi)} \pI  \lbeq{5.13n}
\eeqn

\bigskip
\bigskip

{\bf Numerical Test: Energies vs.~Benchmark Data} 

\bigskip
Let's make a quick numerical check whether these formulae make sense at all. Instead of minimizing over functions $\phi_j(t)$ and $\xi_j(t)$ we just minimize over scalars 
$\phi$ and $\xi$. We calculate on a $12\times 12$ lattice with periodic boundary conditions. After some experimentation, we make the Ansatz 
\beqn
 \phi_j(t)&:=&(-1)^{j_x+j_y} \cdot \phi\;,\;\;\;\; \phi\;\in\;\R \pS \lbeq{5.14a} \\
  \xi_j(t)&:=&(-1)^{j_x+j_y} \cdot  \xi\;,\;\;\;\;\; \xi\;\in\;\R \pS  \lbeq{5.14n}
\eeqn
and minimize over the scalars $\phi$ and $\xi$ on the right hand side of (\req{5.14a},\req{5.14n}). Let's first have a look at the energies $W(\rho_t(\phi))$ and  
$W(\rho_t(\xi))$. Actually these are even functions, $W(\rho_t(-\phi))=W(\rho_t(\phi))$ and  $W(\rho_t(-\xi))=W(\rho_t(\xi))$, so we can restrict to positve 
values for $\phi$ and $\xi$. For the choices
\beq
\phi&\in&\bigl\{\, 0.0\,,\,0.1\,,\,0.2\,,\,\cdots\,,\,1.9\,,\,2.0\,\bigr\}  \\ 
\xi&\in&\bigl\{\, 0.0\,,\,0.1\,,\,0.2\,,\,\cdots\,,\,1.9\,,\,2.0\,\bigr\} \pS
\eeq
we solved the ODE systems (\req{4.37n}) and (\req{5.11n}), calculated $W(\rho_t(\phi))$ and $W(\rho_t(\xi))$ and plotted these quantities, divided by the number 
of lattice sites $L^2$, as a function of $t\in[0,20]$. For each value of $\phi$ we get a different energy curve $t\to W(\rho_t(\phi))/L^2$, so in total 
there are 21 curves which look as follows: 

\bigskip
\centerline{\includegraphics[width=15.5cm]{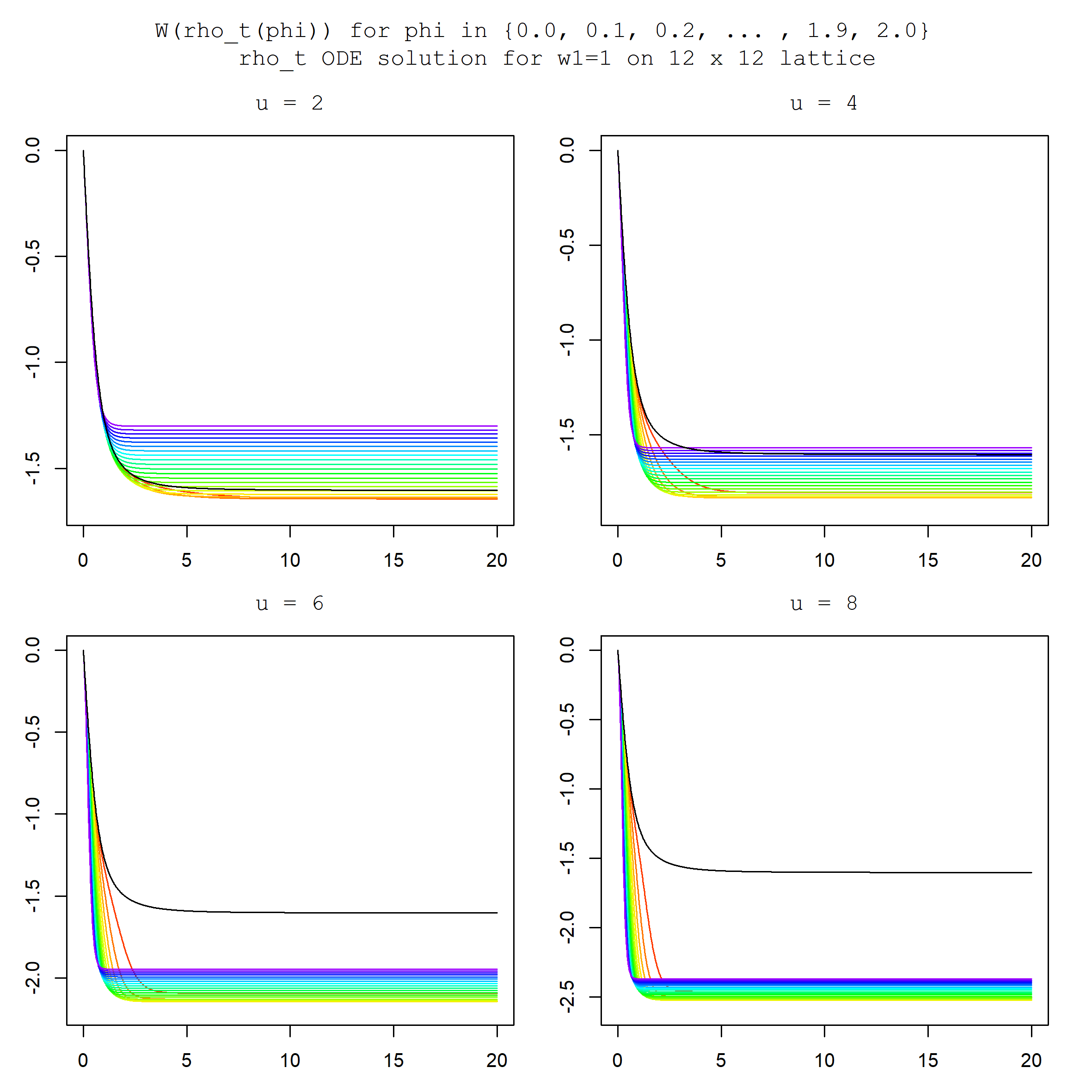}}

\bigskip
The plots for the $W(\rho_t(\xi))$'s  are virtually the same, so we do not display them. Actually the black curve is the $\phi=0$ solution and it is the same for all 
couplings, it coincides with the noninteracting $u=0$ solution. As can be seen, at this point we are only interested in a rough consistency check, the choice of $\beta=20$ 
is sufficient for the calculation of the $V_0(\phi)$ in (\req{5.6n}). We calculated this $V_0(\phi)$ for the couplings
\beq
u&\in&\bigl\{ \; 2\;,\; 4\;,\; 6\;,\; 8\;,\;10\;,\; 12\; \bigr\} \pS
\eeq
and plotted the quantity $V_0(\phi)/L^2$ as a function of $\phi$, now with a more refined choice for the phi's, we used $\,\phi\in\{\,0.00\,,\,0.01\,,\,0.02\,,\,\cdots\,,\,0.99\,,\,1.00\,\}\,$ which are displayed on the horizontal axis and $V_0(\phi)/L^2$ on the vertical axis:

\bigskip
\centerline{\includegraphics[width=15.5cm]{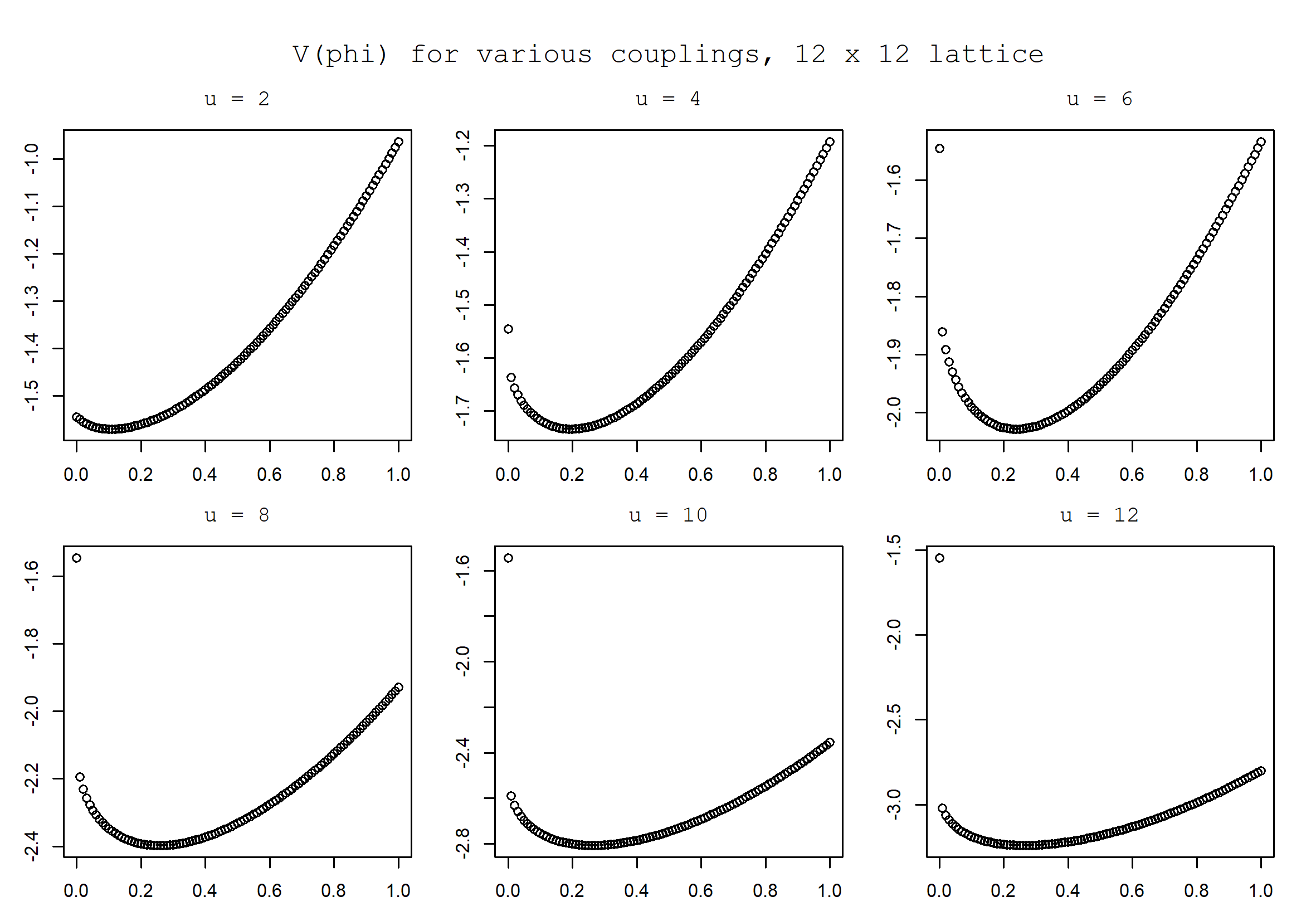}}

\bigskip
This resulted in the following numbers:

\bigskip
\bigskip
\centerline{\includegraphics[width=11.0cm]{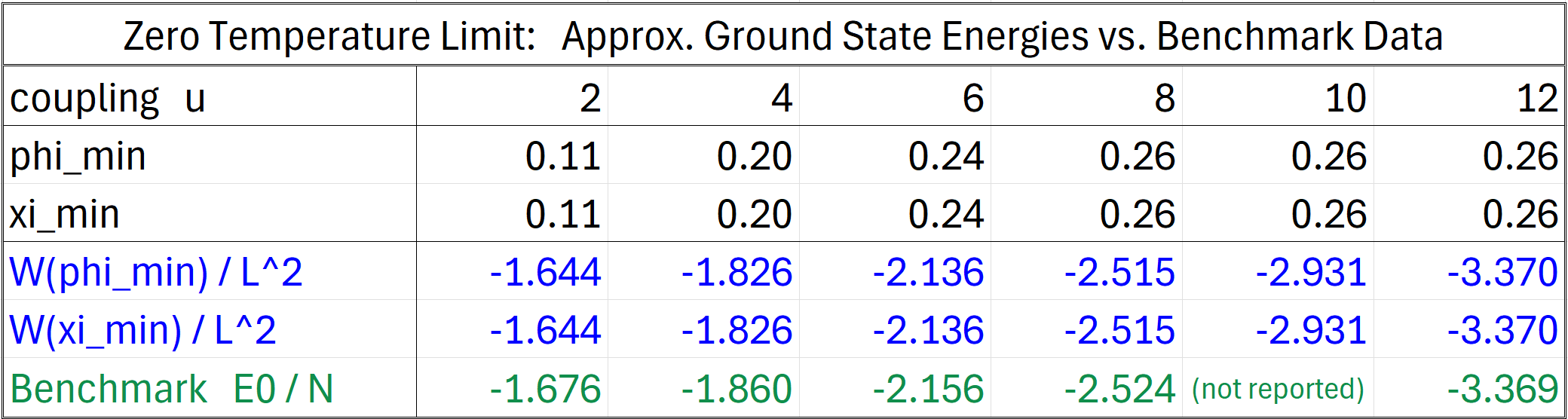}}

\bigskip
\bigskip
Here we have to remark that we solved the ODE systems with a simple Euler discretization with $dt=0.01$ with the exception for the $u=12$ 
values where $dt=0.005$ has been used. If we use there $dt=0.01$ also, we get $\phi_{\rm min}=\xi_{\rm min}=0.27$ and the blue numbers change to $-3.372\;$.  

\bigskip
The benchmark data, the green numbers, have been taken from `Solutions of the Two Dimensional Hubbard Model: Benchmarks and Results from a Wide Range of Numerical Algorithms' [3]. 
Table II on page 23 reports ground state energies for seven different numerical algorithms at half filling. We used the first four rows labelled by AFQMC for Auxiliary-Field 
Quantum Monte Carlo, DMET for Density Matrix Embedding Theory, DMRG for Density Matrix Renormalization Group and FN for Fixed Node Diffusion Monte Carlo, each 
of these values were reported with an error bound, and then we took the arithmetic average of these four values. Finally, since we are considering the 
`symmetrized' Hamiltonian with the one half's subtracted off in the interacting part, we subtracted a $u/4$ to account for that, recall formula (\req{4.27nn}) of 
section 4.2\;. That is, for the green numbers above, we used the following formula:
\beq
\;[\,E_0/N\,]_{\rm benchmark}&:=& {\rm AFQMC\;+\;DMET\;+\;DMRG\;+\;FN \over 4} \;\;-\;\; {u\over 4}  \pS 
\eeq

\goodbreak
\bigskip

Apparently, we obtain some reasonable numbers as far as the energy is concerned. What about correlation functions? 
We can apply the same logic, we take the approximate minimum configurations (\req{5.14a}) and (\req{5.14n}), calculate $\rho$ and $F$ from the ODEs (\req{4.37n}) and (\req{5.11n}) 
and plug the result into the formulae of Theorem 4.3 and 4.4 for the spin-spin and pair-pair correlations. If one does this, one obtains the following values: 
For the spin-spin correlations, for $u=4$, 

\medskip
\bigskip
\centerline{\includegraphics[width=16.5cm]{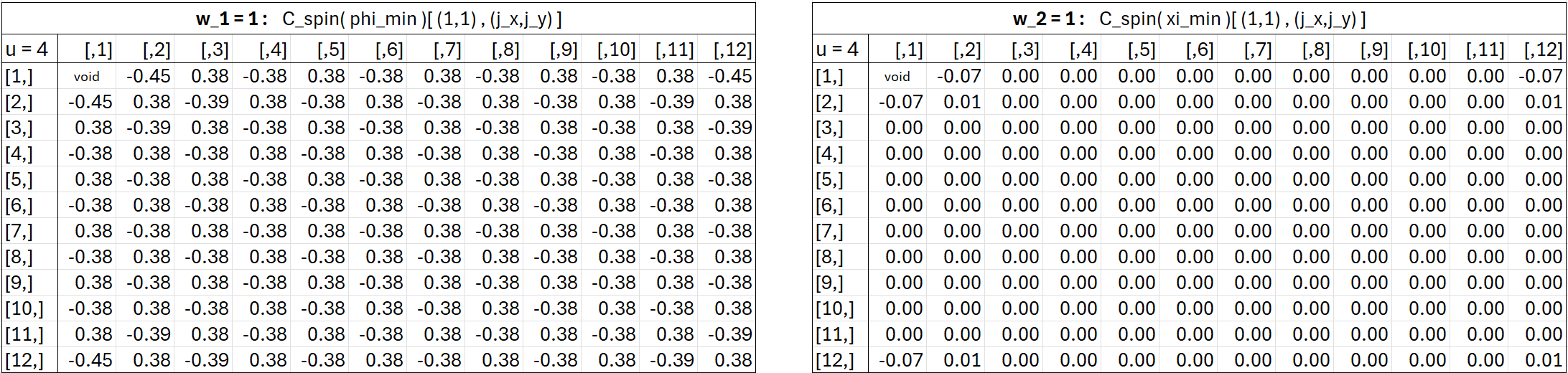}}

\medskip
\bigskip

so basically $\,\pm \,0.38\,$ in the $w_1=1$ representation and just 0 for $w_2=1$. 
And for the pair-pair correlations, for $u=-4$ (for negative values of $u$ the or some approximate minimum configurations are found to be (\req{5.14a}) for the $\phi$'s, 
this does not change, and for the $\xi$'s the Ansatz becomes $\,\xi_j(t)=\xi\in\mathbb R\,$, without the alternating signs, then the energies are found to be identical 
to those reported in the table above, in line with $\la H(-u)\ra=\la H(u)\ra$), 

\medskip
\bigskip
\centerline{\includegraphics[width=16.5cm]{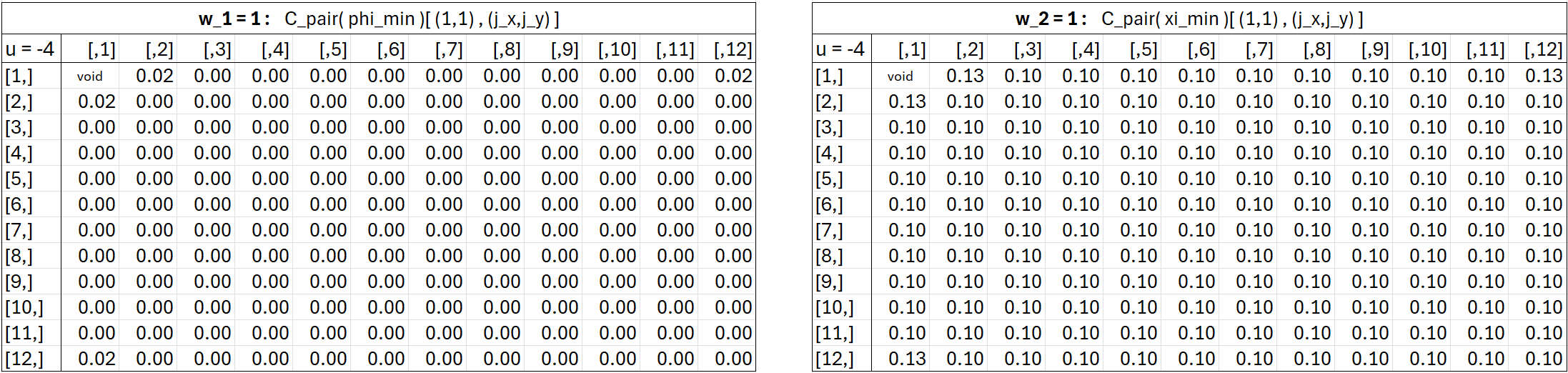}}

\medskip
\bigskip

so basically $\,+\,0.1\,$ in the $w_2=1$ representation and just 0 for $w_1=1$. For larger values of $\sabs u\sabs$, the zero values stay zero while the non-zero values increase. Thus, the results do so to speak `maximally' differ, $w_1=1$ representation vs.~$w_2=1$ representation: If we evaluate by the recipe `just plug in the approximate minimum configurations', then in one representation we get zeros while the other representation produces numbers which are significantly larger than the true values, in [17]\, Scalettar 
and coworkers calculated the spin-spin correlation on a $24\times 24$ lattice at $\beta=24$ and found average values around $\,\pm\,0.09\,$ for $u=4$. Let us also recall that 
in section 4.3 we made some numerical effort with 10 million Monte Carlo simulations to confirm that, when the full expectation values are taken in each representation, 
then these full expectation values do in fact coincide.     

\medskip
In principle, there could be two reasons for these differences. One reason could be, the approximate minimum configurations (\req{5.14a}) and (\req{5.14n}) are too crude, 
and when the exact minima and in particular all of the exact minima are taken into account (which probably include configurations which produce different values 
for the spin-spin correls while having the same minimum value for the energy, within the same representation), then the differences go away. A second reason could be, 
the analogs of (\req{4.40n}) and (\req{5.13n}) for the spin-spin and pair-pair correlations simply do not hold (and also for the energy) and also in the limit 
$\,\beta\to\infty$, Monte Carlo configurations around the exact minima have to be taken into account. For now, all what we can say is that when these correlations are 
calculated with plain Monte Carlo, they show a very high sensitivity with respect to Monte Carlo noise.

\vfill
\newpage

\bigskip
\bigskip
\bigskip
\bigskip
{\large \bf 5. \;Proof of Main Theorem} 
\numberwithin{equation}{section}
\renewcommand\thesection{5}
\setcounter{equation}{0}

\bigskip
\bigskip
\bigskip
{\bf 5.1 \;Pfaffian Quantum Monte Carlo Representation}

\bigskip
\bigskip
The starting point for the proof is the pfaffian quantum Monte Carlo representation given in Theorem 5.2 below. This representation is the generalization of the standard 
determinant quantum Monte Carlo representation for the case that also pairing and exchange terms are present in the quadratic part of the Hamiltonian. 
To calculate traces of products 
of exponentials of arbitrary quadratic operators, we can use a very recent result of Han, Wan and Yao [2]. To write it down, let us temporarily change the notation 
and put all the Majorana operators $a_1,\cdots,a_{2n}$, $b_1,\cdots,b_{2n}$ with $n:=\G$, $\,2n=|\Gamma\times\hbox{\footnotesize$\{\up,\down\}$}|$, into a single 
\beq
a&:=&(a_1,\cdots,a_{2n},a_{2n+1},\cdots,a_{4n})\;\;:=\;\;(a_1,\cdots,a_{2n},b_{1},\cdots,b_{2n}) \pI 
\eeq
Furthermore, we let 
\beqn
H_\ell &:=&{\ts {1\over 4}}\,\summ_{x,y}\, a_x\,h_{x,y}^\ell\,a_y \;\;=:\;\;{\ts {1\over 4}}\,a\,h_\ell\,a  \pI \lbeq{3.2}
\eeqn
with a skew symmetric $h_\ell\,\in\,\mathbb C^{4n\times 4n}$, $\;h_\ell^T\,=\,-\,h_\ell\,$, and we denote by 
\beqn
\cF&=&\Bigl\{\;\;\prodd_{j\sigma}\,(c_{j\sigma}^+)^{n_{j\sigma}}\,1\;\;\;\;\bigr|\;\;\; n_{j\sigma}\;\in\;\{0,1\} \;\; \Bigr\} \pI  \lbeq{3.1}
\eeqn
the grand canonical Fock space, written in terms of the original complex fermion annihilation and creation operators 
$\,c_{j\sigma}\,=\,{\ts{1\over 2}}\,(\,a_{j\sigma}\,+\,i\,b_{j\sigma}\,)\,$ and 
$\,c_{j\sigma}^+\,=\,{\ts{1\over 2}}\,(\,a_{j\sigma}\,-\,i\,b_{j\sigma}\,)\,$. Then there is the following

\goodbreak
\medskip
\bigskip
\bigskip
{\bf Theorem 5.1 (Han, Wan, Yao):} \;Let the $H_\ell$ be given by (\req{3.2}) and abbreviate 
\medskip
\beqn
\la a_x a_y\ra_k\;\;:=\;\;{ \ds Tr_\cF\,\Bigl[ \,a_x\,a_y\,\pro_{\ell=1}^k \,e^{\,-\,H_\ell}\,\Bigr] \over \ds Tr_\cF\,\Bigl[ \,\pro_{\ell=1}^k \,e^{\,-\,H_\ell}\,\Bigr]   } 
\;,\;\;\;\;\;\;\;\;  Z_k\;\;:=\;\;Tr_\cF\,\Bigl[ \,\pro_{\ell=1}^k \,e^{\,-\,H_\ell}\,\Bigr] \pI 
\eeqn

\medskip
Define the skew symmetric $4n\times 4n$ matrix 
\beqn
G_k(x,y)&:=&\begin{cases} \;\la a_x a_y\ra_k\;\; &{\rm for} \;\; x\ne y \\ \;\;\;\;0 &{\rm for} \;\; x=y  \end{cases}
\eeqn
Then there are the following formulae:
\medskip
\beqn
G_k(x,y)&=& 2\,[\,Id\,+\,U_k\,]^{-1}\,-\,Id \;\;=\;\;{ Id\,-\,U_k \over Id\,+\,U_k} \pI 
\eeqn

\medskip
with thermodynamic evolution matrix
\beqn
U_k&:=&\pro_{\ell=1}^k\, e^{\,-\,h_\ell} \;\;\in\;\;SO(4n,\mathbb C)  \pI \lbeq{5.6nn}
\eeqn
and $Z_k$ can be obtained recursively through the equation 
\medskip
\beqn
Z_k&=&Z_{k-1}\;\cdot\;{\rm Pf}\bmat G_{k-1} & -Id \\ +Id & g_k \emat\;\cdot\;{\rm Pf}\bmat \sqrt{2}\;\sinh(h_k/4) & -Id \\ +Id & \sqrt{2}\;\sinh(h_k/4) \emat  \pI\pI
\eeqn

\medskip
with $\,Z_0:=Tr_\cF[\, Id\,]\,$ and $4n\times 4n$ skew symmetric matrix
\beqn
g_k&:=&2\,[\,Id\,+\,e^{\,-\,h_k} \,]^{-1}\,-\,Id\;\;=\;\;{ Id\,-\,e^{\,-\,h_k} \over Id\,+\,e^{\,-\,h_k}}\;\;=\;\;\tanh(h_k/2)\pI\pI 
\eeqn

\medskip
and Pf denoting the Pfaffian of a skew symmetric matrix.

\medskip
\bigskip
{\bf Proof:}\, see [2].\; $\blacksquare$

\bigskip
\bigskip
\bigskip
In our case, the product of exponentials of quadratic operators $\,\pro_\ell e^{-H_\ell}\,$ is given by the representation (\req{2.19}). Thus, in our case 
the matrix $U_k$ in (\req{5.6nn}) is given by
\beqn
U_k\;\;=\;\;U_k(\phi,\xi,\theta)&=&\prodd_{\ell=1}^k\,\Bigl(\, e^{\,-\,i\,dt\,h_0}\;e^{\,-\,i\,dh_1(\phi_\ell)}\, e^{\,-\,i\,dh_2(\xi_\ell)}\, 
  e^{\,-\,i\,dh_3(\theta_\ell)}\,   \Bigr)  \pI \lbeq{3.9}
\eeqn
with $h_0$ given by (\req{2.30}) and (the prefactor $1/4$ is separate and does not move into the $dh_i$)
\medskip
\beq
dh_1(\phi)&=& \sqrt{u}\, \sqrt{w_1\eps_1}\, 
\bmat 0 & 0           & +dx & 0           \\ 
      0 & 0           & 0   & -\eps_1\,dx \\ 
    -dx & 0           & 0   & 0           \\ 
      0 & +\eps_1\,dx & 0   & 0           \emat \;\;=:\;\; \sqrt{u}\;\nu_1\, dB^x  \pI
\eeq
\beq
dh_2(\xi)&=& \sqrt{u}\, \sqrt{w_2\eps_2}\, 
\bmat 0 & 0           & 0           & +dy \; \\ 
      0 & 0           & +\eps_2\,dy & 0    \\ 
      0 & -\eps_2\,dy & 0           & 0    \\ 
    -dy & 0           & 0           & 0    \emat \;\;=:\;\; \sqrt{u}\;\nu_2\, dB^y    \pI
\eeq
\beqn
dh_3(\theta)&=& \sqrt{u}\, \sqrt{w_3\eps_3}\, 
\bmat 0 & +dz          & 0           & 0           \\ 
    -dz & 0           & 0           & 0           \\ 
      0 & 0           & 0           & +\eps_3\,dz \\ 
      0 & 0           & -\eps_3\,dz & 0           \emat \;\;=:\;\; \sqrt{u}\;\nu_3\, dB^z    \pI  \lbeq{3.10}
\eeqn

\medskip
with
\beqn
\nu_i&:=&\sqrt{w_i\eps_i} \pS 
\eeqn
Then straightforward application of Theorem 5.1 leads to the following pfaffian quantum Monte Carlo representation which in the terminology of section 1 would be the 
`untransformed representation' since it is the analog of (\req{1.18}) in section 1.

\bigskip
\bigskip
\bigskip
{\bf Theorem 5.2 (Pfaffian Quantum Monte Carlo Representation):} Let $H$ be the Hubbard-Hamiltonian (\req{Hmain}).  
For some inverse temperature $\beta=kdt$, define the matrix $C$ of correlations in terms of Majorana operators (\req{n2.7}) through 
\medskip
\beqn
C_\beta\;\;=\;\;\bmat C^{aa} & C^{ab} \\ C^{ba} & C^{bb} \emat&:=&
  \bmat \la a_{i\sigma}a_{j\tau}\ra_\beta & \la a_{i\sigma}b_{j\tau}\ra_\beta \\ \la b_{i\sigma}a_{j\tau}\ra_\beta & \la b_{i\sigma}b_{j\tau}\ra_\beta \emat 
  \;-\;\Idv \;\;\in \;\; \mathbb C^{4\G\times 4\G}  \pI \pI \lbeq{3.12}
\eeqn
with 
\beqn
\la A\ra_\beta&:=& Tr_\cF\,A\,e^{-\beta H}\;\bigr/\;Tr_\cF \,e^{-\beta H}  \pS
\eeqn
Then, with $\beta=kdt$,  
\beqn
C_\beta&=&\la \,G_k\,\ra \;\;:=\;\;{  \phantom{\Bigr|} \int_{\R^{3k\G}}\,G_k(\phi,\xi,\theta)\; Z_k(\phi,\xi,\theta)\; dP_k(\phi,\xi,\theta)  \phantom{\Bigr|} \over 
   \phantom{\Bigr|} \int_{\R^{3k\G}}\; Z_k(\phi,\xi,\theta)\; dP_k(\phi,\xi,\theta)  \phantom{\Bigr|} }  \pI \lbeq{3.14}
\eeqn

\medskip
where $G_k$ is given by 
\medskip
\beqn
G_k&=&2\,[\,Id\,+\,U_k\,]^{-1} \,-\,Id  \;\;=\;\;{ Id\,-\,U_k \over Id\,+\,U_k} \;\;\in\, \;\; {\rm SkewSym}\bigl(\,4\G\,,\mathbb C\,\bigr)  \pI \lbeq{3.15}
\eeqn

\medskip
with thermodynamic evolution matrix $U_k$ given by (\req{3.9}), 
\medskip
\beq
U_k\;\;=\;\;U_k(\phi,\xi,\theta)&=&\prodd_{\ell=1}^k\,\Bigl(\, e^{\,-\,i\,dt\,h_0}\;e^{\,-\,i\,dh_1(\phi_\ell)}\, e^{\,-\,i\,dh_2(\xi_\ell)}\, 
  e^{\,-\,i\,dh_3(\theta_\ell)}\,   \Bigr) \;\;\in\, \;\; SO\bigl(\,4\G\,,\mathbb C\,\bigr)  \pI 
\eeq

\medskip
the $dh_i$ given by the skew symmetric matrices (\req{3.10}) above and $dP_k$ given by (\req{2.21}). The function $Z_k=Z_k(\phi,\xi,\theta)$ can be obtained from the recursion 
{\footnotesize (\,with $Id=\Idv$\,)}
\medskip
\beqn
Z_k&=&Z_{k-1}\;\cdot\;{\rm Pf}\bmat G_{k-1} & -Id \\ +Id & \tanh(dh^k/2) \emat\;\cdot\;
    {\rm Pf}\bmat \sqrt{2}\;\sinh(dh^k/4) & -Id \\ +Id & \sqrt{2}\;\sinh(dh^k/4) \emat  \pI\pI \;\;\lbeq{3.16}
\eeqn
with
\beqn
dh^k\;\;\equiv\;\;dh(\phi_k,\xi_k,\theta_k)&:=&i\,dt\,h_0\,+\,i\,dh_1(\phi_k)\,+\,i\,dh_2(\xi_k)\,+\,i\,dh_3(\theta_k) \pI
\eeqn

\medskip
and $\,Z_0:=Tr_\cF[\,Id\,]\,$, $G_0:=0$. Finally, the partition function itself is given by 
\medskip
\beqn
Tr_\cF \,e^{-\beta H} &=&\ts\int_{\R^{3k\G}}\; Z_k(\phi,\xi,\theta)\; dP_k(\phi,\xi,\theta) \;\times\; e^{\,-\,\beta{u\over 4}w\eps\G}  \pI \lbeq{3.18}
\eeqn

\bigskip
{\bf Proof:} Follows immediately from (\req{2.19}) and Theorem 5.1 of Han, Wan and Yao. The fact that we can calculate $Z_k$ from a 
one step recursion instead of a four step recursion, there are four exponentials per time step $dt$ in the $U_k$, we justify at the beginning of the next subsection, 
at a fixed time step, we can actually add the exponentials.\;\;$\blacksquare$ 

\bigskip
\bigskip
{\bf Remarks:} \;{\bf (i)} We subtracted off the identity matrix in (\req{3.12}) and (\req{3.15}) since then both $C_\beta$ and $G_k$ are skew symmetric matrices 
with zeros on the diagonal. 
For $C_\beta$, this follows directly from the definition and for $G_k$, this follows from the orthogonality of the $U\equiv U_k\in SO(4\G,\mathbb C)$, 
\beqn
G^T&=&{ Id-U^T \over Id+U^T}\;\;=\;\;{ Id-U^{-1} \over Id+U^{-1}}\;\;=\;\;{ U-Id \over U+Id}\;\;=\;\; -\,G  \pI 
\eeqn

\medskip
{\bf (ii)} Let us also recall remark (vi) at the end of section 2 concerning the notation: The $G$'s used in this section 5 correspond to expectations $\la a a\ra$, $\la a b\ra$ 
and $\la b b\ra$. In section 5.4, we introduce a ${\ti G}\,:=\,i\,G$ which corresponds to expectations $i\la a a\ra$, $i\la a b\ra$ 
and $i\la b b\ra$. The $G$ in the main theorem is actually a $\ti G$, to keep the notation simple we omitted the tilde in the formulation of the main theorem. 
Recall that for Majorana operators $a,b$ the combination $iab$ is self adjoint.


\bigskip
\bigskip
\bigskip
{\bf 5.2 \;Stochastic Differential Equation Representation}

\bigskip
\bigskip
The next step in the proof of the Main Theorem is to write down stochastic differential equations for the quantities $U_{t_k}$, $G_{t_k}$ and $Z_{t_k}$ which 
show up in the pfaffian quantum Monte Carlo representation of Theorem 5.2. They are summarized in Theorem 5.3 below. 

\bigskip
Let us first write down a stochastic differential equation for the thermodynamic evolution matrix (\req{3.9}),
\beq
U_k&=&\prodd_{\ell=1}^k\,\Bigl(\, e^{\,-\,i\,dt\,h_0}\;e^{\,-\,i\,dh_1(\phi_\ell)}\, e^{\,-\,i\,dh_2(\xi_\ell)}\, 
  e^{\,-\,i\,dh_3(\theta_\ell)}\,   \Bigr)  \pI 
\eeq
Apparently, we have the exact recursion
\beqn
U_k&=&U_{k-1}\; e^{\,-\,i\,dt\,h_0}\,e^{\,-\,i\,dh_1(\phi_k)}\, e^{\,-\,i\,dh_2(\xi_k)}\, e^{\,-\,i\,dh_3(\theta_k)}  \pI  \lbeq{3.20}
\eeqn
Recall the formulae (\req{3.10}) for the $dh_i$, 
\beq
dh_1(\phi)&=& \sqrt{u}\;\nu_1\, dB^x  \pS\\
dh_2(\xi)&=&  \sqrt{u}\;\nu_2\, dB^y   \\
dh_3(\theta)&=& \sqrt{u}\;\nu_3\, dB^z   \pS 
\eeq
with $\nu_i:=\sqrt{w_i\eps_i}\,$. From the calculation rules for Brownian motions (see the appendix of [1] for more background),
\beq
dx_i\,dx_j&=&\delta_{i,j}\; dt \pI
\eeq
for arbitrary lattice site $i,j\in\Gamma$, we get
\beq
(dB^x)^2\;\;=\;\;(dB^y)^2\;\;=\;\;(dB^z)^2\;\;=\;\;-\,dt\;\Idv  \pI 
\eeq
with $\Idv$ being the $4\G\times 4\G$ identity matrix. Also, since  $dx_i\,dy_j=dx_i\,dz_j=dy_i\,dz_j=0$,  
\beq
dB^x\,dB^y\;\;=\;\;dB^x\,dB^z\;\;=\;\;dB^y\,dB^z\;\;=\;\;0  \pS
\eeq
and in the last equation the zero on the right hand side of course means the $4\G\times 4\G$ zero matrix. 
Thus, up to terms $O(\,dt^{3/2}\,)$, we get
\beqn
U_k&=&\ts U_{k-1}\,\bigl(\,1-i\,dt\,h_0\,\bigr)\bigl(\,1-i\,dh_1-{(dh_1)^2\over 2}\,\bigr)
  \bigl(\,1-i\,dh_2-{(dh_2)^2\over 2}\,\bigr)\bigl(\,1-i\,dh_3-{(dh_3)^2\over 2}\,\bigr) \pI \nn  \\ 
&=&\ts U_{k-1}\,\Bigl(\,1\,-\,i\,dt\,h_0\,-\,i\sqrt{u}\,\bigl[\,\nu_1\,dB^x\,+\,\nu_2\,dB^y\,+\,\nu_3\,dB^z\,\bigr]
  \,+\,{udt\over 2}\,w\eps\,\Idv\,\Bigr)  \pI \nn \\ 
&=:&\ts U_{k-1}\,\Bigl(\,1\,-\,i\,dt\,h_0\,-\,i\sqrt{u}\,dB \,+\,{udt\over 2}\,w\eps\,\Idv\,\Bigr)  \pI  \lbeq{3.24}
\eeqn
with
\beq
w\eps&=&w_1\eps_1+w_2\eps_2+w_3\eps_3\;\;=\;\;\nu_1^2+\nu_2^2+\nu_3^2\;\;=\;\;\nu^2 \pM
\eeq
and matrix of Brownian motions
\beq
dB&:=& \nu_1\,dB^x\,+\,\nu_2\,dB^y\,+\,\nu_3\,dB^z  \pI 
\eeq
In continuous time notation, (\req{3.24}) reads as follows:
\beq
dU&=&\ts U\,\Bigl(\,-\,i\,dt\,h_0\,-\,i\sqrt{u}\,dB\,+\,{udt\over 2}\,w\eps\,\Idv \,\Bigr)  \pI
\eeq
We remark that if we put
\beq
dh\;\;=\;\;dh(\phi,\xi,\theta)&:=&i\,dt\,h_0\,+\,i\,dh_1(\phi)\,+\,i\,dh_2(\xi)\,+\,i\,dh_3(\theta) \pI \nn\\ 
&=&i\,dt\,h_0\,+\,i\sqrt{u}\;dB \pS 
\eeq
the above SDE can be written more compactly as 
\beq
dU&=&\ts U\,\bigl(\,-\,dh\,+\,{1\over 2}\,(dh)^2\, \bigr) \pI  
\eeq
which reflects the fact that at a fixed time step we can add the exponents in (\req{3.20}), the recursion for $U_k$ can equivalently be written as 
\beq
U_k&=&U_{k-1}\; e^{\,-\,dh(\phi_k,\xi_k,\theta_k)} \pI
\eeq
Thus we have shown part (a) of the following

\goodbreak

\medskip
\bigskip
\bigskip
{\bf Theorem 5.3 (Untransformed SDE Representations):} 
\medskip
\begin{itemize}
\item[{\bf a)}] Let $\,U_t\,=\,U_{t_k}\,=\,U_{kdt}\,$ be the thermodynamic evolution matrix given by (\req{3.9}). Then $U$ can be obtained from the SDE 
\beqn
dU&=&\ts U\,\Bigl(\,-\,i\,dt\,h_0\,+\,{udt\over 2}\,w\eps\,\Idv\,-\,i\sqrt{u}\,dB \,\Bigr) \;\;=\;\; U\,\Bigl(\,-\,dh\;+\;{1\over 2}\,(dh)^2\; \Bigr)  \pI\pI\pI
\eeqn
with $U_{t=0}=Id\,$. Here, $dh$ is given by
\beqn
dh&=&i\,dt\,h_0\,+\,i\sqrt{u}\;dB \pM \lbeq{6.23}
\eeqn
with $h_0$ given by (\req{2.30}) and 
\medskip
\beqn
dB&=& 
\bmat 0 & +\nu_3\,dz           & +\nu_1\,dx & +\nu_2\,dy           \\ 
      -\nu_3\,dz & 0           & +\nu_2\eps_2\,dy   & -\nu_1\eps_1\,dx \\ 
    -\nu_1\,dx & -\nu_2\eps_2\,dy     & 0   & +\nu_3\eps_3\,dz            \\ 
      -\nu_2\,dy & +\nu_1\eps_1\,dx & -\nu_3\eps_3\,dz    & 0           \emat  \;\;\in\;\; \mathbb C^{4\G\times 4\G}  \pI\pI \lbeq{3.31}
\eeqn

\smallskip
with diagonal matrices of Brownian motions $dx,dy,dz\in\R^{\G\times\G}$ given by (\req{2.22}). 

\bigskip
\item[{\bf b)}] Let $A=(A_{ij})_{i,j\in\Gamma}$ be an arbitrary matrix and let $dx=(\,\delta_{i,j}\,dx_j\,)_{i,j\in\Gamma}$ be a diagonal matrix of independent 
Brownian motions such that $\,dx_i\,dx_j=\delta_{ij}\,dt\,$. Then
\beqn
dx\,A\,dx&=&{\rm diag}[A]\;dt\;\;:=\;\;\bigl(\,\delta_{i,j}A_{jj}\,\bigr)_{i,j\in\Gamma} \;dt  \pI \lbeq{3.32}
\eeqn

\item[{\bf c)}] The skew symmetric matrix
\beqn
G_t&=& 2\,[\,Id\,+\,U_t\,]^{-1} \,-\,Id   \pS 
\eeqn
can be obtained from the SDE 
\smallskip
\beqn
dG&=&\ts +\,{1\over 2}\,(Id-G)\, \Bigl[\;dh\,-\,{1\over 2}\,dh\,G\,dh\; \Bigr] \, (Id+G)  \pI  \nn\\ 
&=&\ts +\,{1\over 2}\,(Id-G)\, \Bigl[\;i\,h_0\,dt\,+\,i\sqrt{u}\,dB \,+\,{u\over 2}\,dB\,G\,dB\; \Bigr] \, (Id+G)  \pI\pI \lbeq{3.34}
\eeqn

\smallskip
with $G_{t=0}:=0$. If we write the $4\G\times4\G$ matrix $G$ in terms of $\G\times\G$ block matrices as follows, 
\beqn
G&=&\bmat G^{aa} & G^{ab} \\ G^{ba} & G^{bb} \emat \;\;=\;\;
\bmat G^{aa}_{\su\su} & G^{aa}_{\su\sd} & G^{ab}_{\su\su} & G^{ab}_{\su\sd}  \\ 
      G^{aa}_{\sd\su} & G^{aa}_{\sd\sd} & G^{ab}_{\sd\su} & G^{ab}_{\sd\sd}  \\ 
      G^{ba}_{\su\su} & G^{ba}_{\su\sd} & G^{bb}_{\su\su} & G^{bb}_{\su\sd}  \\  
      G^{ba}_{\sd\su} & G^{ba}_{\sd\sd} & G^{bb}_{\sd\su} & G^{bb}_{\sd\sd}  \emat \pI 
\eeqn  

\medskip
then the quantity $dB\,G\,dB$ in the SDE (\req{3.34}) is given by 
\beqn
dB\,G\,dB&=& \bigl[\,w_1\,D_1G\;+\;w_2\,D_2G\;+\; w_3\,D_3G \,\bigr]\, dt \pI 
\eeqn
with matrices
\medskip
\beq
D_1G&:=&
\bmat 
      0                                   & +{\rm diag}[\,G^{bb}_{\su\sd}\,]   & -\eps_1\,{\rm diag}[\,G^{ab}_{\su\su}\,]  & +{\rm diag}[\,G^{ab}_{\sd\su}\,]  \\  
      -{\rm diag}[\,G^{bb}_{\su\sd}\,]    & 0                                  & +{\rm diag}[\,G^{ab}_{\su\sd}\,]          & -\eps_1\,{\rm diag}[\,G^{ab}_{\sd\sd}\,]  \\
 +\eps_1\,{\rm diag}[\,G^{ab}_{\su\su}\,] & -{\rm diag}[\,G^{ab}_{\su\sd}\,]   & 0                                         & +{\rm diag}[\,G^{aa}_{\su\sd}\,] \\ 
 -{\rm diag}[\,G^{ab}_{\sd\su}\,]         & +\eps_1\,{\rm diag}[\,G^{ab}_{\sd\sd}\,] & -{\rm diag}[\,G^{aa}_{\su\sd}\,]    & 0   \emat\pI\pI
\eeq

\medskip
\beq
D_2G&:=&
\bmat
                      0          & +{\rm diag}[\,G^{bb}_{\su\sd}\,]   & -{\rm diag}[\,G^{ab}_{\sd\sd}\,]                & -\eps_2\,{\rm diag}[\,G^{ab}_{\su\sd}\,]  \\ 
-{\rm diag}[\,G^{bb}_{\su\sd}\,] &           0                        & -\eps_2\,{\rm diag}[\,G^{ab}_{\sd\su}\,]        & -{\rm diag}[\,G^{ab}_{\su\su}\,]\\ 
+{\rm diag}[\,G^{ab}_{\sd\sd}\,] & +\eps_2\,{\rm diag}[\,G^{ab}_{\sd\su}\,]         &            0                      & +{\rm diag}[\,G^{aa}_{\su\sd}\,]\\ 
+\eps_2\,{\rm diag}[\,G^{ab}_{\su\sd}\,]         & +{\rm diag}[\,G^{ab}_{\su\su}\,] & -{\rm diag}[\,G^{aa}_{\su\sd}\,]  &      0                          \emat\pI\pI
\eeq

\medskip
\beqn
D_3G&:=&
\bmat 
       0                         & -\eps_3\,{\rm diag}[\,G^{aa}_{\su\sd}\,] & -{\rm diag}[\,G^{ab}_{\sd\sd}\,] & +{\rm diag}[\,G^{ab}_{\sd\su}\,]  \\ 
+\eps_3\,{\rm diag}[\,G^{aa}_{\su\sd}\,] &          0                & +{\rm diag}[\,G^{ab}_{\su\sd}\,] & -{\rm diag}[\,G^{ab}_{\su\su}\,]  \\ 
+{\rm diag}[\,G^{ab}_{\sd\sd}\,] & -{\rm diag}[\,G^{ab}_{\su\sd}\,]  &        0                         & -\eps_3\,{\rm diag}[\,G^{bb}_{\su\sd}\,]  \\  
-{\rm diag}[\,G^{ab}_{\sd\su}\,] & +{\rm diag}[\,G^{ab}_{\su\su}\,]  & +\eps_3\,{\rm diag}[\,G^{bb}_{\su\sd}\,] & 0                       \emat \pI\pI \lbeq{3.37}
\eeqn

\smallskip
\bigskip
\item[ {\bf d)}] The quantity $\,Z_t\,=\,Z_{t_k}\,=\,Z_{kdt}\,$ can be obtained from the SDE 
\medskip
\beqn
dZ&=&\ts \Bigl\{\; {1\over 4}\; Tr[\,G\,dh\,]\;+\;{1\over 32}\; \bigl(\,Tr[\,G\,dh\,]\,\bigr)^2 
  \;-\;{1\over 16}\; Tr[\,G\,dh\,G\,dh\,] \;+\; {1\over 16}\;Tr[\,(dh)^2\,] \;\Bigr\}\; Z  \nn \\
&&  \lbeq{3.38}
\eeqn
with initial value $\,Z_{t=0}:=Tr_\cF[\,Id\,]\,$ and $dh$ given by (\req{6.23}) above.  
\end{itemize}

\goodbreak
\medskip
\bigskip
\bigskip
{\bf Remark:} Let us point out that in the untransformed SDE for $G$ given by (\req{3.34}) above not just diffusive part, the $dB$-part, depends on the details of 
the Hubbard-Stratonovich transformation, but also the drift part, the $dt$-part: the quantity 
\beqn
dB\,G\,dB&=& \bigl[\,w_1\,D_1G\;+\;w_2\,D_2G\;+\; w_3\,D_3G \,\bigr]\, dt \pI 
\eeqn
obviously depends on the choices for the $w_i$ and $\eps_i$. It is quite remarkable that through the Girsanov transformation, we consider this in the next section, 
the drift part gets altered in precisely such a way that it becomes independent of the choices for the $w_i$ and $\eps_i\,$. 

\bigskip
\bigskip
In the remaining part of this subsection 5.2 we prove Theorem 5.3. We start with the following two elementary lemmata:

\bigskip
\bigskip
{\bf Lemma 5.1:} Let $A=A_t\in \mathbb C^{n\times n}$ be an arbitrary family of $n\times n$ matrices indexed by 
some time (or inverse temperature) variable $t\in \R\,$. Let us temporarily use the notation 
\beq
\ts {1\over A}&:=& A^{-1} \pI 
\eeq
for the matrix inverse. Then there are the following formulae:
\begin{itemize}
\item[{\bf a)}] For arbitrary indices $s,t$ we have
\beqn
\ts {1\over A_t}\,-\,{1\over A_s}\;\;=\;\;{1\over A_t}\,(A_s-A_t)\,{1\over A_s}  \pI 
\eeqn
\item[{\bf b)}] For arbitrary indices $s,t$ we also have
\beqn
\ts {1\over A_t}\,-\,{1\over A_s}\;\;=\;\;{1\over A_t}\,(A_s-A_t)\,{1\over A_s}\,(A_s-A_t)\,{1\over A_s}
     \;+\;{1\over A_s}\,(A_s-A_t)\,{1\over A_s} \pI 
\eeqn
\item[{\bf c)}] We put $\,s:=t-dt\,$ and $\,dA:=A_t-A_{t-dt}\,$ in part (b). Then in the limit $\,dt\to 0\,$,  
\beqn
\ts d\,\bigl({1\over A}\bigr)&=&\ts \;-\;{1\over A}\, dA\,{1\over A} \,\;+\,\; {1\over A} \,\bigl(\,dA\,{1\over A}\,\bigr)^2 \;\;+\;\;
  O\bigl(\,(dA)^3\,\bigr)  \pI 
\eeqn
\end{itemize} 

\bigskip
\noindent{\bf Proof:} Part (a) is obvious. To obtain (b), we use (a) and write 
\beq
\ts {1\over A_t}\,-\,{1\over A_s}&=&\ts {1\over A_t}\,(A_s-A_t)\,{1\over A_s}  \pI \\ 
&=&\ts \bigl({1\over A_t}-{1\over A_s}\bigr)\,(A_s-A_t)\,{1\over A_s} \;+\;{1\over A_s}\,(A_s-A_t)\,{1\over A_s}  \pI \\ 
&\buildrel {\rm(a)}\over=&\ts {1\over A_t}\,(A_s-A_t)\,{1\over A_s}\,(A_s-A_t)\,{1\over A_s} \;+\;{1\over A_s}\,(A_s-A_t)\,{1\over A_s}  \pI 
\eeq
Finally, to obtain part (c) we make another iteration using (b) and (a) and write 
\beq
\ts {1\over A_t}\,-\,{1\over A_s}&\buildrel {\rm(b)}\over=&\ts {1\over A_t}\,(A_s-A_t)\,{1\over A_s}\,(A_s-A_t)\,{1\over A_s}
     \;+\;{1\over A_s}\,(A_s-A_t)\,{1\over A_s} \pI 
\eeq
\beq
&=&\ts \bigl({1\over A_t}-{1\over A_s}\bigr)\,(A_s-A_t)\,{1\over A_s}\,(A_s-A_t)\,{1\over A_s} \pI \\ 
&&\ts   \;+\;{1\over A_s}\,(A_s-A_t)\,{1\over A_s}\,(A_s-A_t)\,{1\over A_s}
     \;+\;{1\over A_s}\,(A_s-A_t)\,{1\over A_s} \pI \\
&\buildrel {\rm(a)}\over=&\ts {1\over A_t}\,(A_s-A_t)\,{1\over A_s} \,(A_s-A_t)\,{1\over A_s}\,(A_s-A_t)\,{1\over A_s} \pI \\ 
&&\ts   \;+\;{1\over A_s}\,(A_s-A_t)\,{1\over A_s}\,(A_s-A_t)\,{1\over A_s}
     \;+\;{1\over A_s}\,(A_s-A_t)\,{1\over A_s} \pI
\eeq
With $\,s:=t-dt\,$ and $\,dA=dA_t:=A_t-A_{t-dt}\,$ then the first term is $\,O\bigl(\,(dA)^3\,\bigr) \,$ and the lemma is proven. \;$\blacksquare$

\bigskip
\bigskip
\bigskip
{\bf Lemma 5.2:} Let $A=(a_{ij})$ be an arbitrary skew symmetric matrix, $A^T=-A$, and let ${\rm Pf}A$ denote the Pfaffian of $A$. Then: 
\smallskip
\begin{itemize}
\item[{\bf a)}] There are the following exact identities 
\beqn
 {1\over {\rm Pf}A}\; {\pt {\rm Pf}A \over \pt a_{ij}} &=&\ts   {1\over 2}\; [A^{-1}]_{ji}\;\;=\;\;  {1\over 2}\; [\,(A^{-1})^T]_{ij}  \pI \\ 
&&\nn\\
 {1\over {\rm Pf}A}\;{\pt^2 {\rm Pf}A \over \pt a_{k\ell}\,\pt a_{ij}} &=&\ts   {1\over 4}\; A^{-1}_{ji} \; A^{-1}_{\ell k} 
  \;-\;{1\over 2}\; A^{-1}_{jk}\; A^{-1}_{\ell i} \pI
\eeqn

\smallskip
\item[{\bf b)}] Up to terms $O\bigl[\,(dA)^3\,\bigr]$, we have 
\beqn
\lefteqn{
{\rm Pf}(A+dA)\;\;=\;\;{\rm Pf}A\;\times\;  } \pS \\ 
&&\ts \Bigl\{\; 1\;+\; {1\over 2}\;Tr\bigl[ A^{-1} dA\,\bigr]\;+\;{1\over 8}\;\bigl(\,Tr\bigl[ A^{-1} dA\,\bigr]\,\bigr)^2 
  \;-\;{1\over 4}\;Tr\bigl[\,A^{-1}dA \,A^{-1} dA\,\bigr] \; \Bigr\} \pS \nn
\eeqn
\end{itemize}

\bigskip
{\bf Proof:} Follows by differentiating the formula
\beq
({\rm Pf}A)^2&=&\det A  \pS 
\eeq
with respect to $a_{ij}$ and $a_{k\ell}$\,. \;$\blacksquare$

\bigskip
\bigskip
\bigskip
{\bf Proof of Theorem 5.3\,:}  \;Part (a) was shown at the beginning of this subsection. 

\medskip
{\bf Part b)} This follows immediately from the Brownian motion calculation rule
\beq
dx_i\,dx_j&=&\delta_{i,j}\; dt \pS
\eeq 
Since this is the basic mechanism of how the $\,{\rm diag}[G]\,$ matrices in part (c) and also in the Main Theorem are generated, 
we have put this identity (\req{3.32}) into an extra part (b) of the theorem to make it more visible, although it is quite obvious.   

\bigskip 
{\bf Part c)} We use part (c) of Lemma 5.1 and write 
\beq
\ts {1\over 2}\, dG&=&d\bigl(\,[\,Id\,+\,U\,]^{-1}\,\bigr) \pS \\ 
&=&-\,[\,Id\,+\,U\,]^{-1}\,dU\,[\,Id\,+\,U\,]^{-1}\;+\; \bigl(\, [\,Id\,+\,U\,]^{-1}\,dU\,\bigr)^2\,[\,Id\,+\,U\,]^{-1} \pS 
\eeq
Using the formulae
\beq
[Id+U]^{-1} &=&\ts {1\over 2}\,(Id+G)  \pS \\ 
\phantom{.}[Id+U]^{-1}\,U&=&\ts {1\over 2}\,(Id-G)  \pS
\eeq
we get
\beq
[\,Id\,+\,U\,]^{-1}\,dU&=&\ts [\,Id\,+\,U\,]^{-1} \,U\,\bigl[\,-\,dh\, +\,{1\over 2}\, (dh)^2\,\bigr]  \pS \\ 
&=&\ts  {1\over 2}\,(Id-G)\, \bigl[\,-\,dh\, +\,{1\over 2}\, (dh)^2\,\bigr] \pS
\eeq
From that, we obtain
\beq
\bigl(\,[\,Id\,+\,U\,]^{-1}\,dU\,\bigr)^2&=&\ts {1\over 4}\, (Id-G)\,dh\,(Id-G)\,dh   \pI 
\eeq
Hence we arrive at 
\beq
\ts {1\over 2}\, dG&=&-\,\ts {1\over 4}\,(Id-G)\, \bigl[\,-\,dh\, +\,{1\over 2}\, (dh)^2\,\bigr] \, (Id+G)  
   \,+\,\ts {1\over 8}\,(Id-G)\, dh\, (Id-G)\,dh\,(Id+G)    \pI
\eeq
or
\beq
dG&=&\ts +\,{1\over 2}\,(Id-G)\, \Bigl[\;dh\,-\,{1\over 2}\,dh\,G\,dh\; \Bigr] \, (Id+G)  \pI
\eeq
with 
\beq
-\,dh\,G\,dh&=&+\,u\,dB\,G\,dB\;\;=\;\;+\,u\,\bigl(\,\nu_1^2\,dB^x\,G\,dB^x\,+\,\nu_2^2\,dB^y\,G\,dB^y\,+\,\nu_3^2\,dB^z\,G\,dB^z\,\bigr)  \pI 
\eeq
where we used $\,dx\,dy\,=\,dx\,dz\,=\,dy\,dz\,=\,0\,$. Now, using the structure of the matrices $dB^x,dB^y,dB^z$ as given in (\req{3.10}) above 
and recalling that $\nu_i^2=w_i\eps_i$, we obtain the matrices 
\beq
D_1G&=& \eps_1\,dB^x\,G\,dB^x \pS \\ 
D_2G&=& \eps_2\,dB^y\,G\,dB^y     \\ 
D_3G&=& \eps_3\,dB^z\,G\,dB^z \pS 
\eeq
as specified in (\req{3.37}). For example, 
\medskip
\beq
\eps_1\,dB^x\,G\,dB^x&=&dt
\bmat 
      -\eps_1\,{\rm diag}[\,G^{bb}_{\su\su}\,]          & +{\rm diag}[\,G^{bb}_{\su\sd}\,]   & +\eps_1\,{\rm diag}[\,G^{ba}_{\su\su}\,]        & -{\rm diag}[\,G^{ba}_{\su\sd}\,]  \\  
      +{\rm diag}[\,G^{bb}_{\sd\su}\,]  & -\eps_1\,{\rm diag}[\,G^{bb}_{\sd\sd}\,]          & -{\rm diag}[\,G^{ba}_{\sd\su}\,] & +\eps_1\,{\rm diag}[\,G^{ba}_{\sd\sd}\,]  \\
      +\eps_1\,{\rm diag}[\,G^{ab}_{\su\su}\,]          & -{\rm diag}[\,G^{ab}_{\su\sd}\,]  & -\eps_1\,{\rm diag}[\,G^{aa}_{\su\su}\,]         & +{\rm diag}[\,G^{aa}_{\su\sd}\,] \\ 
      -{\rm diag}[\,G^{ab}_{\sd\su}\,]  & +\eps_1\,{\rm diag}[\,G^{ab}_{\sd\sd}\,]          & +{\rm diag}[\,G^{aa}_{\sd\su}\,] & -\eps_1\,{\rm diag}[\,G^{aa}_{\sd\sd}\,]    \emat
\eeq

\medskip
and using the skew symmetry of $G$, 
\beq
{\rm diag}[\,G^{aa}_{\su\su}\,]\;\;=\;\;{\rm diag}[\,G^{aa}_{\sd\sd}\,] \;\;=\;\;{\rm diag}[\,G^{bb}_{\su\su}\,]\;\;=\;\;{\rm diag}[\,G^{bb}_{\sd\sd}\,] \;\;=\;\;0 
\eeq
and
\beq
{\rm diag}[\,G^{aa}_{\sd\su}\,]&=&-\,{\rm diag}[\,G^{aa}_{\su\sd}\,] \pS \\
{\rm diag}[\,G^{bb}_{\sd\su}\,]&=&-\,{\rm diag}[\,G^{bb}_{\su\sd}\,]  \\
{\rm diag}[\,G^{ba}_{\su\su}\,]&=&-\,{\rm diag}[\,G^{ab}_{\su\su}\,] \pS \\
{\rm diag}[\,G^{ba}_{\su\sd}\,]&=&-\,{\rm diag}[\,G^{ab}_{\sd\su}\,] 
\eeq
this can be written as 
\medskip
\beq
\eps_1\,dB^x\,G\,dB^x&=&dt
\bmat 
      0                                   & +{\rm diag}[\,G^{bb}_{\su\sd}\,]   & -\eps_1\,{\rm diag}[\,G^{ab}_{\su\su}\,]  & +{\rm diag}[\,G^{ab}_{\sd\su}\,]  \\  
      -{\rm diag}[\,G^{bb}_{\su\sd}\,]    & 0                                  & +{\rm diag}[\,G^{ab}_{\su\sd}\,]          & -\eps_1\,{\rm diag}[\,G^{ab}_{\sd\sd}\,]  \\
 +\eps_1\,{\rm diag}[\,G^{ab}_{\su\su}\,] & -{\rm diag}[\,G^{ab}_{\su\sd}\,]   & 0                                         & +{\rm diag}[\,G^{aa}_{\su\sd}\,] \\ 
 -{\rm diag}[\,G^{ab}_{\sd\su}\,]         & +\eps_1\,{\rm diag}[\,G^{ab}_{\sd\sd}\,] & -{\rm diag}[\,G^{aa}_{\su\sd}\,]    & 0   \emat
\eeq
which coincides with the first matrix in (\req{3.37}). 

\medskip
\bigskip
{\bf Part d)} We use the recursion (\req{3.16}) for $Z_k$ from Theorem 5.2 which was given by (let us write $dh_k$ instead of $dh^k$)
\medskip
\beq
Z_k&=&Z_{k-1}\;\cdot\;{\rm Pf}\bmat G_{k-1} & -Id \\ +Id & \tanh(dh_k/2) \emat\;\cdot\;
    {\rm Pf}\bmat \sqrt{2}\;\sinh(dh_k/4) & -Id \\ +Id & \sqrt{2}\;\sinh(dh_k/4) \emat  \pI 
\eeq

\medskip
and apply part (b) of Lemma 5.2,
\beq
\lefteqn{
{\rm Pf}(A+dA)\;\;=\;\;{\rm Pf}A\;\times\;  } \pI \\ 
&&\ts \Bigl\{\; 1\;+\; {1\over 2}\;Tr\bigl[ A^{-1} dA\,\bigr]\;+\;{1\over 8}\;\bigl(\,Tr\bigl[ A^{-1} dA\,\bigr]\,\bigr)^2 
  \;-\;{1\over 4}\;Tr\bigl[\,A^{-1}dA \,A^{-1} dA\,\bigr] \; \Bigr\} \pS
\eeq

\medskip
to the 2 Pfaffians on the right hand side in the recursion above. First let's consider 
\medskip
\beq
{\rm Pf}\bmat G_{k-1} & -Id \\ +Id & dg_k \emat &=& {\rm Pf}\biggl[\;\bmat G_{k-1} & -Id \\ +Id & 0 \emat \;+\; 
 \bmat 0 & 0 \\ 0 & dg_k \emat  \;\biggr] \;\;=:\;\;{\rm Pf}(A+dA)  \pI
\eeq
with 
\beq
dg_k&:=& \tanh(dh_k/2)\;\;=\;\;dh_k/2\;+\;O\bigl[\,(dh_k)^3\,\bigr] \pI 
\eeq
We have with $n:=\G$
\medskip
\beq
{\rm Pf}\bmat G_{k-1} & -Id \\ +Id & 0 \emat\;\;=\;\; {\rm Pf}\bmat 0 & -Id \\ +Id & 0 \emat \;\;=\;\;
(-1)^{4n(4n-1)/2}\;\det[\,-Id] \;\;=\;\;+1  \pI
\eeq

\medskip
Furthermore, since
\beq
\bmat G_{k-1} & -Id \\ +Id & 0 \emat^{-1}&=& \bmat 0  & +Id \\ -Id &  G_{k-1} \emat \pI 
\eeq
we have
\beq
A^{-1}dA&=&\bmat 0  & +Id \\ -Id &  G_{k-1} \emat \bmat 0 & 0 \\ 0 & dh_k/2 \emat  \;\;=\;\; \bmat 0 & dh_k/2 \\ 0 &G_{k-1}\,dh_k/2 \emat \pI
\eeq
and
\beq
[\,A^{-1}dA\,]^2&=& \bmat \;0\; & \;dh_k\,G_{k-1}\,dh_k/4 \;\\\; 0\; &\; G_{k-1}\,dh_k\,G_{k-1}\,dh_k/4\; \emat \pI
\eeq

\medskip
Thus we arrive at 
\medskip
\beq
\lefteqn{
\ts {\rm Pf}\bmat G_{k-1} & -Id \\ +Id & dg_k \emat \;\;=\;\;
1\;+\;  {1\over 2}\;Tr\bigl[ A^{-1} dA\,\bigr]\;+\;{1\over 8}\;\bigl(\,Tr\bigl[ A^{-1} dA\,\bigr]\,\bigr)^2 
  \;-\;{1\over 4}\;Tr\bigl[\,A^{-1}dA \,A^{-1} dA\,\bigr] } \pI \\ 
&&\\
&=&\ts  1\;+\; {1\over 4}\; Tr[\,G_{k-1}\,dh_k\,]\;+\;{1\over 32}\; \bigl(\,Tr[\,G_{k-1}\,dh_k\,]\,\bigr)^2 
   \;-\;{1\over 16}\; Tr[\,G_{k-1}\,dh_k\,G_{k-1}\,dh_k\,]  \pI
\eeq

\medskip
Now let's consider the second pfaffian, 
\medskip
\beq
\lefteqn{ 
{\rm Pf}\bmat \sqrt{2}\;\sinh(dh_k/4) & -Id \\ +Id & \sqrt{2}\;\sinh(dh_k/4) \emat\;\;=\;\;  } \pI \\ 
&&\phantom{I}\\
&& {\rm Pf}\biggl[\;\bmat 0 & -Id \\ +Id & 0 \emat \;+\; 
    \bmat \sqrt{2}\;\sinh(dh_k/4) & 0 \\ 0 & \sqrt{2}\;\sinh(dh_k/4)  \emat \;\biggr] \;\;=:\;\;{\rm Pf}(A+dA) \pI 
\eeq
with 
\beq
\sinh(dh_k/4)&=& dh_k/4\;+\;O(dh_k^3) \pI 
\eeq
We have 
\beq
A^{-1}&=&\bmat 0 & +Id \\ -Id & 0 \emat \pI 
\eeq
and
\beq
A^{-1}dA&=&\bmat 0 & +Id \\ -Id & 0 \emat \bmat \sqrt{2}\;dh_k/4 & 0 \\ 0 & \sqrt{2}\;dh_k/4  \emat 
  \;\;=\;\; \bmat 0 & \sqrt{2}\;dh_k/4  \\  -\sqrt{2}\;dh_k/4 & 0  \emat  \pI 
\eeq
and
\beq
[\,A^{-1}dA\,]^2&=&\bmat -(dh_k)^2/8 & 0\\ 0 & -(dh_k)^2/8 \emat \pI 
\eeq

\medskip
and we get 
\beq
{\rm Pf}(A+dA)&=&\ts {\rm Pf}A\,\Bigl\{\;1\;+\; {1\over 2}\;Tr\bigl[ A^{-1} dA\,\bigr]\;+\;{1\over 8}\;\bigl(\,Tr\bigl[ A^{-1} dA\,\bigr]\,\bigr)^2 
  \;-\;{1\over 4}\;Tr\bigl[\,A^{-1}dA \,A^{-1} dA\,\bigr] \; \Bigr\} \pI \\  
&=&\ts 1\;\;+\; {1\over 16}\;Tr[\,(dh_k)^2\,] \pI
\eeq
Thus, for the product of both pfaffians, we obtain 
\medskip
\beq
\lefteqn{
{\rm Pf}\bmat G_{k-1} & -Id \\ +Id & dg_k \emat\;\times\;{\rm Pf}\bmat \sqrt{2}\;\sinh(dh_k/4) & -Id \\ +Id & \sqrt{2}\;\sinh(dh_k/4) \emat \pI }  \\ 
\phantom{.}&&\\ 
&=&\ts  \Bigl\{\; 1\;+\; {1\over 4}\; Tr[\,G_{k-1}\,dh_k\,]\;+\;{1\over 32}\; \bigl(\,Tr[\,G_{k-1}\,dh_k\,]\,\bigr)^2 
     \;-\;{1\over 16}\; Tr[\,G_{k-1}\,dh_k\,G_{k-1}\,dh_k\,] \; \Bigr\} \;\times \pI  \\ 
&&\ts \Bigl\{\;1\;+\; {1\over 16}\;Tr[\,(dh_k)^2\,] \;\Bigr\} \;\;+\;O\bigl(\,(dh_k)^3\,\bigr)\pI \\ 
&=&\ts 1\;+\; {1\over 4}\; Tr[\,G_{k-1}\,dh_k\,]\;+\;{1\over 32}\; \bigl(\,Tr[\,G_{k-1}\,dh_k\,]\,\bigr)^2 
  \;-\;{1\over 16}\; Tr[\,G_{k-1}\,dh_k\,G_{k-1}\,dh_k\,] \pI \\ 
&&\ts \phantom{1}\;+\; {1\over 16}\;Tr[\,(dh_k)^2\,]  \pI
\eeq
This completes the proof of Theorem 5.3. \;$\blacksquare$

\goodbreak

\bigskip
\bigskip
\bigskip
\bigskip
{\bf 5.3 \;Girsanov Transformation}

\bigskip
\bigskip
Recall the pfaffian Monte Carlo representation of Theorem 5.2: For $\beta=t_k=kdt$, correlation functions are given by $\,C_\beta=C_{t_k}=\la \,G_{t_k}\,\ra\,$ with
\medskip
\beq
\la \,G_{t_k}\,\ra &=& {  \phantom{\Bigr|} \int_{\R^{3k\G}}\,G_{t_k}(\phi,\xi,\theta)\; Z_{t_k}(\phi,\xi,\theta)\; dP_k(\phi,\xi,\theta)  \phantom{\Bigr|} \over 
   \phantom{\Bigr|} \int_{\R^{3k\G}}\; Z_{t_k}(\phi,\xi,\theta)\; dP_k(\phi,\xi,\theta)  \phantom{\Bigr|} }  \pI 
\eeq

\medskip
Recall also the SDE (\req{3.38}) for $Z=Z_t=Z_t(\phi,\xi,\theta)\,$ from part (d) of Theorem 5.3, 
\medskip
\beq
dZ&=&\ts \Bigl\{\; {1\over 4}\; Tr[\,G\,dh\,]\;+\;{1\over 32}\; \bigl(\,Tr[\,G\,dh\,]\,\bigr)^2 
  \;-\;{1\over 16}\; Tr[\,G\,dh\,G\,dh\,] \;+\; {1\over 16}\;Tr[\,(dh)^2\,] \;\Bigr\}\; Z \pI
\eeq

\medskip
As in (\req{1.19}), we have for nonzero $Z$ 
\medskip
\beq
d\log Z&=&\ts {dZ \over Z} \;-\; {1\over 2}\,\bigl( {dZ\over Z}\bigr)^2 \pS  \nn \\ 
&=&\ts  {1\over 4}\; Tr[\,G\,dh\,]  \;-\;{1\over 16}\; Tr[\,G\,dh\,G\,dh\,] \;+\; {1\over 16}\;Tr[\,(dh)^2\,] \pI\pI\pI
\eeq
Thus, as long as there is no fermionic sign problem, we can write 
\medskip
\beq
Z_t&=&\ts Z_0\;\exp\Bigl\{\; \int_0^t \,\Bigl[\,  {1\over 4}\; Tr[\,G\,dh\,]
  \;-\;{1\over 16}\; Tr[\,G\,dh\,G\,dh\,] \;+\; {1\over 16}\;Tr[\,(dh)^2\,] \,\Bigr] \; \Bigr\} \pI\pI
\eeq

\medskip
We recall (\req{6.23}), 
\beq
dh&=& i\,h_0\, dt  \;+\; i\,  \sqrt{u} \; dB   \pS 
\eeq
and write
\beqn
Z_t&=&\ts Z_0\;\exp\Bigl\{\; \int_0^t \,\Bigl[\,  {i\over 4}\; Tr[\,G\,h_0\,]\,dt 
  \;-\;{1\over 16}\; Tr[\,G\,dh\,G\,dh\,] \;+\; {1\over 16}\;Tr[\,(dh)^2\,] \,\Bigr] \; \Bigr\} \;\times\; \pI \nn \\ 
&&\ts \phantom{Z_0}\; \exp\Bigl\{\;  {i\over 4}\,\sqrt{u}\;\int_0^t \,  Tr[\,G\,dB\,]\; \Bigr\} \pI \lbeq{3.48}
\eeqn

\goodbreak

with
\beqn
\ts Tr[\,G\,dB\,] &=&\ts \nu_1\,Tr[\,G\,dB^x\,] \,+\, \nu_2\,Tr[\,G\,dB^y\,] \,+\, \nu_3\,Tr[\,G\,dB^z\,]  \pI \lbeq{3.49}
\eeqn
At a given time step $\,t=t_\ell=\ell dt\,$, this is a linear combination of the integration variables $\phi_{j,\ell}\,$, $\xi_{j,\ell}$ and $\theta_{j,\ell}\,$. 
To obtain the exact coefficients, let us introduce the $4\G\times 4\G$ skew symmetric matrices 
\medskip
\beq
D_j&:=& \nu_1 \bmat      0     &  0           & +1_j    &  0             \\ 
                         0     &  0           &  0      &  -\eps_1\,1_j  \\
                         -1_j  &  0           &  0      &  0             \\ 
                         0     & +\eps_1\,1_j &  0      &  0             \emat  \pI
\eeq
\beq
E_j&:=& \nu_2 \bmat 0 &  0           & 0            & +1_j  \\ 
                    0 &  0           & +\eps_2\,1_j &  0    \\
                    0 & -\eps_2\,1_j &  0           &  0    \\ 
                 -1_j &            0 &  0           &  0    \emat  \pI 
\eeq
\beqn
F_j&:=& \nu_3 \bmat   0   & +1_j  &  0           &  0            \\ 
                    -1_j &  0    &  0           &  0            \\
                     0   &  0    &  0           & +\eps_3\,1_j  \\ 
                     0   &  0    & -\eps_3\,1_j &  0            \emat    \pI  \lbeq{3.50}
\eeqn

\medskip
where for a given lattice site $j\in\Gamma$ the quantity $\,1_j\,$ is the $\G\times \G$ matrix which has exactly one 1 on the diagonal at the $(j,j)$-entry and the 
remaining $\G^2-1$ matrix elements are all zero. That is, 
\beqn
1_{j_0}&:=&\bigl(\,\delta_{i,j}\,\delta_{j,j_0}\,\bigr)_{i,j\in\Gamma} \;\;\in\;\;\R^{\G\times\G}  \pI \lbeq{3.51}
\eeqn
With that notation, we can write 
\beqn
dB&=&\summ_j\,\bigl(\, D_j\,dx_j\,+\, E_j\,dy_j\,+\, F_j\,dz_j\, \bigr)  \pI  \lbeq{3.52}
\eeqn
and the contribution (\req{3.49}), now at some concrete time step $t_\ell=\ell dt$, can be written as
\beqn
\lefteqn{
 Tr[\,G_{\ell-1}\,dB_\ell\,] \;\;=\;\;
  \summ_j\,\Bigl[\; Tr[\,G_{\ell-1}\,D_j\,]\,dx_{j,\ell} \,+\, Tr[\,G_{\ell-1}\,E_j\,]\,dy_{j,\ell} \,+\,Tr[\,G_{\ell-1}\,F_j\,]\,dz_{j,\ell} \; \Bigr]  }\pI \nn \\ 
&=&\sqrt{dt}\,
  \summ_j \Bigl[\; Tr[\,G_{\ell-1}\,D_j\,]\,\phi_{j,\ell} \,+\, Tr[\,G_{\ell-1}\,E_j\,]\,\xi_{j,\ell} \,+\,Tr[\,G_{\ell-1}\,F_j\,]\,\theta_{j,\ell} \; \Bigr] \pI\pI\lbeq{3.53}
\eeqn
Here we recall again that, as in section 1, it is crucial that the $G$ in (\req{3.53}) above depends only on time $t_{\ell-1}$ but not on time $t_\ell$. This means that 
the integration variables which may occur in these $G$'s are
\beq
G_{\ell-1}&=&G_{\ell-1}\Bigl(\,\bigl\{\,\phi_{j,m},\xi_{j,m},\theta_{j,m}\,\bigr\}_{j\in\Gamma,\,1\le m\le \ell-1}\,\Bigr) \;, \pI
\eeq
but there are no $\,\phi_{j,\ell},\,\xi_{j,\ell},\,\theta_{j,\ell}\,$ in such a $G_{\ell-1}$. Exactly this feature allows us to absorb the contributions (\req{3.53}) 
or (\req{3.49}) into the integration measure $dP_k$ just by completing the square which then is called a Girsanov transformation in the mathematics literature. 
Collecting the quadratic terms from $dP_k(\phi,\xi,\theta)$ and the linear terms from $Tr[\,Gdh\,]$, we get the following terms in the exponential:
\medskip
\beqn
\lefteqn{
\ts -\,{1\over 2}\,\bigl(\phi_{j,\ell}^2+\xi_{j,\ell}^2+\theta_{j,\ell}^2\bigr) 
   \;+\; {i\over 4}\,\sqrt{udt}\,\Bigl(\, Tr[\,G_{\ell-1}D_j\,]\,\phi_{j,\ell} \,+\,Tr[\,G_{\ell-1}E_j\,]\,\xi_{j,\ell}
   \,+\,Tr[\,G_{\ell-1}F_j\,]\,\theta_{j,\ell} \,\Bigr)   }  \phantom{m} \pI \nn\\ 
&=&\ts -\,{1\over 2}\,\Bigl[\, \phi_{j,\ell}  \;-\;  {i\over 4}\,\sqrt{udt}\; Tr[\,G_{\ell-1}D_j\,]\, \Bigr]^2 \;+\;
   {1\over 2}\,\bigl(\,{i\over 4}\,\sqrt{udt}\; Tr[\,G_{\ell-1}D_j\,]\, \bigr)^2 \phantom{mmmm} \pI \nn \\ 
&&\ts -\,{1\over 2}\,\Bigl[\, \xi_{j,\ell}  \;-\;  {i\over 4}\,\sqrt{udt}\; Tr[\,G_{\ell-1}E_j\,]\, \Bigr]^2 \;+\;
   {1\over 2}\,\bigl(\,{i\over 4}\,\sqrt{udt}\; Tr[\,G_{\ell-1}E_j\,]\, \bigr)^2  \pI  \lbeq{3.54} \\
&&\ts -\,{1\over 2}\,\Bigl[\, \theta_{j,\ell}  \;-\;  {i\over 4}\,\sqrt{udt}\; Tr[\,G_{\ell-1}F_j\,]\, \Bigr]^2 \;+\;
   {1\over 2}\,\bigl(\,{i\over 4}\,\sqrt{udt}\; Tr[\,G_{\ell-1}F_j\,]\, \bigr)^2  \pI \nn  \\  
&=&\ts -\,{1\over 2}\,\bigl(\ti\phi_{j,\ell}^2+\ti\xi_{j,\ell}^2+\ti\theta_{j,\ell}^2\bigr) 
  \;-\;{udt\over 32}\, \Bigl[\,(\,Tr[\,GD_j\,]\,)^2 \,+\,(\,Tr[\,GE_j\,]\,)^2\,+\,(\,Tr[\,GF_j\,]\,)^2\,\Bigr] \pI \nn
\eeqn

\medskip
with the Girsanov transformed variables (analog for $\xi$ and $\theta$)
\beqn
\ti\phi_{j,\ell}&:=&\ts\phi_{j,\ell} \;-\;  {i\over 4}\,\sqrt{udt}\; Tr[\,G_{\ell-1}D_j\,] \pI \nn  \\ 
d\ti x_{j,\ell}&:=&\ts dx_{j,\ell} \;-\;  {i\over 4}\,\sqrt{u}\; Tr[\,G_{\ell-1}D_j\,] \;dt \pS \lbeq{3.55}
\eeqn

\medskip
Let us summarize in the following intermediate or preparatory

\goodbreak

\medskip
\bigskip
\bigskip
{\bf Proposition 5.1 (introduction Girsanov transformation):} With the following substitution of variables 
\medskip
\beqn
\ti\phi_{j,\ell}&:=&\ts\phi_{j,\ell} \;-\;  {i\over 4}\,\sqrt{udt}\; Tr[\,G_{\ell-1}\,D_j\,]  \nn \\ 
\ti\xi_{j,\ell}&:=&\ts\xi_{j,\ell} \;-\;  {i\over 4}\,\sqrt{udt}\; Tr[\,G_{\ell-1}\,E_j\,] \pI  \lbeq{3.56} \\ 
\ti\theta_{j,\ell}&:=&\ts\theta_{j,\ell} \;-\;  {i\over 4}\,\sqrt{udt}\; Tr[\,G_{\ell-1}\,F_j\,] \nn  
\eeqn 

\medskip
with $4\G\times 4\G$ skew symmetric matrices $D_j$, $E_j$ and $F_j$ given by (\req{3.50}) above, the pfaffian Monte Carlo representation (\req{3.14}) from Theorem 5.2 
changes to 
\medskip
\beqn
\la \,G_{t_k}\,\ra &=& {  \phantom{\Bigr|} \int_{\R^{3k\G}}\,G_{t_k}(\ti\phi,\ti\xi,\ti\theta)\; \exp\bigl\{\,{\ds\summ_{\ell=1}^k}\,V(G_{t_{\ell-1}})\,dt\,\bigr\}\; 
 dP_k(\ti\phi,\ti\xi,\ti\theta)  \phantom{\Bigr|} \over 
   \phantom{\Bigr|} \int_{\R^{3k\G}}\;\exp\bigl\{\,{\ds\summ_{\ell=1}^k}\,V(G_{t_{\ell-1}})\,dt\,\bigr\}\; dP_k(\ti\phi,\ti\xi,\ti\theta)  \phantom{\Bigr|} }  \pI \lbeq{3.57}
\eeqn

\medskip
with 
\beqn
V(G)&:=&  {\ts  {i\over 4}}\; Tr[\,G\,h_0\,]
   \;+\;{\ts{u\over 16}}\; \summ_j\,\Bigl[\, Tr[\,G\,D_j\,G\,D_j\,]\,+\,Tr[\,G\,E_j\,G\,E_j\,]\,+\,Tr[\,G\,F_j\,G\,F_j\,]\, \Bigr] \pI \nn \\
&& \phantom{\ts  {i\over 4}\; Tr[\,G\,h_0\,]} \; -\;{\ts {u\over 32}}\,\summ_j \ts\,  \Bigl[\,(\,Tr[\,G\,D_j\,]\,)^2 
  \,+\,(\,Tr[\,G\,E_j\,]\,)^2\,+\,(\,Tr[\,G\,F_j\,]\,)^2\,\Bigr]  \pI  \lbeq{3.58}
\eeqn

\medskip
and $G$ given by the SDE 
\beqn
dG&=&\ts +\,{1\over 2}\,(Id-G)\, \Bigl[\;i\,h_0\,dt\,+\,i\sqrt{u}\,d\ti B \,+\,{\rm newdrift}\,\cdot\,dt\; \Bigr] \, (Id+G)  \pI \lbeq{3.60}
\eeqn
with 
\beqn
{\rm newdrift} &=&{\ts {u\over 2}}\,\summ_j\,\Bigl[\, D_j\,G\,D_j\,+\,E_j\,G\,E_j\,+\,F_j\,G\,F_j\, \Bigr] \pI \nn \\
&&\;-\; {\ts {u\over 4}}\,\summ_j \Bigl[ \;D_j\, Tr[\,G\,D_j\,]\,+\,E_j\, Tr[\,G\,E_j\,]\,+\, F_j\, Tr[\,G\,F_j\,]\; \Bigr] \;\; \pI \lbeq{3.61}
\eeqn
The partition function itself is given by 
\beqn
Tr_{\cF}\bigl[\,e^{\,-\,t_kH}\,\bigr]\;\bigr/\;Tr_{\cF}[\,Id\,]
  &=&\ts  \int_{\R^{3k\G}}\;\exp\bigl\{\,{\ds\summ_{\ell=1}^k}\,V(G_{t_{\ell-1}})\,dt\,\bigr\}\; dP_k(\ti\phi,\ti\xi,\ti\theta)    \pI\pI 
\eeqn
\goodbreak
\medskip
\bigskip
{\bf Proof:}  Recall the formula for $Z_t$ in (\req{3.48}). We have  
\beq
dB&=&\summ_j\,\bigl(\, D_j\,dx_j\,+\, E_j\,dy_j\,+\, F_j\,dz_j\, \bigr)  \pI 
\eeq
which gives
\beq
dB\,G\,dB&=& dt\,\summ_j\,\bigl\{\, D_j\,G\,D_j\,+\,E_j\,G\,E_j\,+\,F_j\,G\,F_j\, \bigr\}  \pS 
\eeq
and
\beq
Tr[\,G\, dB\,G\,dB\,] &=&dt\,\summ_j\,\Bigl[\, Tr[\,G\,D_j\,G\,D_j\,]\,+\,Tr[\,G\,E_j\,G\,E_j\,]\,+\,Tr[\,G\,F_j\,G\,F_j\,]\, \Bigr]  \pI
\eeq
which results in the formula for the $V$ in the exponential. Furthermore, since
\beq
d\ti x_{j,\ell}&=&\ts dx_{j,\ell} \;-\;  {i\over 4}\,\sqrt{u}\; Tr[\,G_{\ell-1}\,D_j\,] \;dt \\
d\ti y_{j,\ell}&=&\ts dy_{j,\ell} \;-\;  {i\over 4}\,\sqrt{u}\; Tr[\,G_{\ell-1}\,E_j\,] \;dt \pI \\
d\ti z_{j,\ell}&=&\ts dz_{j,\ell} \;-\;  {i\over 4}\,\sqrt{u}\; Tr[\,G_{\ell-1}\,F_j\,] \;dt 
\eeq
we have
\beq
dB &=&d\ti B\;+\; i\sqrt{u}\; {\ts {dt\over 4}}\,\summ_j \Bigl[ \,D_j\, Tr[\,GD_j\,]\,+\,E_j\, Tr[\,GE_j\,]\,+\, F_j\, Tr[\,GF_j\,]\, \Bigr]  \pS \lbeq{3.62n}
\eeq
from which we get the formula for the new drift. Finally, recall from (\req{3.18}) of Theorem 5.2 that the partition function was given by 
\beq
Tr_\cF \,e^{-\beta H} &=&\ts\int_{\R^{3k\G}}\; Z_k(\phi,\xi,\theta)\; dP_k(\phi,\xi,\theta) \;\times\; e^{\,-\,\beta{u\over 4}w\eps\G} \;\; \pI
\eeq
Since 
\beq
(dB)^2&=& (\nu_1\,dB^x+\nu_2\,dB^y+\nu_3\,dB^z)^2\;\;=\;\;\nu_1^2\,(dB^x)^2 +\nu_2^2\,(dB^y)^2+\nu_3^2\,(dB^z)^2 \pS \\
&=&-\,\nu^2\,\Idv\,dt \;\;=\;\; -\,w\eps\,\Idv\,dt \pS
\eeq
we have
\beq
(dh)^2&=& (\,i\,h_0\, dt  + i\,  \sqrt{u} \; dB\,)^2\;\;=\;\; (\, i\,  \sqrt{u} \; dB\,)^2 \;\;=\;\;-\,u\,(dB)^2 
\;\;=\;\; +\,u\,w\eps\,\Idv \,dt \pS 
\eeq
and
\beq
\summ_{\ell=1}^k \ts \, {1\over 16}\;Tr[\,(dh_\ell)^2\,] &=&\ts  {u\over 16}\; w\eps\,4\G\, k\,dt \;\;=\;\; +\,\beta\,{u\over 4}\,w\eps\,\G \pS
\eeq
Thus the last trace term in the first line of (\req{3.48}) cancels exactly the explicit $\,e^{\,-\,\beta{u\over 4}w\eps\G}\,$ factor 
in the partition function (\req{3.18})\,. 
\;\;$\blacksquare$ 

\bigskip
\bigskip
To complete the proof of the Main Theorem, all what is left now is to calculate the terms in the exponential and the new drift 
in the Girsanov transformed SDE for $G$. The latter we do in the following Theorem 5.4 and the exponential we calculate in Theorem 5.5 below.

\bigskip
\bigskip
\bigskip
{\bf Theorem 5.4 (Girsanov transformed SDE for the density matrix $\bm{G}$):} The SDE (\req{3.60}) with new drift (\req{3.61}) simplifies to  
\beqn
dG&=&\ts +\,{1\over 2}\,(Id-G)\, \Bigl[\;i\,h_0\,dt\,+\,i\sqrt{u}\,d\ti B \,+\,{udt \over 2}\,DG\; \Bigr] \, (Id+G)  \pI \lbeq{3.62}
\eeqn
with an $(w_i,\eps_i)$ independent matrix 
\medskip
\beqn
DG&:=&+\,\bmat 
       0                         & +{\rm diag}[\,G^{bb}_{\su\sd}\,] & -{\rm diag}[\,G^{ab}_{\sd\sd}\,] & +{\rm diag}[\,G^{ab}_{\sd\su}\,]  \\ 
-{\rm diag}[\,G^{bb}_{\su\sd}\,] &          0                & +{\rm diag}[\,G^{ab}_{\su\sd}\,] & -{\rm diag}[\,G^{ab}_{\su\su}\,]  \\ 
+{\rm diag}[\,G^{ab}_{\sd\sd}\,] & -{\rm diag}[\,G^{ab}_{\su\sd}\,]  &        0                         & +{\rm diag}[\,G^{aa}_{\su\sd}\,]  \\  
-{\rm diag}[\,G^{ab}_{\sd\su}\,] & +{\rm diag}[\,G^{ab}_{\su\su}\,]  & -{\rm diag}[\,G^{aa}_{\su\sd}\,] & 0                       \emat  \pI\pI  \lbeq{3.63}
\eeqn

\medskip
which has been obtained through the combination 
\beqn
DG&=&\summ_j \, \Bigl\{\, D_jGD_j\,+\,E_jGE_j\,+\,F_jGF_j\, \Bigr\}  \pS  \nn \\ 
&&  \,-\,{\ts {1\over 2}}\,\summ_j \, \Bigl\{ \, D_j \,Tr[\,G D_j\,]\,+\,E_j \,Tr[\,G E_j\,]\,+\, F_j \,Tr[\,G F_j\,]\, \Bigr\}   \pS \pS
\eeqn
or more compactly 
\beqn
DG\,dt&=&\ts dB\,G\,dB\,-\,{1\over 2}\,Tr[\,G\,dB\,]\,dB \pS\pS  \lbeq{3.65}
\eeqn
and $d\ti B$ is the same as $dB$, just with the new variables. That is, with $\nu_i=\sqrt{w_i\eps_i}\,$, 
\beqn
d\ti B&=& 
\bmat 0 & +\nu_3\,d\ti z           & +\nu_1\,d\ti x & +\nu_2\,d\ti y           \\ 
      -\nu_3\,d\ti z & 0           & +\nu_2\eps_2\,d\ti y   & -\nu_1\eps_1\,d\ti x \\ 
    -\nu_1\,d\ti x & -\nu_2\eps_2\,d\ti y     & 0   & +\nu_3\eps_3\,d\ti z            \\ 
      -\nu_2\,d\ti y & +\nu_1\eps_1\,d\ti x & -\nu_3\eps_3\,d\ti z    & 0           \emat \;\;\;   \pI
\eeqn

\medskip
\bigskip
Apparently, because of equation (\req{3.65}), one may speculate whether the following, actually quite loosely formulated, conjecture holds:

\bigskip
\bigskip
{\bf Conjecture:} For an arbitrary Hubbard-Stratonovich factorization 
\beq
e^{\,-\,dt H_\rint}&=&\ts \int \, e^{\,-\,dQ(B)}\; dP(B)  \pI 
\eeq
with quadratic operator
\beq
dQ&=& i\, {\ts {\sqrt{u}\over 4}}\,  \bmat a_\su & a_\sd & b_\su & b_\sd \emat  \;dB\;  \bmat a_\su \\ a_\sd \\ b_\su \\ b_\sd \emat  \pI
\eeq
the quantity 
\beq
\ts dB\,G\,dB\,-\,{1\over 2}\,Tr[\,G\,dB\,]\,dB \;\;=:\;\; DG\,dt \pI
\eeq
is always given by (\req{3.63}).

\bigskip
\bigskip
{\bf Proof of Theorem 5.4:} In part (c) of Theorem 5.3 we calculated 
\beq
dB\,G\,dB&=& \bigl[\,w_1\,D_1G\;+\;w_2\,D_2G\;+\; w_3\,D_3G \,\bigr]\, dt \pI 
\eeq
with matrices $D_iG$ explicitely given in (\req{3.37}). From Proposition 5.1 we have 
\beq
i\sqrt{u}\,dB &=&i\sqrt{u}\,d\ti B \;-\; {\ts {udt\over 4}}\,\summ_j \Bigl[ \,D_j\, Tr[\,GD_j\,]\,+\,E_j\, Tr[\,GE_j\,]\,+\, F_j\, Tr[\,GF_j\,]\, \Bigr]  \pI
\eeq
Straightforward calculation gives
\medskip
\beq
Tr[\,GD_j\,]&=&\nu_1\,\bigl(\,G^{ba}_{\su\su,jj}\,-\,\eps_1\, G^{ba}_{\sd\sd,jj} \,-\,G^{ab}_{\su\su,jj} \,+\,\eps_1\, G^{ab}_{\sd\sd,jj} \, \bigr) 
\;\;=\;\;+\,2\,\nu_1\,(\,\eps_1\, G^{ab}_{\sd\sd,jj} \,-\,G^{ab}_{\su\su,jj} \,) \pS \\ 
Tr[\,GE_j\,]&=&\nu_2\,\bigl(\,G^{ba}_{\sd\su,jj} \,+\, \eps_2\,G^{ba}_{\su\sd,jj} \,-\, \eps_2\,G^{ab}_{\sd\su,jj} \,-\, G^{ab}_{\su\sd,jj} \,\bigr)
  \;\;=\;\; -\,2\,\nu_2\,(\,G^{ab}_{\su\sd,jj} \,+\, \eps_2\,G^{ab}_{\sd\su,jj} \,)  \pS \\
Tr[\,GF_j\,]&=&\nu_3\,\bigl(\,G^{aa}_{\sd\su,jj} \,-\,G^{aa}_{\su\sd,jj} \,+\,\eps_3\,G^{bb}_{\sd\su,jj}\,-\,\eps_3\,G^{bb}_{\su\sd,jj} \,\bigr)
\;\;=\;\;-\,2\,\nu_3\,( \,G^{aa}_{\su\sd,jj} \,+\,\eps_3\,G^{bb}_{\su\sd,jj}\, )  \pS
\eeq

\medskip
where we used the skew symmetry of $G$ in the second equal signs. Thus,
\medskip
\beq
{\ts {u dt\over 4}}\,\summ_j \, D_j \,Tr[\,GD_j\,]\;\;=\;\; {\ts {u dt\over 2}}\,\nu_1^2\,\summ_j \,(\,\eps_1\, G^{ab}_{\sd\sd,jj} \,-\,G^{ab}_{\su\su,jj} \,) 
\bmat      0     &  0           & +1_j    &  0             \\ 
                         0     &  0           &  0      &  -\eps_1\,1_j  \\
                         -1_j  &  0           &  0      &  0             \\ 
                         0     & +\eps_1\,1_j &  0      &  0             \emat  \pI 
\eeq
\beq
\;\;=\;\; {\ts {u dt\over 2}}\,w_1\,
\bmat      0     &  0           & +{\rm diag}[\, G^{ab}_{\sd\sd}  \,]    &  0             \\ 
                         0     &  0           &  0      &  +{\rm diag}[\,G^{ab}_{\su\su} \,]  \\
                         -{\rm diag}[\, G^{ab}_{\sd\sd}  \,]  &  0           &  0      &  0             \\ 
                         0     & -{\rm diag}[\, G^{ab}_{\su\su} \,] &  0      &  0             \emat  \pI  
\eeq
\beq
+\, {\ts {u dt\over 2}}\,w_1\,\eps_1\,
\bmat      0     &  0           & -{\rm diag}[\, G^{ab}_{\su\su} \,]    &  0             \\ 
                         0     &  0           &  0      &  -{\rm diag}[\, G^{ab}_{\sd\sd}  \,]  \\
                         +{\rm diag}[\, G^{ab}_{\su\su} \,]  &  0           &  0      &  0             \\ 
                         0     & +{\rm diag}[\, G^{ab}_{\sd\sd}  \,] &  0      &  0             \emat  \pI
\eeq

\medskip
Now the last matrix with the $\eps_1$ in front of it cancels exactly the $\eps_1$ entries in the $w_1\,D_1G$ matrix, we have 
\beq
+\,{\ts {u dt\over 2}}\,w_1\,D_1G  \,-\,{\ts {u dt\over 4}}\,\summ_j \, D_j \,Tr[G D_j]   \pI 
\eeq
\beq
\;\;=\;\;+\,{\ts {u dt\over 2}}\,w_1
\bmat 
      0                                   & +{\rm diag}[\,G^{bb}_{\su\sd}\,]   & -\eps_1\,{\rm diag}[\,G^{ab}_{\su\su}\,]  & +{\rm diag}[\,G^{ab}_{\sd\su}\,]  \\  
      -{\rm diag}[\,G^{bb}_{\su\sd}\,]    & 0                                  & +{\rm diag}[\,G^{ab}_{\su\sd}\,]          & -\eps_1\,{\rm diag}[\,G^{ab}_{\sd\sd}\,]  \\
 +\eps_1\,{\rm diag}[\,G^{ab}_{\su\su}\,] & -{\rm diag}[\,G^{ab}_{\su\sd}\,]   & 0                                         & +{\rm diag}[\,G^{aa}_{\su\sd}\,] \\ 
 -{\rm diag}[\,G^{ab}_{\sd\su}\,]         & +\eps_1\,{\rm diag}[\,G^{ab}_{\sd\sd}\,] & -{\rm diag}[\,G^{aa}_{\su\sd}\,]    & 0   \emat \pI 
\eeq
\beq
\;-\; {\ts {u dt\over 2}}\,w_1\,
\bmat      0     &  0           & +{\rm diag}[\, G^{ab}_{\sd\sd}  \,]    &  0             \\ 
                         0     &  0           &  0      &  +{\rm diag}[\,G^{ab}_{\su\su} \,]  \\
                         -{\rm diag}[\, G^{ab}_{\sd\sd}  \,]  &  0           &  0      &  0             \\ 
                         0     & -{\rm diag}[\, G^{ab}_{\su\su} \,] &  0      &  0             \emat  \pI  
\eeq
\beq
\;-\; {\ts {u dt\over 2}}\,w_1\,\eps_1\,
\bmat      0     &  0           & -{\rm diag}[\, G^{ab}_{\su\su} \,]    &  0             \\ 
                         0     &  0           &  0      &  -{\rm diag}[\, G^{ab}_{\sd\sd}  \,]  \\
                         +{\rm diag}[\, G^{ab}_{\su\su} \,]  &  0           &  0      &  0             \\ 
                         0     & +{\rm diag}[\, G^{ab}_{\sd\sd}  \,] &  0      &  0             \emat  \pI
\eeq

\beq
\;\;=\;\;+\,{\ts {u dt\over 2}}\,w_1
\bmat 
      0                                   & +{\rm diag}[\,G^{bb}_{\su\sd}\,]   & -{\rm diag}[\,G^{ab}_{\sd\sd}\,]  & +{\rm diag}[\,G^{ab}_{\sd\su}\,]  \\  
      -{\rm diag}[\,G^{bb}_{\su\sd}\,]    & 0                                  & +{\rm diag}[\,G^{ab}_{\su\sd}\,]          & -{\rm diag}[\,G^{ab}_{\su\su}\,]  \\
 +{\rm diag}[\,G^{ab}_{\sd\sd}\,] & -{\rm diag}[\,G^{ab}_{\su\sd}\,]   & 0                                         & +{\rm diag}[\,G^{aa}_{\su\sd}\,] \\ 
 -{\rm diag}[\,G^{ab}_{\sd\su}\,]         & +{\rm diag}[\,G^{ab}_{\su\su}\,] & -{\rm diag}[\,G^{aa}_{\su\sd}\,]    & 0   \emat 
\;\;=\;\; +\,{\ts {u dt\over 2}}\,w_1\, DG\pI 
\eeq

\bigskip
Quite remarkably, the same effect occurs for the $E$ and $F$ matrices and we always end up with the same matrix $DG$: We have
\medskip
\beq
-\,{\ts {u dt\over 4}}\,\summ_j \, E_j \,Tr[G E_j] &=&+\,{\ts {u dt\over 2}}\,\summ_j \, \nu_2^2 \,(\,G^{ab}_{\su\sd,jj} \,+\, \eps_2\,G^{ab}_{\sd\su,jj} \,) 
\bmat 0 &  0           & 0            & +1_j  \\ 
      0 &  0           & +\eps_2\,1_j &  0    \\
      0 & -\eps_2\,1_j &  0           &  0    \\ 
   -1_j &            0 &  0           &  0    \emat  \pI
\eeq
\beq 
&=&+\,{\ts {u dt\over 2}}\,w_2
\bmat 0 &  0           & 0            & +{\rm diag}[\,G^{ab}_{\sd\su}\,]  \\ 
      0 &  0           & +{\rm diag}[\,G^{ab}_{\su\sd}\,] &  0    \\
      0 & -{\rm diag}[\,G^{ab}_{\su\sd}\,] &  0           &  0    \\ 
   -{\rm diag}[\,G^{ab}_{\sd\su}\,] &            0 &  0           &  0    \emat  \pI\\
&&\phantom{I} \\ 
&&+\,{\ts {u dt\over 2}}\,w_2\,\eps_2
\bmat 0 &  0           & 0            & +{\rm diag}[\,G^{ab}_{\su\sd}\,]  \\ 
      0 &  0           & +{\rm diag}[\,G^{ab}_{\sd\su}\,] &  0    \\
      0 & -{\rm diag}[\,G^{ab}_{\sd\su}\,] &  0           &  0    \\ 
   -{\rm diag}[\,G^{ab}_{\su\sd}\,] &            0 &  0           &  0    \emat  \pI
\eeq

\medskip
which gives
\beq
+\,{\ts {u dt\over 2}}\,w_2\,D_2G \;-\;{\ts {u dt\over 4}}\,\summ_j \, E_j \,Tr[G E_j]  \pI 
\eeq
\beq
\;=\;+\,{\ts {u dt\over 2}}\,w_2
\bmat
                      0          & +{\rm diag}[\,G^{bb}_{\su\sd}\,]   & -{\rm diag}[\,G^{ab}_{\sd\sd}\,]                & -\eps_2\,{\rm diag}[\,G^{ab}_{\su\sd}\,]  \\ 
-{\rm diag}[\,G^{bb}_{\su\sd}\,] &           0                        & -\eps_2\,{\rm diag}[\,G^{ab}_{\sd\su}\,]        & -{\rm diag}[\,G^{ab}_{\su\su}\,]\\ 
+{\rm diag}[\,G^{ab}_{\sd\sd}\,] & +\eps_2\,{\rm diag}[\,G^{ab}_{\sd\su}\,]         &            0                      & +{\rm diag}[\,G^{aa}_{\su\sd}\,]\\ 
+\eps_2\,{\rm diag}[\,G^{ab}_{\su\sd}\,]         & +{\rm diag}[\,G^{ab}_{\su\su}\,] & -{\rm diag}[\,G^{aa}_{\su\sd}\,]  &      0                          \emat \pI
\eeq
\beq
\;+\;{\ts {u dt\over 2}}\,w_2
\bmat 0 &  0           & 0            & +{\rm diag}[\,G^{ab}_{\sd\su}\,]  \\ 
      0 &  0           & +{\rm diag}[\,G^{ab}_{\su\sd}\,] &  0    \\
      0 & -{\rm diag}[\,G^{ab}_{\su\sd}\,] &  0           &  0    \\ 
   -{\rm diag}[\,G^{ab}_{\sd\su}\,] &            0 &  0           &  0    \emat  \pI
\eeq
\beq
\;+\;{\ts {u dt\over 2}}\,w_2\,\eps_2
\bmat 0 &  0           & 0            & +{\rm diag}[\,G^{ab}_{\su\sd}\,]  \\ 
      0 &  0           & +{\rm diag}[\,G^{ab}_{\sd\su}\,] &  0    \\
      0 & -{\rm diag}[\,G^{ab}_{\sd\su}\,] &  0           &  0    \\ 
   -{\rm diag}[\,G^{ab}_{\su\sd}\,] &            0 &  0           &  0    \emat  \pI
\eeq
\beq
\;=\;+\,{\ts {u dt\over 2}}\,w_2
\bmat
                      0          & +{\rm diag}[\,G^{bb}_{\su\sd}\,]   & -{\rm diag}[\,G^{ab}_{\sd\sd}\,]                & +{\rm diag}[\,G^{ab}_{\sd\su}\,]  \\ 
-{\rm diag}[\,G^{bb}_{\su\sd}\,] &           0                        & +{\rm diag}[\,G^{ab}_{\su\sd}\,]        & -{\rm diag}[\,G^{ab}_{\su\su}\,]\\ 
+{\rm diag}[\,G^{ab}_{\sd\sd}\,] & -{\rm diag}[\,G^{ab}_{\su\sd}\,]         &            0                      & +{\rm diag}[\,G^{aa}_{\su\sd}\,]\\ 
-{\rm diag}[\,G^{ab}_{\sd\su}\,]         & +{\rm diag}[\,G^{ab}_{\su\su}\,] & -{\rm diag}[\,G^{aa}_{\su\sd}\,]  &      0                          \emat 
\;=\;+\,{\ts {u dt\over 2}}\,w_2\,DG \pI
\eeq

\medskip
Finally, 
\beq
-\,{\ts {u dt\over 4}}\,\summ_j \, F_j \,Tr[\,GF_j\,] \;\;=\;\;+\,{\ts {u dt\over 2}}\,\summ_j \, \nu_3^2 \,( \,G^{aa}_{\su\sd,jj} \,+\,\eps_3\,G^{bb}_{\su\sd,jj}\, ) 
\bmat   0   & +1_j  &  0           &  0            \\ 
       -1_j &  0    &  0           &  0            \\
        0   &  0    &  0           & +\eps_3\,1_j  \\ 
        0   &  0    & -\eps_3\,1_j &  0            \emat   \phantom{mmm} \pI 
\eeq
\beq 
&=&+\,{\ts {u dt\over 2}}\, w_3
\bmat   0   & +{\rm diag}[\,G^{bb}_{\su\sd}\,]  &  0           &  0            \\ 
       -{\rm diag}[\,G^{bb}_{\su\sd}\,] &  0    &  0           &  0            \\
        0   &  0    &  0           & +{\rm diag}[\,G^{aa}_{\su\sd}\,]  \\ 
        0   &  0    & -{\rm diag}[\,G^{aa}_{\su\sd}\,] &  0            \emat  \pI \\ 
&&\phantom{I} \\ 
&&+\,{\ts {u dt\over 2}}\, w_3\eps_3
\bmat   0   & +{\rm diag}[\,G^{aa}_{\su\sd}\,]  &  0           &  0            \\ 
       -{\rm diag}[\,G^{aa}_{\su\sd}\,] &  0    &  0           &  0            \\
        0   &  0    &  0           & +{\rm diag}[\,G^{bb}_{\su\sd}\,]  \\ 
        0   &  0    & -{\rm diag}[\,G^{bb}_{\su\sd}\,] &  0            \emat  \pI
\eeq

\medskip
which gives 
\beq
+\,{\ts {u dt\over 2}}\,w_3\,D_3G \,-\,{\ts {u dt\over 4}}\,\summ_j \, F_j \,Tr[G F_j] \;\;=\;\;
\eeq
\beq
+\,{\ts {u dt\over 2}}\,w_3
\bmat 
       0                         & -\eps_3\,{\rm diag}[\,G^{aa}_{\su\sd}\,] & -{\rm diag}[\,G^{ab}_{\sd\sd}\,] & +{\rm diag}[\,G^{ab}_{\sd\su}\,]  \\ 
+\eps_3\,{\rm diag}[\,G^{aa}_{\su\sd}\,] &          0                & +{\rm diag}[\,G^{ab}_{\su\sd}\,] & -{\rm diag}[\,G^{ab}_{\su\su}\,]  \\ 
+{\rm diag}[\,G^{ab}_{\sd\sd}\,] & -{\rm diag}[\,G^{ab}_{\su\sd}\,]  &        0                         & -\eps_3\,{\rm diag}[\,G^{bb}_{\su\sd}\,]  \\  
-{\rm diag}[\,G^{ab}_{\sd\su}\,] & +{\rm diag}[\,G^{ab}_{\su\su}\,]  & +\eps_3\,{\rm diag}[\,G^{bb}_{\su\sd}\,] & 0                       \emat 
\eeq
\beq
+\,{\ts {u dt\over 2}}\, w_3
\bmat   0   & +{\rm diag}[\,G^{bb}_{\su\sd}\,]  &  0           &  0            \\ 
       -{\rm diag}[\,G^{bb}_{\su\sd}\,] &  0    &  0           &  0            \\
        0   &  0    &  0           & +{\rm diag}[\,G^{aa}_{\su\sd}\,]  \\ 
        0   &  0    & -{\rm diag}[\,G^{aa}_{\su\sd}\,] &  0            \emat  \pI 
\eeq
\beq
+\,{\ts {u dt\over 2}}\, w_3\,\eps_3
\bmat   0   & +{\rm diag}[\,G^{aa}_{\su\sd}\,]  &  0           &  0            \\ 
       -{\rm diag}[\,G^{aa}_{\su\sd}\,] &  0    &  0           &  0            \\
        0   &  0    &  0           & +{\rm diag}[\,G^{bb}_{\su\sd}\,]  \\ 
        0   &  0    & -{\rm diag}[\,G^{bb}_{\su\sd}\,] &  0            \emat  \pI
\eeq
\beq
\;\;=\;\;+\,{\ts {u dt\over 2}}\,w_3
\bmat 
       0                         & +{\rm diag}[\,G^{bb}_{\su\sd}\,] & -{\rm diag}[\,G^{ab}_{\sd\sd}\,] & +{\rm diag}[\,G^{ab}_{\sd\su}\,]  \\ 
-{\rm diag}[\,G^{bb}_{\su\sd}\,] &          0                & +{\rm diag}[\,G^{ab}_{\su\sd}\,] & -{\rm diag}[\,G^{ab}_{\su\su}\,]  \\ 
+{\rm diag}[\,G^{ab}_{\sd\sd}\,] & -{\rm diag}[\,G^{ab}_{\su\sd}\,]  &        0                         & +{\rm diag}[\,G^{aa}_{\su\sd}\,]  \\  
-{\rm diag}[\,G^{ab}_{\sd\su}\,] & +{\rm diag}[\,G^{ab}_{\su\su}\,]  & -{\rm diag}[\,G^{aa}_{\su\sd}\,] & 0                       \emat 
\;\;=\;\;+\,{\ts {u dt\over 2}}\,w_3\,DG
\eeq

\bigskip
This completes the proof of Theorem 5.4. \;\;$\blacksquare$

\goodbreak
\bigskip
\bigskip
\bigskip
{\bf Theorem 5.5 (remaining exponential is the energy):} The effective potential $V(G)$ given by (\req{3.58}) of Proposition 5.1,
\beq
V(G)&:=&  {\ts  {i\over 4}}\; Tr[\,G\,h_0\,]
   \;+\;{\ts{u\over 16}}\; \summ_j\,\Bigl[\, Tr[\,G\,D_j\,G\,D_j\,]\,+\,Tr[\,G\,E_j\,G\,E_j\,]\,+\,Tr[\,G\,F_j\,G\,F_j\,]\, \Bigr] \pI \\
&& \phantom{\ts  {i\over 4}\; Tr[\,G\,h_0\,]} \; -\;{\ts {u\over 32}}\,\summ_j \ts\,  \Bigl[\,(\,Tr[\,G\,D_j\,]\,)^2 
  \,+\,(\,Tr[\,G\,E_j\,]\,)^2\,+\,(\,Tr[\,G\,F_j\,]\,)^2\,\Bigr]  \pI
\eeq
simplifies to 
\beqn
V(G)&=& {\ts{i\over 4}}\; Tr[\,G\,h_0\,]
 \;+\; {\ts {u\over 4}}\,\summ_j\,\bigl(\,G^{ab}_{\su\su,jj}\;G^{ab}_{\sd\sd,jj} 
   \,-\,G^{ab}_{\su\sd,jj}\;G^{ab}_{\sd\su,jj} \, -\,G^{aa}_{\su\sd,jj}\;G^{bb}_{\su\sd,jj} \,\bigr)   \pI\pI \lbeq{3.67}
\eeqn

\bigskip
{\bf Proof:} In the proof of Theorem 5.4 we have shown the following identities:
\medskip
\beq
\summ_j \, D_j \,G\, D_j  \,-\,{\ts {1\over 2}}\,\summ_j \, D_j \,Tr[\,G\, D_j\,] &=&w_1\,DG   \\ 
\summ_j \, E_j \,G\, E_j  \;-\;{\ts {1\over 2}}\,\summ_j \, E_j \,Tr[\,G\, E_j\,] &=&w_2\,DG  \\ 
\summ_j \, F_j \,G\, F_j  \;\,-\;\,{\ts {1\over 2}}\,\summ_j \, F_j \,Tr[\,G\, F_j\,] &=&w_3\,DG   
\eeq

\medskip
where the $DG$ on the right is the matrix (\req{3.63}) of Theorem 5.4. We multiply with $G$ from the left and take the trace. We obtain
\medskip
\beq
\summ_j \,Tr[\,G \, D_j \,G\, D_j\,]  \,-\,{\ts {1\over 2}}\,\summ_j \,Tr[\,G \, D_j\,] \,Tr[\,G\, D_j\,] &=&w_1\,Tr[\,G\,DG \,]  \\ 
\summ_j \,Tr[\,G \, E_j \,G\, E_j\,]  \;-\;{\ts {1\over 2}}\,\summ_j \,Tr[\,G \, E_j\,] \,Tr[\,G\, E_j\,] &=&w_2\,Tr[\,G\,DG \,] \\ 
\summ_j \,Tr[\,G \, F_j \,G\, F_j\,]  \;\,-\;\,{\ts {1\over 2}}\,\summ_j \,Tr[\,G \, F_j\,] \,Tr[\,G\, F_j\,] &=&w_3\,Tr[\,G\,DG \,]  
\eeq

\medskip

which gives, using $\,w_1+w_2+w_3=1\,$, 
\beq
 \summ_j\,\Bigl\{\, Tr[\,G\,D_j\,G\,D_j\,]\,+\,Tr[\,G\,E_j\,G\,E_j\,]\,+\,Tr[\,G\,F_j\,G\,F_j\,]\, \Bigr\} && \pI \\
 \; -\;{\ts {1\over 2}}\,\summ_j \ts\,  \Bigl[\,(\,Tr[\,G\,D_j\,]\,)^2 
  \,+\,(\,Tr[\,G\,E_j\,]\,)^2\,+\,(\,Tr[\,G\,F_j\,]\,)^2\,\Bigr] &=& Tr[\,G\,DG \,]  \pI 
\eeq
with the trace on the right hand side given by 
\medskip
\beq
Tr\bigl[\; G\, DG \;\bigr]\;\;=\;\;\phantom{mmmmmmmmmmmmmmmmmmmmmmmmmmmmmmmmmm} \pI\\
 Tr 
\bmat G^{aa}_{\su\su} & G^{aa}_{\su\sd} & G^{ab}_{\su\su} & G^{ab}_{\su\sd}  \\ 
      G^{aa}_{\sd\su} & G^{aa}_{\sd\sd} & G^{ab}_{\sd\su} & G^{ab}_{\sd\sd}  \\ 
      G^{ba}_{\su\su} & G^{ba}_{\su\sd} & G^{bb}_{\su\su} & G^{bb}_{\su\sd}  \\  
      G^{ba}_{\sd\su} & G^{ba}_{\sd\sd} & G^{bb}_{\sd\su} & G^{bb}_{\sd\sd}  \emat 
\bmat 
      0                                   & +{\rm diag}[\,G^{bb}_{\su\sd}\,]   & -{\rm diag}[\,G^{ab}_{\sd\sd}\,]   & +{\rm diag}[\,G^{ab}_{\sd\su}\,]  \\  
      -{\rm diag}[\,G^{bb}_{\su\sd}\,]    & 0                                  & +{\rm diag}[\,G^{ab}_{\su\sd}\,]   & -{\rm diag}[\,G^{ab}_{\su\su}\,]  \\
 +{\rm diag}[\,G^{ab}_{\sd\sd}\,]         & -{\rm diag}[\,G^{ab}_{\su\sd}\,]   & 0                                  & +{\rm diag}[\,G^{aa}_{\su\sd}\,] \\ 
 -{\rm diag}[\,G^{ab}_{\sd\su}\,]         & +{\rm diag}[\,G^{ab}_{\su\su}\,]   & -{\rm diag}[\,G^{aa}_{\su\sd}\,]   & 0   \emat  
\eeq

\medskip
That is, 
\beq
Tr\bigl[\; G\, DG \;\bigr] &=&\;+\;Tr\Bigl\{\, -\,G^{aa}_{\su\sd}\;{\rm diag}[\,G^{bb}_{\su\sd}\,]\,+\,G^{ab}_{\su\su}\;{\rm diag}[\,G^{ab}_{\sd\sd}\,]
   \,-\,G^{ab}_{\su\sd}\;{\rm diag}[\,G^{ab}_{\sd\su}\,] \;\Bigr\} \pI \\ 
&&\;+\;Tr\Bigl\{\,+\,G^{aa}_{\sd\su}\;{\rm diag}[\,G^{bb}_{\su\sd}\,] \,-\, G^{ab}_{\sd\su}\;{\rm diag}[\,G^{ab}_{\su\sd}\,]
     \,+\,G^{ab}_{\sd\sd}\;{\rm diag}[\,G^{ab}_{\su\su}\,] \; \Bigr\} \pI \\ 
&&\;+\;Tr\Bigl\{\,-\,G^{ba}_{\su\su}\;{\rm diag}[\,G^{ab}_{\sd\sd}\,]  \,+\, G^{ba}_{\su\sd}\;{\rm diag}[\,G^{ab}_{\su\sd}\,]
     \,-\, G^{bb}_{\su\sd} \;{\rm diag}[\,G^{aa}_{\su\sd}\,] \;\Bigr\} \pI \\ 
&&\;+\;Tr\Bigl\{\,+\,G^{ba}_{\sd\su}\;{\rm diag}[\,G^{ab}_{\sd\su}\,] \,-\, G^{ba}_{\sd\sd}\;{\rm diag}[\,G^{ab}_{\su\su}\,] 
     \,+\, G^{bb}_{\sd\su}\; {\rm diag}[\,G^{aa}_{\su\sd}\,]  \; \Bigr\}  \pI
\eeq
or
\beq
Tr\bigl[\; G\, DG \;\bigr] &=&\;+\;\summ_j\,\Bigl\{\, -\,G^{aa}_{\su\sd,jj}\;G^{bb}_{\su\sd,jj}\,+\,G^{ab}_{\su\su,jj}\;G^{ab}_{\sd\sd,jj} 
   \,-\,G^{ab}_{\su\sd,jj}\;G^{ab}_{\sd\su,jj} \;\Bigr\} \pI \\ 
&&\;+\;\summ_j\,\Bigl\{\,+\,G^{aa}_{\sd\su,jj}\;G^{bb}_{\su\sd,jj} \,-\, G^{ab}_{\sd\su,jj}\;G^{ab}_{\su\sd,jj}
     \,+\,G^{ab}_{\sd\sd,jj}\; G^{ab}_{\su\su,jj} \; \Bigr\} \pI \\ 
&&\;+\;\summ_j\,\Bigl\{\,-\,G^{ba}_{\su\su,jj}\;G^{ab}_{\sd\sd,jj}  \,+\, G^{ba}_{\su\sd,jj}\;G^{ab}_{\su\sd,jj}
     \,-\, G^{bb}_{\su\sd,jj} \;G^{aa}_{\su\sd,jj} \;\Bigr\} \pI \\ 
&&\;+\;\summ_j\,\Bigl\{\,+\,G^{ba}_{\sd\su,jj}\;G^{ab}_{\sd\su,jj} \,-\, G^{ba}_{\sd\sd,jj}\;G^{ab}_{\su\su,jj}  
     \,+\, G^{bb}_{\sd\su,jj}\; G^{aa}_{\su\sd,jj}  \; \Bigr\}  \pI
\eeq
or, using the skew symmetry of $G$, 
\beq
Tr\bigl[\; G\, DG \;\bigr] &=&\;+\;\summ_j\,\Bigl\{\, -\,G^{aa}_{\su\sd,jj}\;G^{bb}_{\su\sd,jj}\,+\,G^{ab}_{\su\su,jj}\;G^{ab}_{\sd\sd,jj} 
   \,-\,G^{ab}_{\su\sd,jj}\;G^{ab}_{\sd\su,jj} \;\Bigr\} \pI \\ 
&&\;+\;\summ_j\,\Bigl\{\,-\,G^{aa}_{\su\sd,jj}\;G^{bb}_{\su\sd,jj} \,-\, G^{ab}_{\sd\su,jj}\;G^{ab}_{\su\sd,jj}
     \,+\,G^{ab}_{\sd\sd,jj}\; G^{ab}_{\su\su,jj} \; \Bigr\} \pI \\ 
&&\;+\;\summ_j\,\Bigl\{\,+\,G^{ab}_{\su\su,jj}\;G^{ab}_{\sd\sd,jj}  \,-\, G^{ab}_{\sd\su,jj}\;G^{ab}_{\su\sd,jj}
     \,-\, G^{bb}_{\su\sd,jj} \;G^{aa}_{\su\sd,jj} \;\Bigr\} \pI \\ 
&&\;+\;\summ_j\,\Bigl\{\,-\,G^{ab}_{\su\sd,jj}\;G^{ab}_{\sd\su,jj} \,+\, G^{ab}_{\sd\sd,jj}\;G^{ab}_{\su\su,jj}  
     \,-\, G^{bb}_{\su\sd,jj}\; G^{aa}_{\su\sd,jj}  \; \Bigr\}  \pI
\eeq
which is the same as 
\beq
Tr\bigl[\; G\, DG \;\bigr] &=& 4\,\summ_j\,\Bigl\{\,+\,G^{ab}_{\su\su,jj}\;G^{ab}_{\sd\sd,jj} 
   \,-\,G^{ab}_{\su\sd,jj}\;G^{ab}_{\sd\su,jj} \, -\,G^{aa}_{\su\sd,jj}\;G^{bb}_{\su\sd,jj} \;\Bigr\} \pI
\eeq
Taking into account the prefactor of $u/16$ in (\req{3.58}), the theorem follows. \;\;$\blacksquare$

\goodbreak
\bigskip
\bigskip
\bigskip
{\bf 5.4 \;Correlation Functions}

\bigskip
\bigskip
Now we can finalize the proof of the Main Theorem. Recall the identities (\req{n2.11}) and (\req{n2.12}), 
\beq
i\;a_{i\sigma}\,b_{j\tau}&=&\ts c^+_{i\sigma}\,c_{j\tau} \, +\,c^+_{j\tau} \,c_{i\sigma} \;+\; c_{i\sigma}\,c_{j\tau}  \, + \,c^+_{j\tau}\, c^+_{i\sigma}
   \;-\;\delta_{i\sigma,j\tau} \pS \nn \\ 
\ts {i\over 2}\;(\,a_{i\sigma}\,b_{j\tau}\,+\,a_{j\tau}\,b_{i\sigma}\,)&=&\ts c^+_{i\sigma}\,c_{j\tau} \, + \,c^+_{j\tau}\,c_{i\sigma} \,-\,\delta_{i\sigma,j\tau} \pS \nn \\ 
\ts {i\over 2}\;(\,a_{i\sigma}\,b_{j\tau}\,-\,a_{j\tau}\,b_{i\sigma}\,)&=&\ts c_{i\sigma}\,c_{j\tau} \, + \, c^+_{j\tau}\,c^+_{i\sigma}    \pS 
\eeq
from which we get the correlations (\req{2.34}-\req{2.36}),
\beq
\ts \la c_{j\su}^+c_{j\su}\ra\;-\;{1\over 2} &=&\ts +\,{i\over 2}\,\la\, G^{ab}_{\su\su,jj}\, \ra  \pI \\ 
\la\,c_{j\up}^+\,c_{j\down}\,+\,c_{j\down}^+\,c_{j\up}\,\ra &=&+\,{\ts{i\over 2}}\,\bigl\la \,G^{ab}_{\su\sd,jj} \,+\,G^{ab}_{\sd\su,jj}  \,\bigr\ra   \\
\la c_{j\up}^+\,c^+_{j\down}\,+\,c_{j\down}\,c_{j\up}\ra&=&-\,{\ts{i\over 2}}\,\bigl\la\,G^{ab}_{\su\sd,jj} \,-\,G^{ab}_{\sd\su,jj}  \,\bigr\ra \pI \pI 
\eeq
Apparently, it makes a certain sense to absorb the $\,i=\sqrt{-1}\,$ into the definition of $G$, so let us define a $\tilde G$ through 
\beqn
\ti G&:=& +\,i\, G \pS \nn \\
G&=&-\,i\,\ti G   \lbeq{3.75}
\eeqn
As already remarked earlier, the $G$ of the Main Theorem is exactly this $\ti G$, for notational simplicity the tilde was omitted in the formulation of the Main Theorem. 
The Girsanov transformed SDE (\req{3.62}) for $G$ in Theorem 5.4 changes to 
\beqn
d\ti G&=&\ts +\,{1\over 2}\,(\,\ti G -i\,Id\,)\, \Bigl[\;-\,h_0\,dt\,-\,\sqrt{u}\,d\ti B \,+\,{udt \over 2}\,D\ti G\; \Bigr] \, (\,\ti G+i\,Id\,)  \pI
\eeqn
This is identical to the SDE (\req{2.29}) in the Main Theorem. And the effective potential (\req{3.67}) of Theorem 5.5 becomes 
\beqn
V(G)&=&V(-i\,\ti G)  \pI \nn \\ 
&=& +\,{\ts{1\over 4}}\; Tr[\,\ti G\,h_0\,]
 \;-\; {\ts {u\over 4}}\,\summ_j\,\bigl(\,\ti G^{ab}_{\su\su,jj}\;\ti G^{ab}_{\sd\sd,jj} 
   \,-\,\ti G^{ab}_{\su\sd,jj}\;\ti G^{ab}_{\sd\su,jj} \, -\,\ti G^{aa}_{\su\sd,jj}\;\ti G^{bb}_{\su\sd,jj} \,\bigr)   \pI \pI
\eeqn
Recall 
\beq
h_0&=&\bmat      0 & 0           & \vep-\mu    & s-r  \\ 
           0 & 0           & s+r     & \vep-\mu    \\ 
 -(\vep-\mu) & -(s+r)    & 0           & 0    \\ 
    -(s-r) & -(\vep-\mu) & 0           & 0    \emat \;\;=:\;\; \bmat 0 & A \\ -A^T & 0 \emat  \pI 
\eeq

\medskip
We get 
\beq
Tr[\,\ti G\,h_0\,]&=& Tr\bmat \ti G^{aa} & \ti G^{ab} \\ \ti G^{ba} & \ti G^{bb} \emat \bmat 0& A \\ -A^T & 0 \emat  \\ 
&=&Tr[\,-\,\ti G^{ab}A^T\,] \;+\; Tr[\,\ti G^{ba}A\,] \;\;=\;\;-\,Tr[\,\ti G^{ab}A^T\,] \;-\; Tr[\, (\ti G^{ab})^TA\,]  \pI \\ 
&=&-\,Tr[\,\ti G^{ab}A^T\,] \;-\; Tr[\, A^T \ti G^{ab}\,]  \;\;=\;\; -\,2\;Tr[\,A^T\ti G^{ab}\,]     
\eeq
or
\beq
Tr[\,\ti G\,h_0\,]&=&-\,2\;Tr\bmat \vep -\mu & s+r \\ s-r & \vep -\mu \emat 
  \bmat \ti G^{ab}_{\su\su} & \ti G^{ab}_{\su\sd} \\ \ti G^{ab}_{\sd\su} & \ti G^{ab}_{\sd\sd}  \emat  \pI \\ 
&& \\ 
&=&-\,2\;Tr\Bigl[\, (\vep-\mu)\,\ti G^{ab}_{\su\su}\,+\,(s+r)\,\ti G^{ab}_{\sd\su}\,+\,(\vep-\mu)\,\ti G^{ab}_{\sd\sd} \,+\,(s-r)\,\ti G^{ab}_{\su\sd} \,\Bigr] \pI
\eeq

\medskip
Thus we have
\beqn
W(\ti G)&:=&-\,V(G)\;\;=\;\;-\,V(-i\,\ti G) \pI \nn \\ 
&=& -\,{\ts{1\over 4}} \;  Tr[\,\ti G\,h_0\,]
  \;+\;{\ts { u\over 4}}\;\summ_j\,\Bigl\{\,\ti G^{ab}_{\su\su,jj}\;\ti G^{ab}_{\sd\sd,jj} 
   \,-\,\ti G^{ab}_{\su\sd,jj}\;\ti G^{ab}_{\sd\su,jj} \, -\,\ti G^{aa}_{\su\sd,jj}\;\ti G^{bb}_{\su\sd,jj} \,\Bigr\}  \pI \nn \\ 
&=& + \, {\ts {1\over 2}}\, Tr\Bigl[\, (\vep-\mu)\,(\ti G^{ab}_{\su\su}+\ti G^{ab}_{\sd\sd}) \,+\,s\,(\ti G^{ab}_{\sd\su}+\ti G^{ab}_{\su\sd}) 
  \,+\,r\,(\ti G^{ab}_{\sd\su}-\ti G^{ab}_{\su\sd})  \,\Bigr] \pI \nn \\
&&\phantom{mmm}  \;+\; {\ts {u\over 4}}\,\summ_j\,\Bigl\{\,\ti G^{ab}_{\su\su,jj}\;\ti G^{ab}_{\sd\sd,jj} 
   \,-\,\ti G^{ab}_{\su\sd,jj}\;\ti G^{ab}_{\sd\su,jj} \, -\,\ti G^{aa}_{\su\sd,jj}\;\ti G^{bb}_{\su\sd,jj} \,\Bigr\} \pI
\eeqn

\medskip
This is identical to the $W(G)$ given by (\req{2.28}) in the Main Theorem. Finally, recall that 
\beq
H_0&=&\ts {i\over 2}\, \bigl\{\,a_\su(\vep-\mu)b_\su\,+\,a_\sd(\vep-\mu)b_\sd\,+\,s\,(a_\su b_\sd+a_\sd b_\su)\,-\,r\,(a_\su b_\sd-a_\sd b_\su)\,\bigr\} \pI
\eeq
Thus, 
\beqn
\la H_0\ra_\beta&=&\biggl\la\;{\ts {i\over 2}}\; \Bigl\{\;\summ_{i,j}(\vep-\mu)_{i,j}\,(G^{ab}_{\su\su,ij}+G^{ab}_{\sd\sd,ij}) 
  \,+\,s\,\summ_j\,(G^{ab}_{\su\sd,jj}+G^{ab}_{\sd\su,jj}) \pI \\ 
&&\phantom{mmmmmm}   \,-\,r\,\summ_j \,(G^{ab}_{\su\sd,jj}-G^{ab}_{\sd\su,jj}) (a_\su b_\sd-a_\sd b_\su)\,\Bigr\} \;\biggr\ra  \pI \nn \\ 
&=&\biggl\la\;{\ts {1\over 2}}\, Tr\Bigl[\, (\vep-\mu)\,(\ti G^{ab}_{\su\su}+\ti G^{ab}_{\sd\sd}) \,+\,s\,(\ti G^{ab}_{\sd\su}+\ti G^{ab}_{\su\sd}) 
  \,+\,r\,(\ti G^{ab}_{\sd\su}-\ti G^{ab}_{\su\sd})  \,\Bigr]\;\biggr\ra    \nn \pI
\eeqn
Furthermore, since 
\beq
H_\rint&=&  +\,{\ts {u\over 4}}\,\summ_j \,a_{j\su}\,a_{j\sd}\,b_{j\su}\,b_{j\sd} \pI
\eeq
we have
\beqn
\la H_\rint\ra&=& +\,{\ts {u\over 4}}\,\summ_j \,\Bigl\la \, G^{aa}_{\su\sd,jj}G^{bb}_{\su\sd,jj}\,-\,G^{ab}_{\su\su,jj}G^{ab}_{\sd\sd,jj} 
  \,+\,G^{ab}_{\su\sd,jj}G^{ab}_{\sd\su,jj}  \,\Bigr\ra \pI  \nn \\ 
&=& +\,{\ts {u\over 4}}\,\summ_j \,\Bigl\la \, \ti G^{ab}_{\su\su,jj}\;\ti G^{ab}_{\sd\sd,jj} 
   \,-\,\ti G^{ab}_{\su\sd,jj}\;\ti G^{ab}_{\sd\su,jj} \, -\,\ti G^{aa}_{\su\sd,jj}\;\ti G^{bb}_{\su\sd,jj}  \,\Bigr\ra \pI 
\eeqn
Thus,  $W(\ti G)$ is exactly the energy, we have 
\medskip
\beqn
\la\, W(\ti G)\,\ra &=& \biggl\la\; {\ts {1\over 2}}\, Tr\Bigl[\, (\vep-\mu)\,(\ti G^{ab}_{\su\su}+\ti G^{ab}_{\sd\sd}) 
  \,+\,s\,(\ti G^{ab}_{\sd\su}+\ti G^{ab}_{\su\sd}) \,+\,r\,(\ti G^{ab}_{\sd\su}-\ti G^{ab}_{\su\sd})  \,\Bigr] \pI \nn \\
&&\phantom{mmmmm}  \;+\; {\ts {u\over 4}}\,\summ_j\,\Bigl[\,\ti G^{ab}_{\su\su,jj}\;\ti G^{ab}_{\sd\sd,jj} 
   \,-\,\ti G^{ab}_{\su\sd,jj}\;\ti G^{ab}_{\sd\su,jj} \, -\,\ti G^{aa}_{\su\sd,jj}\;\ti G^{bb}_{\su\sd,jj} \,\Bigr] \,\;\biggr\ra \pI \nn \\ 
&& \nn \\ 
&=&  \bigl\la\; H_0+H_\rint\;\bigr\ra  \pI 
\eeqn
and this completes the proof of the Main Theorem of section 2. \;\,$\blacksquare$

\newpage

{\bf \large Appendix: Proofs of Theorems of Section 4}
\numberwithin{equation}{section}
\renewcommand\thesection{A}
\setcounter{equation}{0}

\bigskip
\bigskip
\bigskip
{\bf A.1 \;Proof of Theorem 4.1} 

\bigskip
\bigskip
We proceed by induction on time steps $t_k=k\,dt$. At $k=0$, we have $t_0=0$ and 
\beq
0\;\;=\;\;G_{t_0=0}&=&\bmat i\, F_0 & \rho_0 \\ -\rho_0^T & i\,\ti F_0 \emat \;\;=\;\; \bmat 0 & 0 \\ 0 & 0 \emat   \pI 
\eeq
so the equations (\req{n4.6}-\req{n4.9}) are apparently fulfilled. Suppose they hold at some time step $t_k=k\,dt$. Then 
\beq
\rho_{t_{k+1}}&=&\rho_{t_k} \;+\; d\rho_{t_{k+1}}  \\
F_{t_{k+1}}&=&F_{t_k} \;+\; dF_{t_{k+1}}  
\eeq
with
\beq
d\rho_{t_{k+1}} &=&\ts {1\over 2}\,\Bigl[\; (Id-F_{t_k})\,dh_{t_{k+1}}\,(Id+\ti F_{t_k})\;-\; \rho_{t_k}\,dh^T_{t_{k+1}} \rho_{t_k}  \;\Bigr]  \pI \\ 
&&  \ts \;-\;{udt\over 4}\,\Bigl[\; (Id-F)\, DF\,\rho \;+\;\rho\,DF\,(Id+ \ti F ) \;\Bigr]_{t_k} \\
dF_{t_{k+1}}&=&\ts {1\over 2}\,\Bigl[\; [\,\rho_{t_k}\,dh^T_{t_{k+1}}(Id+ F_{t_k})\,]^T \,-\,\rho_{t_k}\,dh^T_{t_{k+1}}(Id+ F_{t_k})  \; \Bigr]  \pI \\ 
&& \ts \;-\; {udt\over 4}\,\Bigl[\; (Id- F)\,DF\,(Id+F) \,+\,\rho\,DF\, \rho^T   \;\Bigr]_{t_k}        
\eeq
and, for $\mu=0$, 
\beq
dh_{t_{k+1}}&=&\ts -\,dt\,\bmat \vep & s-r \\  s+r  & \vep \emat  \;\;+\;\;{udt\over 2}\,D\rho_{t_k} 
  \;\;-\;\;\sqrt{\sabs u\sabs \,dt}\,\bmat  0 & \xi_{k+1} \\ \eps_u\,\xi_{k+1} & 0 \emat  \pI 
\eeq
with a diagonal matrix of standard normal random numbers 
\beq
\xi_{k+1}&=&\bigl(\, \delta_{i,j}\,\xi_{j,k+1}\,\bigr)_{i,j\in\Gamma} \;\;\in\;\;\R^{\G\times\G}  \pI 
\eeq
By the induction hypothesis for equation (\req{n4.9}), we have 
\beq
DF_{t_k}&=& \bmat  0  & +{\rm diag}[\,F_{\su\sd}(t_k)\,] \; \\ -{\rm diag}[\,F_{\su\sd}(t_k)\,] &  0 \emat \;\;=\;\; \bmat 0&0\\0&0\emat \pI 
\eeq  

\medskip
Thus, the equations for $d\rho$ and $dF$ simplify to 
\beq
2\,d\rho_{t_{k+1}} &=&\ts  (Id-F_{t_k})\,dh_{t_{k+1}}\,(Id+\ti F_{t_k})\;-\; \rho_{t_k}\,dh^T_{t_{k+1}}\, \rho_{t_k}   \pI \\ 
2\,dF_{t_{k+1}}&=&\ts  [\,\rho_{t_k}\,dh^T_{t_{k+1}}(Id+ F_{t_k})\,]^T \,-\,\rho_{t_k}\,dh^T_{t_{k+1}}(Id+ F_{t_k})   
\eeq

\medskip
Using the induction hypothesis again, this time for equation (\req{n4.6}), we get 
\medskip
\beq
D\rho_{t_k}\;\;=\;\;\bmat  -{\rm diag}[\,\rho_{\sd\sd}(t_k)\,] & +{\rm diag}[\,\rho_{\sd\su}(t_k)\,] \; \\ 
  +{\rm diag}[\,\rho_{\su\sd}(t_k)\,] & -{\rm diag}[\,\rho_{\su\su}(t_k)\,]  \; \emat\;\;=\;\;
\bmat  0 & {\rm diag}[\,\rho_{\sd\su}(t_k)\,] \; \\  {\rm diag}[\,\rho_{\su\sd}(t_k)\,] & 0 \; \emat  
\eeq

\medskip
and we may write 
\beq
dh_{t_{k+1}}&=&  -\bmat \vep & 0 \\  0  & \vep \emat dt \;\;+\;\;  \bmat 0 & dD_1(t_{k+1}) \\  dD_2(t_{k+1})  & 0 \emat \pI 
\eeq
with diagonal matrices 
\medskip
\beq
dD_1(t_{k+1})&:=&\ts  -(s-r)Id \;dt \;+\; {udt\over 2}\,{\rm diag}[\,\rho_{\sd\su}(t_k)\,] \;-\;\sqrt{\sabs u\sabs \,dt}\; \xi_{k+1}  \pS\\ 
dD_2(t_{k+1})&:=&\ts  -(s+r)Id \;dt \;+\; {udt\over 2}\,{\rm diag}[\,\rho_{\su\sd}(t_k)\,] \;-\;\sqrt{\sabs u\sabs \,dt}\,\eps_u\; \xi_{k+1}  \pS
\eeq

\medskip
We obtain, using $\vep^T=\vep$ and $\ti F=PFP$,
\medskip
\beq
2\,d\rho&=&\bmat Id-F_{\su\su} & -F_{\su\sd} \\ -F_{\sd\su} & Id-F_{\sd\sd} \emat \bmat -\vep dt & dD_1 \\ dD_2 & -\vep dt \emat 
  \bmat Id+ F_{\sd\sd} & + F_{\sd\su} \\ + F_{\su\sd} & Id+ F_{\su\su} \emat  \pI \\ 
&&\phantom{I} \\ 
&&\;-\; \bmat \rho_{\su\su} & \rho_{\su\sd} \\ \rho_{\sd\su} & \rho_{\sd\sd} \emat  \bmat -\vep dt & dD_2 \\ dD_1 & -\vep dt \emat 
  \bmat \rho_{\su\su} & \rho_{\su\sd} \\ \rho_{\sd\su} & \rho_{\sd\sd} \emat   \pI
\eeq

\medskip
Straightforward calculation gives
\beq
2\,d\rho_{\su\su}&=&(Id-F_{\su\su})\,(-\vep dt)\,(Id+ F_{\sd\sd})\;+\;F_{\su\sd}\,\vep dt\,F_{\su\sd} 
                \;+\;\rho_{\su\su}\,\vep dt\, \rho_{\su\su} \;+\;\rho_{\su\sd}\,\vep dt\, \rho_{\sd\su}  \pI \\ 
&&\;+\;(Id-F_{\su\su})\,dD_1\,F_{\su\sd}\;-\;F_{\su\sd}\,dD_2\,(Id+ F_{\sd\sd})\;-\;\rho_{\su\su}\,dD_2\,\rho_{\sd\su}\;-\;\rho_{\su\sd}\,dD_1\,\rho_{\su\su} \\
2\,d\rho_{\sd\sd}&=&(Id-F_{\sd\sd})\,(-\vep dt)\,(Id+ F_{\su\su})\;+\;F_{\sd\su}\,\vep dt\,F_{\sd\su} 
                \;+\;\rho_{\sd\sd}\,\vep dt\, \rho_{\sd\sd} \;+\;\rho_{\sd\su}\,\vep dt\, \rho_{\su\sd}  \pI \\ 
&&\;+\;(Id-F_{\sd\sd})\,dD_2\,F_{\sd\su}\;-\;F_{\sd\su}\,dD_1\,(Id+ F_{\su\su})\;-\;\rho_{\sd\sd}\,dD_1\,\rho_{\su\sd}\;-\;\rho_{\sd\su}\,dD_2\,\rho_{\sd\sd} 
\eeq
and
\beq
2\,d\rho_{\su\sd}&=&(Id-F_{\su\su})\,(-\vep dt)\,F_{\sd\su}\;+\;F_{\su\sd}\,\vep dt\,(Id+F_{\su\su})\;+\;
   \rho_{\su\su}\,\vep dt\,\rho_{\su\sd}\;+\;\rho_{\su\sd}\,\vep dt\,\rho_{\sd\sd}  \pI \\ 
&&\;+\;(Id-F_{\su\su})\,dD_1\,(Id+F_{\su\su})\;-\;F_{\su\sd}\,dD_2\,F_{\sd\su}\;-\;\rho_{\su\su}\,dD_2\,\rho_{\sd\sd}\;-\;\rho_{\su\sd}\,dD_1\,\rho_{\su\sd}  \\
2\,d\rho_{\sd\su}&=&(Id-F_{\sd\sd})\,(-\vep dt)\,F_{\su\sd}\;+\;F_{\sd\su}\,\vep dt\,(Id+F_{\sd\sd})\;+\;
   \rho_{\sd\sd}\,\vep dt\,\rho_{\sd\su}\;+\;\rho_{\sd\su}\,\vep dt\,\rho_{\su\su}  \pI \\ 
&&\;+\;(Id-F_{\sd\sd})\,dD_2\,(Id+F_{\sd\sd})\;-\;F_{\sd\su}\,dD_1\,F_{\su\sd}\;-\;\rho_{\sd\sd}\,dD_1\,\rho_{\su\su}\;-\;\rho_{\sd\su}\,dD_2\,\rho_{\sd\su}
\eeq

\medskip
We have to prove, with the notation introduced at the beginning of section 4, 
\beq
\;[d\rho_{\su\su}]^{\rm on}&=&0  \\
\;[d\rho_{\su\sd}]^{\rm off}&=&0 \\
\;[dF_{\su\su}]^{\rm off}&=&0 \\
\;[dF_{\su\sd}]^{\rm on}&=&0     
\eeq
Now let us call some $\G\times\G$ matrix $M$ 
\beq
\hbox{\rm an on\,-\,matrix\,:}\;\; &\buildrel{\rm Def. }\over \Leftrightarrow& \;\; M^{\rm off}\;\;=\;\;0  \pS\pS \\ 
\hbox{\rm an off\,-\,matrix\,:}\;\; &\buildrel{\rm Def. }\over \Leftrightarrow& \;\; M^{\rm on}\;\;=\;\;0  
\eeq
Then the following properties are easily checked: The matrix product of two on-matrices or two off-matrices is an on-matrix. The matrix product of an on-matrix 
and an off-matrix or the product of an off-matrix with an on-matrix is an off-matrix. For short, 
\beq
{\rm on} \cdot {\rm on} &=& {\rm on} \\ 
{\rm off} \cdot {\rm off} &=& {\rm on} \\ 
{\rm on} \cdot {\rm off} &=& {\rm off} \\ 
{\rm off} \cdot {\rm on} &=& {\rm off} 
\eeq
If we denote the set of all on-matrices and off-matrices with ${\cal M}^{\rm on}$ and ${\cal M}^{\rm off}$, 
then we have by assumption and by the induction hypothesis 
\beqn
\vep&\in&{\cal M}^{\rm off} \nn\\ 
dD_{1,2}&\in& {\cal M}^{\rm on}\nn \\ 
Id\pm F_{\sigma\sigma}&\in& {\cal M}^{\rm on}  \lbeq{n4.14} \\ 
F_{\su\sd},\,F_{\sd\su}&\in& {\cal M}^{\rm off} \nn\\ 
\rho_{\sigma\sigma}&\in& {\cal M}^{\rm off} \nn\\ 
\rho_{\su\sd},\,\rho_{\sd\su}&\in& {\cal M}^{\rm on} \nn
\eeqn
Thus, $d\rho_{\su\su}$ is of type 
\beq
2\,d\rho_{\su\su}&=&(Id-F_{\su\su})\,(-\vep dt)\,(Id+ F_{\sd\sd})\;+\;F_{\su\sd}\,\vep dt\,F_{\su\sd} 
                \;+\;\rho_{\su\su}\,\vep dt\, \rho_{\su\su} \;+\;\rho_{\su\sd}\,\vep dt\, \rho_{\sd\su}  \pI \\ 
&&\;+\;(Id-F_{\su\su})\,dD_1\,F_{\su\sd}\;-\;F_{\su\sd}\,dD_2\,(Id+ F_{\sd\sd})\;-\;\rho_{\su\su}\,dD_2\,\rho_{\sd\su}\;-\;\rho_{\su\sd}\,dD_1\,\rho_{\su\su} \\
&\in&{\cal M}^{\rm on}\,{\cal M}^{\rm off}\,{\cal M}^{\rm on}\;+\;{\cal M}^{\rm off}\,{\cal M}^{\rm off}\,{\cal M}^{\rm off}
  \;+\;{\cal M}^{\rm off}\,{\cal M}^{\rm off}\,{\cal M}^{\rm off} \;+\;{\cal M}^{\rm on}\,{\cal M}^{\rm off}\, {\cal M}^{\rm on}  \pI \\ 
&&\;+\;{\cal M}^{\rm on}\,{\cal M}^{\rm on}\,{\cal M}^{\rm off}\;-\;{\cal M}^{\rm off}\,{\cal M}^{\rm on}\,{\cal M}^{\rm on}
  \;-\;{\cal M}^{\rm off}\,{\cal M}^{\rm on}\,{\cal M}^{\rm on}\;-\;{\cal M}^{\rm on}\,{\cal M}^{\rm on}\,{\cal M}^{\rm off} \\ 
&=&{\cal M}^{\rm off} \pI
\eeq
which completes the induction step for $\rho_{\su\su}(t_{k+1})$. For $d\rho_{\su\sd}$ we get 
\beq
2\,d\rho_{\su\sd}&=&(Id-F_{\su\su})\,(-\vep dt)\,F_{\sd\su}\;+\;F_{\su\sd}\,\vep dt\,(Id+F_{\su\su})\;+\;
   \rho_{\su\su}\,\vep dt\,\rho_{\su\sd}\;+\;\rho_{\su\sd}\,\vep dt\,\rho_{\sd\sd}  \pI \\ 
&&\;+\;(Id-F_{\su\su})\,dD_1\,(Id+F_{\su\su})\;-\;F_{\su\sd}\,dD_2\,F_{\sd\su}\;-\;\rho_{\su\su}\,dD_2\,\rho_{\sd\sd}\;-\;\rho_{\su\sd}\,dD_1\,\rho_{\su\sd}  \\
&\in&{\cal M}^{\rm on}\,{\cal M}^{\rm off}\,{\cal M}^{\rm off}\;+\;{\cal M}^{\rm off}\,{\cal M}^{\rm off}\,{\cal M}^{\rm on}\;+\;
   {\cal M}^{\rm off}\,{\cal M}^{\rm off}\,{\cal M}^{\rm on}\;+\;{\cal M}^{\rm on}\,{\cal M}^{\rm off}\,{\cal M}^{\rm off}  \pI \\ 
&&\;+\;{\cal M}^{\rm on}\,{\cal M}^{\rm on}\,{\cal M}^{\rm on}\;-\;{\cal M}^{\rm off}\,{\cal M}^{\rm on}\,{\cal M}^{\rm off}\;-\;{\cal M}^{\rm off}\,{\cal M}^{\rm on}\,{\cal M}^{\rm off}\;-\;{\cal M}^{\rm on}\,{\cal M}^{\rm on}\,{\cal M}^{\rm on} \\ 
&=& {\cal M}^{\rm on} \pI
\eeq
which completes the induction step for $\rho_{\su\sd}(t_{k+1})$. It remains to check the $dF$. We have 
\beq
2\,dF&=&\ts  [\,\rho\,dh^T(Id+ F)\,]^T \,-\,\rho\,dh^T(Id+ F) \;\;=:\;\;[d\alpha]^T\;-\;d\alpha  \pI 
\eeq
with
\beq
d\alpha\;\;=\;\;\rho\,dh^T(Id+ F)&=& \bmat \rho_{\su\su} & \rho_{\su\sd} \\ \rho_{\sd\su} & \rho_{\sd\sd} \emat  \bmat -\vep dt & dD_2 \\ dD_1 & -\vep dt \emat 
  \bmat Id+F_{\su\su} & +F_{\su\sd} \\ +F_{\sd\su} & Id+F_{\sd\sd} \emat
\eeq
Straightforward calculation gives 
\beq
d\alpha_{\su\su}&=&\;-\,\rho_{\su\su}\,\vep dt\,(Id+F_{\su\su}) \;-\;\rho_{\su\sd}\,\vep dt\,F_{\sd\su} 
   \;+\; \rho_{\su\su}\,dD_2\,F_{\sd\su}\;+\;\rho_{\su\sd}\,dD_1\,(Id+F_{\su\su}) \pS \\ 
d\alpha_{\sd\sd}&=&\;-\,\rho_{\sd\sd}\,\vep dt\,(Id+F_{\sd\sd}) \;-\;\rho_{\sd\su}\,\vep dt\,F_{\su\sd} 
   \;+\; \rho_{\sd\sd}\,dD_1\,F_{\su\sd}\;+\;\rho_{\sd\su}\,dD_2\,(Id+F_{\sd\sd}) \pS
\eeq
and
\beq
d\alpha_{\su\sd}&=&\;-\,\rho_{\su\su}\,\vep dt\,F_{\su\sd}\;-\;\rho_{\su\sd}\,\vep dt\,(Id+F_{\sd\sd}) 
    \;+\;\rho_{\su\su}\,dD_2\,(Id+F_{\sd\sd})\;+\;\rho_{\su\sd}\,dD_1\,F_{\su\sd} \pS \\ 
d\alpha_{\sd\su}&=&\;-\,\rho_{\sd\sd}\,\vep dt\,F_{\sd\su}\;-\;\rho_{\sd\su}\,\vep dt\,(Id+F_{\su\su}) 
    \;+\;\rho_{\sd\sd}\,dD_1\,(Id+F_{\su\su})\;+\;\rho_{\sd\su}\,dD_2\,F_{\sd\su} \pS
\eeq

\medskip
Using (\req{n4.14}) again, we find
\beq
d\alpha_{\su\su}&\in& {\cal M}^{\rm off}\,{\cal M}^{\rm off}\,{\cal M}^{\rm on} \;-\;{\cal M}^{\rm on}\,{\cal M}^{\rm off}\,{\cal M}^{\rm off}
  \;+\;{\cal M}^{\rm off}\,{\cal M}^{\rm on}\,{\cal M}^{\rm off}\;+\;{\cal M}^{\rm on}\,{\cal M}^{\rm on}\,{\cal M}^{\rm on} \pS \\
&=&{\cal M}^{\rm on}  \pS \\
d\alpha_{\su\sd}&\in& {\cal M}^{\rm off}\,{\cal M}^{\rm off}\,{\cal M}^{\rm off}\;-\;{\cal M}^{\rm on}\,{\cal M}^{\rm off}\,{\cal M}^{\rm on}
  \;+\;{\cal M}^{\rm off}\,{\cal M}^{\rm on}\,{\cal M}^{\rm on}\;+\;{\cal M}^{\rm on}\,{\cal M}^{\rm on}\,{\cal M}^{\rm off} \pS \\ 
&=&{\cal M}^{\rm off} \pS
\eeq
with analog calculations for $d\alpha_{\sd\sd}$ and $d\alpha_{\sd\su}\,$. This results in 
\beq
2\,dF_{\su\su}&=&[d\alpha_{\su\su}]^T \;-\;d\alpha_{\su\su} \;\;\in\;\; {\cal M}^{\rm on} \pS \\ 
2\,dF_{\su\sd}&=&[d\alpha_{\sd\su}]^T \;-\;d\alpha_{\su\sd} \;\;\in\;\; {\cal M}^{\rm off}  \pS
\eeq
and completes the proof of Theorem 4.1\,. \; $\blacksquare$

\bigskip
\bigskip
\bigskip
{\bf A.2 \;Proof of Theorem 4.2} 

\bigskip
\bigskip
We use the on- and off-definitions and notations from the proof of Theorem 4.1. 

\bigskip
{\bf Part a)} \;From part (a) of Theorem 3.2, we have $\,F^b=F^a\,$,  $\,\rho^T= \rho\,$ and with $F\equiv F^a$, 
\beq
\rho_{\su\sd}\;\;=\;\;\rho_{\sd\su}&=& 0   \nn \\  
F_{\su\sd}\;\;=\;\;F_{\sd\su}&=& 0   
\eeq
The $\rho_{\sigma\sigma}$ and $F_{\sigma\sigma}$ are given by the SDE system 
\beq
d\rho_{\sigma\sigma}&=&\ts {1\over 2}\,\Bigl[\; (Id-F_{\sigma\sigma})\,dh_{\sigma\sigma}\,(Id+F_{\sigma\sigma})\;-\; 
   \rho_{\sigma\sigma}\,dh_{\sigma\sigma} \rho_{\sigma\sigma}    \;\Bigr] \pI \nn \\ 
dF_{\sigma\sigma}&=&\ts {1\over 2}\,\Bigl[\; (Id-F_{\sigma\sigma})\,dh_{\sigma\sigma}\,\rho_{\sigma\sigma} \;-\; 
  \rho_{\sigma\sigma}\,dh_{\sigma\sigma}\,(Id+F_{\sigma\sigma})  \; \Bigr] 
\eeq
with 
\beq
dh_{\su\su}&=& -\,\vep\,dt \;-\;\ts {udt \over 2} \,{\rm diag}[\,\rho_{\sd\sd}\,] \;-\; \sqrt{\sabs u\sabs}\; dx \pS \nn \\ 
dh_{\sd\sd}&=& -\,\vep\,dt \;-\;\ts {udt \over 2} \,{\rm diag}[\,\rho_{\su\su}\,] \;+\; \eps_u \sqrt{\sabs u\sabs}\; dx  \pS
\eeq
or
\beq
dh_{\su\su}&=& -\,\vep\,dt \;+\;\ts {|u|dt \over 2} \;{\rm diag}[\;-\eps_u\,\rho_{\sd\sd}^{\rm on}\;] \;-\; \sqrt{\sabs u\sabs}\; dx
  \;\;=:\;\; -\,\vep\,dt \;+\;dD_1 \pS \nn \\ 
dh_{\sd\sd}&=& -\,\vep\,dt \;-\;\eps_u\,\ts {|u|dt \over 2} \;{\rm diag}[\;\rho_{\su\su}^{\rm on}\;] \;+\; \eps_u \sqrt{\sabs u\sabs}\; dx  
   \;\;=:\;\; -\,\vep\,dt \;-\;\eps_u\, dD_2  \pS
\eeq
with diagonal matrices 
\beq
dD_1&:=&\ts {|u|dt \over 2} \;{\rm diag}[\;-\eps_u\,\rho_{\sd\sd}^{\rm on}\;] \;-\; \sqrt{\sabs u\sabs}\; dx  \pS \\ 
dD_2&:=&\ts {|u|dt \over 2} \;{\rm diag}[\;\rho_{\su\su}^{\rm on}\;] \;-\;  \sqrt{\sabs u\sabs}\; dx \pS
\eeq
Thus, 
\beq
d\rho_{\su\su}&=&\ts {1\over 2}\,\Bigl[\; (Id-F_{\su\su})\,(-\vep dt)\,(Id+F_{\su\su})\;-\; \rho_{\su\su}\,(-\vep dt)\, \rho_{\su\su}    \;\Bigr] \pI \nn \\ 
&&\;+\;\ts {1\over 2}\,\Bigl[\; (Id-F_{\su\su})\,dD_1\,(Id+F_{\su\su})\;-\; \rho_{\su\su}\,dD_1\, \rho_{\su\su}    \;\Bigr] \\
d\rho_{\sd\sd}&=&\ts {1\over 2}\,\Bigl[\; (Id-F_{\sd\sd})\,(-\vep dt)\,(Id+F_{\sd\sd})\;-\; \rho_{\sd\sd}\,(-\vep dt)\, \rho_{\sd\sd}    \;\Bigr] \pI \nn \\ 
&&\;-\;\eps_u\;\ts {1\over 2}\,\Bigl[\; (Id-F_{\sd\sd})\,dD_2\,(Id+F_{\sd\sd})\;-\; \rho_{\sd\sd}\,dD_2\, \rho_{\sd\sd}    \;\Bigr] 
\eeq
and
\beq
dF_{\su\su}&=&\ts {1\over 2}\,\Bigl[\; (Id-F_{\su\su})\,(-\vep dt)\,\rho_{\su\su} \;-\;  \rho_{\su\su}\,(-\vep dt)\,(Id+F_{\su\su})  \; \Bigr] \pI\\ 
&&\;+\;\ts {1\over 2}\,\Bigl[\; (Id-F_{\su\su})\,dD_1\,\rho_{\su\su} \;-\;  \rho_{\su\su}\,dD_1\,(Id+F_{\su\su})  \; \Bigr]  \\ 
dF_{\sd\sd}&=&\ts {1\over 2}\,\Bigl[\; (Id-F_{\sd\sd})\,(-\vep dt)\,\rho_{\sd\sd} \;-\;  \rho_{\sd\sd}\,(-\vep dt)\,(Id+F_{\sd\sd})  \; \Bigr] \pI\\ 
&&\;-\; \eps_u\; \ts {1\over 2}\,\Bigl[\; (Id-F_{\sd\sd})\,dD_2\,\rho_{\sd\sd} \;-\;  \rho_{\sd\sd}\,dD_2\,(Id+F_{\sd\sd})  \; \Bigr] 
\eeq

\medskip
We have to take the on-part and the off-part of these equations. Since $\vep$ is an off-matrix and the diagonal matrices $dD_1$ and $dD_2$ are on-matrices, we get
\beqn
[\,d\rho_{\su\su}\,]^{\rm on}&=&\ts {1\over 2}\,\Bigl[\; (Id-F_{\su\su})^{\rm on}\,(-\vep dt)\,F_{\su\su}^{\rm off}
  \;-\; \rho_{\su\su}^{\rm on}\,(-\vep dt)\, \rho_{\su\su}^{\rm off}    \;\Bigr] \pI \nn \\ 
&&\;+\; \ts {1\over 2}\,\Bigl[\; (-F_{\su\su}^{\rm off})\,(-\vep dt)\,(Id+F_{\su\su})^{\rm on}\;-\; \rho_{\su\su}^{\rm off}\,(-\vep dt)\, \rho_{\su\su}^{\rm on}    \;\Bigr] \nn\\
&&\;+\;\ts {1\over 2}\,\Bigl[\; (Id-F_{\su\su})^{\rm on}\,dD_1\,(Id+F_{\su\su})^{\rm on} \;-\; \rho_{\su\su}^{\rm on}\,dD_1\, \rho_{\su\su}^{\rm on}    \;\Bigr] \pI \nn \\
&&\;+\;\ts {1\over 2}\,\Bigl[\; (-F_{\su\su}^{\rm off})\,dD_1\,F_{\su\su}^{\rm off}\;-\; \rho_{\su\su}^{\rm off}\,dD_1\, \rho_{\su\su}^{\rm off}    \;\Bigr] \lbeq{A.2}
\eeqn
\beqn
[\,d\rho_{\sd\sd}\,]^{\rm on}&=&\ts {1\over 2}\,\Bigl[\; (Id-F_{\sd\sd})^{\rm on}\,(-\vep dt)\,F_{\sd\sd}^{\rm off}
  \;-\; \rho_{\sd\sd}^{\rm on}\,(-\vep dt)\, \rho_{\sd\sd}^{\rm off}    \;\Bigr] \pI \nn \\ 
&&\;+\; \ts {1\over 2}\,\Bigl[\; (-F_{\sd\sd}^{\rm off})\,(-\vep dt)\,(Id+F_{\sd\sd})^{\rm on}\;-\; \rho_{\sd\sd}^{\rm off}\,(-\vep dt)\, \rho_{\sd\sd}^{\rm on}    \;\Bigr]\nn \\
&&\;-\;\eps_u\;\ts {1\over 2}\,\Bigl[\; (Id-F_{\sd\sd})^{\rm on}\,dD_2\,(Id+F_{\sd\sd})^{\rm on} \;-\; \rho_{\sd\sd}^{\rm on}\,dD_2\, \rho_{\sd\sd}^{\rm on}    \;\Bigr] \pI \nn\\
&&\;-\;\eps_u\;\ts {1\over 2}\,\Bigl[\; (-F_{\sd\sd}^{\rm off})\,dD_2\,F_{\sd\sd}^{\rm off}\;-\; \rho_{\sd\sd}^{\rm off}\,dD_2\, \rho_{\sd\sd}^{\rm off}    \;\Bigr] 
\eeqn
The identities we have to prove are given by (\req{4.18nn}). So let's make the induction hypothesis, at time step $t_k=kdt$, 
\beqn
\rho_{\su\su}^{\rm on}\;\;=\;\;-\;\eps_u\,\rho_{\sd\sd}^{\rm on}\;,\;\;\;\;\;&& \rho_{\su\su}^{\rm off}\;\;=\;\;\rho_{\sd\sd}^{\rm off}\; \pS  \nn \\
 F_{\su\su}^{\rm off}\;\;=\;\;-\;\eps_u\,F_{\sd\sd}^{\rm off}\;,\;\;\;\;\;&& F_{\su\su}^{\rm on}\;\;=\;\;F_{\sd\sd}^{\rm on}\; \pS  \lbeq{A.4}
\eeqn
Under (\req{A.4}), we have $dD_1=dD_2$. If we substitute (\req{A.4}) on the right hand side of (\req{A.2}), we get
\beq
[\,d\rho_{\su\su}\,]^{\rm on}&=&\ts {1\over 2}\,\Bigl[\; (Id-F_{\sd\sd})^{\rm on}\,(-\vep dt)\,(-\eps_u F_{\sd\sd}^{\rm off})
  \;-\; (-\eps_u\,\rho_{\sd\sd}^{\rm on})\,(-\vep dt)\, \rho_{\sd\sd}^{\rm off}    \;\Bigr] \pI \nn \\ 
&&\;+\; \ts {1\over 2}\,\Bigl[\; (+\eps_u F_{\sd\sd}^{\rm off})\,(-\vep dt)\,(Id+F_{\sd\sd})^{\rm on}
  \;-\; \rho_{\sd\sd}^{\rm off}\,(-\vep dt)\, (-\eps_u\,\rho_{\sd\sd}^{\rm on})    \;\Bigr] \nn\\
&&\;+\;\ts {1\over 2}\,\Bigl[\; (Id-F_{\sd\sd})^{\rm on}\,dD_2\,(Id+F_{\sd\sd})^{\rm on} \;-\; \rho_{\sd\sd}^{\rm on}\,dD_2\, \rho_{\sd\sd}^{\rm on}    \;\Bigr] \pI \nn \\
&&\;+\;\ts {1\over 2}\,\Bigl[\; (-F_{\sd\sd}^{\rm off})\,dD_2\,F_{\sd\sd}^{\rm off}\;-\; \rho_{\sd\sd}^{\rm off}\,dD_2\, \rho_{\sd\sd}^{\rm off}    \;\Bigr] \\ 
&=& -\,\eps_u\,[\,d\rho_{\sd\sd}\,]^{\rm on}  \pI
\eeq
This verifies the induction hypothesis for the first equation in (\req{A.4}). To verify the second, we have to calculate the off-parts. We get 
\beqn
[\,d\rho_{\su\su}\,]^{\rm off}&=&\ts {1\over 2}\,\Bigl[\; (Id-F_{\su\su})^{\rm on}\,(-\vep dt)\,(Id+F_{\su\su})^{\rm on}
  \;-\; \rho_{\su\su}^{\rm on}\,(-\vep dt)\, \rho_{\su\su}^{\rm on}    \;\Bigr] \pI \nn \\ 
&&\;+\; \ts {1\over 2}\,\Bigl[\; (-F_{\su\su}^{\rm off})\,(-\vep dt)\,F_{\su\su}^{\rm off}\;-\; \rho_{\su\su}^{\rm off}\,(-\vep dt)\, \rho_{\su\su}^{\rm off}    \;\Bigr] \nn\\
&&\;+\;\ts {1\over 2}\,\Bigl[\; (Id-F_{\su\su})^{\rm on}\,dD_1\,F_{\su\su}^{\rm off} \;-\; \rho_{\su\su}^{\rm on}\,dD_1\, \rho_{\su\su}^{\rm off}    \;\Bigr] \pI \nn \\
&&\;+\;\ts {1\over 2}\,\Bigl[\; (-F_{\su\su}^{\rm off})\,dD_1\,(Id+F_{\su\su})^{\rm on}\;-\; \rho_{\su\su}^{\rm off}\,dD_1\, \rho_{\su\su}^{\rm on}    \;\Bigr] \lbeq{A.5}
\eeqn
\beqn
[\,d\rho_{\sd\sd}\,]^{\rm off}&=&\ts {1\over 2}\,\Bigl[\; (Id-F_{\sd\sd})^{\rm on}\,(-\vep dt)\,(Id+F_{\sd\sd})^{\rm on}
  \;-\; \rho_{\sd\sd}^{\rm on}\,(-\vep dt)\, \rho_{\sd\sd}^{\rm on}    \;\Bigr] \pI \nn \\ 
&&\;+\; \ts {1\over 2}\,\Bigl[\; (-F_{\sd\sd}^{\rm off})\,(-\vep dt)\,F_{\sd\sd}^{\rm off}\;-\; \rho_{\sd\sd}^{\rm off}\,(-\vep dt)\, \rho_{\sd\sd}^{\rm off}    \;\Bigr] \nn\\
&&\;-\;\eps_u\;\ts {1\over 2}\,\Bigl[\; (Id-F_{\sd\sd})^{\rm on}\,dD_2\,F_{\sd\sd}^{\rm off} \;-\; \rho_{\sd\sd}^{\rm on}\,dD_2\, \rho_{\sd\sd}^{\rm off}    \;\Bigr] \pI \nn \\
&&\;-\;\eps_u\;\ts {1\over 2}\,\Bigl[\; (-F_{\sd\sd}^{\rm off})\,dD_2\,(Id+F_{\sd\sd})^{\rm on}\;-\; \rho_{\sd\sd}^{\rm off}\,dD_2\, \rho_{\sd\sd}^{\rm on}    \;\Bigr] \lbeq{A.6}
\eeqn

\medskip
We substitute (\req{A.4}) on the right hand side of (\req{A.5}) and obtain, using $dD_1=dD_2$ again, 
\beq
[\,d\rho_{\su\su}\,]^{\rm off}&=&\ts {1\over 2}\,\Bigl[\; (Id-F_{\sd\sd})^{\rm on}\,(-\vep dt)\,(Id+F_{\sd\sd})^{\rm on}
  \;-\; \rho_{\sd\sd}^{\rm on}\,(-\vep dt)\, \rho_{\sd\sd}^{\rm on}    \;\Bigr] \pI \nn \\ 
&&\;+\; \ts {1\over 2}\,\Bigl[\; (-F_{\sd\sd}^{\rm off})\,(-\vep dt)\,F_{\sd\sd}^{\rm off}\;-\; \rho_{\sd\sd}^{\rm off}\,(-\vep dt)\, \rho_{\sd\sd}^{\rm off}    \;\Bigr] \nn\\
&&\;+\;\ts {1\over 2}\,\Bigl[\; (Id-F_{\sd\sd})^{\rm on}\,dD_2\,(-\eps_u F_{\sd\sd}^{\rm off}) 
    \;-\; (-\eps_u\,\rho_{\sd\sd}^{\rm on})\,dD_2\,\rho_{\sd\sd}^{\rm off} \;\Bigr] \pI \nn \\
&&\;+\;\ts {1\over 2}\,\Bigl[\; (-F_{\sd\sd}^{\rm off})\,(-\eps_u\,dD_2)\,(Id+F_{\sd\sd})^{\rm on}
    \;-\; \rho_{\sd\sd}^{\rm off}\,(-\eps_u\,dD_2)\, \rho_{\sd\sd}^{\rm on} \;\Bigr]    \pI \\ 
&=& [\,d\rho_{\sd\sd}\,]^{\rm off} \pI
\eeq
This verifies the induction hypothesis for the second equation in (\req{A.4}). The equations for the $F_{\sigma\sigma}$ can be checked in the same way.  
The energy in the $w_1=1$ representation is given by Theorem 3.2 through 
\beq
W&=&{\ts {1\over 2}}\, \summ_{i,j}\, \vep_{ij}\,(\rho_{\su\su,ji} + \rho_{\sd\sd,ji})    \;+\; {\ts {u\over 4}}\,\summ_j\, \rho_{\su\su,jj}\;\rho_{\sd\sd,jj}   \pS \\ 
&=&{\ts {1\over 2}}\, \summ_{i,j}\, \vep_{ij}\,(\,\rho_{\su\su,ji}^{\rm off} + \rho_{\sd\sd,ji}^{\rm off}\,)  
   \;+\; {\ts {u\over 4}}\,\summ_j\, \rho_{\su\su,jj}^{\rm on}\;\rho_{\sd\sd,jj}^{\rm on}   \pS
\eeq 
With the equations (\req{4.18nn}), this apparently reduces to the expression (\req{4.22n}). 

\bigskip
{\bf Part b)} Let us introduce the $2\G\times 2\G$ matrix 
\beq
Q&:=&\sqrt{\eps_u}\;\bmat 0& Id\\ \eps_u\,Id&0\emat 
\eeq
with, to be definite, $\sqrt{-1}:=+i\,$. For some arbitrary $2\G\times 2\G$ matrix 
\beq
M&=& \bmat A&B\\ C&D\emat \pI 
\eeq
we have the following identity:
\beq
Q\,M\,Q\;\;=\;\; \bmat D& \eps_u\,C \\ \eps_u\,B & A \emat \pI
\eeq
In particular, $\,Q^2=\Idz\,$. Thus, with 
\beq
\rho\;\;=\;\;\bmat \rho_{\su\su} & \rho_{\su\sd} \\ \rho_{\sd\su} & \rho_{\sd\sd} \emat,\;\;\;\; \;\;\;\;\;\;
F\;\;=\;\;\bmat F_{\su\su} & F_{\su\sd} \\ F_{\sd\su} & F_{\sd\sd} \emat \pI
\eeq
the equations (\req{4.23n}) which we have to prove, can equivalently also be written as 
\beqn
Q\,\rho\,Q\;\;=\;\;\rho\;,\;\;\;\;\;\;\;\;\;\; Q\,F\,Q\;\;=\;\;F  \pS  \lbeq{A.7} 
\eeqn
We prove (\req{A.7}) by induction on time steps. Suppose the relations hold at $t_k$, 
\beq
Q\,\rho_{t_k}\,Q &=& \rho_{t_k}  \pS \\ 
Q\,F_{t_k}\,Q &=& F_{t_k}  \pS
\eeq
Then we have to prove
\beq
Q\,(\rho_{t_k}+d\rho_{t_{k+1}})\,Q &=& \rho_{t_k}+d\rho_{t_{k+1}}  \pS \\ 
Q\,(F_{t_k}+dF_{t_{k+1}})\,Q &=& F_{t_k}+dF_{t_{k+1}}   \pS
\eeq
which, using the induction hypothesis, is equivalent to 
\beq
Q\,d\rho\,Q &=& d\rho  \pS \\ 
Q\,dF\,Q &=& dF  \pS
\eeq
with $d\rho$ and $dF$ given by (\req{n4.12}) of Theorem 4.1. We get, using $\,Q^2=Id\,$ and the induction hypothesis,  
\beq
Q\,d\rho\,Q&=&\ts {1\over 2}\,Q\,\Bigl[\;(Id-F)\,dh\,(Id+\ti F) \;-\; \rho\,dh^T \rho  \;\Bigr]\,Q \pI \\
&=&\ts {1\over 2}\,\Bigl[\;Q\,(Id-F)\,Q\;Q\,dh\,Q\;Q\,(Id+\ti F)\,Q \;-\; Q\,\rho\, Q\;Q\,dh^TQ \; Q\,\rho\, Q  \;\Bigr] \pI \\ 
&=&\ts {1\over 2}\,\Bigl[\;(Id-F)\;Q\,dh\,Q\;(Id+Q\ti FQ\,) \;-\; \rho\;Q\,dh^TQ \; \rho  \;\Bigr] \pI
\eeq
We have 
\beq
Q\,\ti F\,Q&=&Q\,\bmat F_{\sd\sd} & F_{\sd\su} \\ F_{\su\sd} & F_{\su\su} \emat\, Q \;\;=\;\; 
\bmat F_{\su\su} & \eps_u\,F_{\su\sd} \\ \eps_u\, F_{\sd\su} & F_{\sd\sd} \emat \;\;\buildrel {\rm ind.hyp.}\over=\;\;
  \bmat F_{\sd\sd} & F_{\sd\su} \\ F_{\su\sd} & F_{\su\su} \emat \;\;=\;\; \ti F  \pI 
\eeq

\medskip
Furthermore, using $\,Q^T=\eps_u\,Q\,$,
\beq
Q\,dh^TQ&=&Q^T dh^TQ^T\;\;=\;\; [Q\,dh\,Q\,]^T  \pS 
\eeq
Finally, with $dh$ given by (\req{4.17nn}) of Theorem 4.1, we get with $\,r=s=0\,$:
\beq
\lefteqn{
Q\,dh\,Q\;\;=\;\;Q\;\Biggl[\; \ts -\,dt\,\bmat \vep & 0 \\ 0  & \vep \emat  
  \;+\;{udt\over 2}\,\bmat  0 & {\rm diag}[\,\rho_{\sd\su}\,] \; \\ {\rm diag}[\,\rho_{\su\sd}\,] & 0  \; \emat
  \;-\;\sqrt{\sabs u\sabs}\;\bmat  0 & dy \\ \eps_u\,dy & 0 \emat  \;\Biggr]\; Q }  \pI \\ 
&& \nn \\ 
&=&  \ts -\,dt\,\bmat \vep & 0 \\ 0  & \vep \emat  
  \;+\;{udt\over 2}\,\bmat  0 & \eps_u\,{\rm diag}[\,\rho_{\su\sd}\,] \; \\ \eps_u\,{\rm diag}[\,\rho_{\sd\su}\,] & 0  \; \emat
  \;-\;\sqrt{\sabs u\sabs}\;\bmat  0 & dy \\ \eps_u\,dy & 0 \emat  \pI \\ 
&& \nn \\ 
&\buildrel {\rm ind.hyp.}\over=&  \ts -\,dt\,\bmat \vep & 0 \\ 0  & \vep \emat  
  \;+\;{udt\over 2}\,\bmat  0 & {\rm diag}[\,\rho_{\sd\su}\,] \; \\ {\rm diag}[\,\rho_{\su\sd}\,] & 0  \; \emat
  \;-\;\sqrt{\sabs u\sabs}\;\bmat  0 & dy \\ \eps_u\,dy & 0 \emat \pI \\ 
&=& dh  \pI
\eeq
This results in 
\beq
Q\,d\rho\,Q&=&\ts {1\over 2}\,\Bigl[\;(Id-F)\,dh\,(Id+\ti F\,) \;-\; \rho\,dh^T \rho  \;\Bigr]  \;\;=\;\; d\rho  \pS
\eeq
In the same way, we get
\beq
Q\,dF\,Q&=&\ts {1\over 2}\,Q\,\Bigl[\; [\,\rho\,dh^T(Id+ F)\,]^T \,-\,\rho\,dh^T(Id+ F)  \; \Bigr] \,Q \pI \\ 
&=&\ts {1\over 2}\,\Bigl[\; [\,Q\,\rho\,dh^T(Id+ F)\,Q\,]^T \,-\,Q\,\rho\,dh^T(Id+ F)\,Q  \; \Bigr]  \pS \\ 
&=&\ts {1\over 2}\,\Bigl[\; [\,\rho\,dh^T(Id+ F)\,]^T \,-\,\rho\,dh^T(Id+ F)  \; \Bigr]  \;\;=\;\; dF \pI
\eeq
The energy $W$ is given by (\req{2.28}) in the Main Theorem. At half filling $\,\mu=0\,$ and for zero external pairing and exchange terms, $\,r=s=0\,$, 
it reduces to   
\beq
W&=&{\ts {1\over 2}}\, Tr_{\mathbb C^\G}\Bigl[\; \vep\,(G^{ab}_{\su\su} + G^{ab}_{\sd\sd})  \;\Bigr] 
  \;+\; {\ts {u\over 4}}\,\summ_j\,\Bigl[\;G^{ab}_{\su\su,jj}\;G^{ab}_{\sd\sd,jj} 
   \,-\,G^{ab}_{\su\sd,jj}\;G^{ab}_{\sd\su,jj} \, -\,G^{aa}_{\su\sd,jj}\;G^{bb}_{\su\sd,jj} \;\Bigr]  \pS \\  
&=&{\ts {1\over 2}}\, Tr_{\mathbb C^\G}\Bigl[\; \vep\,(\rho_{\su\su} + \rho_{\sd\sd})  \;\Bigr] 
  \;+\; {\ts {u\over 4}}\,\summ_j\,\Bigl[\;\rho_{\su\su,jj}\;\rho_{\sd\sd,jj} 
   \,-\,\rho_{\su\sd,jj}\;\rho_{\sd\su,jj} \, -\,iF_{\su\sd,jj}\;i\ti F_{\su\sd,jj} \;\Bigr]  \pS 
\eeq
or 
\beq  
W&=&{\ts {1\over 2}}\, Tr_{\mathbb C^\G}\Bigl[\; \vep\,(\rho_{\su\su} + \rho_{\sd\sd})  \;\Bigr] 
  \;+\; {\ts {u\over 4}}\,\summ_j\,\Bigl[\;\rho_{\su\su,jj}\;\rho_{\sd\sd,jj} 
   \,-\,\rho_{\su\sd,jj}\;\rho_{\sd\su,jj} \, +\,F_{\su\sd,jj}\;F_{\sd\su,jj} \;\Bigr]  \pS
\eeq
Now, using the results of Theorem 4.1 and the equations (\req{4.23n}) which we have just proven, the above expression further compactifies to 
\beq
W&=& Tr_{\mathbb C^\G}\bigl[\; \vep\,\rho_{\su\su} \;\bigr] 
  \;+\; {\ts {u\over 4}}\,\summ_j\,\Bigl[\;0 \,-\,\rho_{\su\sd,jj}\;\rho_{\sd\su,jj} \, +\,0 \;\Bigr]  \pS \\  
&=& Tr_{\mathbb C^\G}\bigl[\; \vep\,\rho_{\su\su} \;\bigr] 
  \;+\; {\ts {u\over 4}}\,\summ_j\,\Bigl[\;0 \,-\,\rho_{\su\sd,jj}\;\eps_u\,\rho_{\su\sd,jj} \, +\,0 \;\Bigr]  \pS \\
&=& Tr_{\mathbb C^\G}\bigl[\; \vep\,\rho_{\su\su} \;\bigr]  \;-\; {\ts {|u|\over 4}}\,\summ_j\,\rho_{\su\sd,jj}^2   \pS
\eeq
and this completes the proof of Theorem 4.2\,. \;$\blacksquare$

\bigskip
\bigskip
\bigskip
{\bf A.3 \;Proof of Theorem 4.3} 

\bigskip
\bigskip
In terms of Majorana fermion operators
\beq
a_{j\sigma}\;\;:=\;\;c_{j\sigma}\,+\,c^+_{j\sigma}  \;,&&\;\;\;\;\;\;  b_{j\sigma}\;\;:=\;\;{\ts {1\over i}}\,(\,c_{j\sigma}\,-\,c^+_{j\sigma}\,)  \pS  \nn \\
c_{j\sigma}\;\;=\;\;{\ts{1\over 2}}\,(\,a_{j\sigma}\,+\,i\,b_{j\sigma}\,)\;,&&\;\;\;\;\;\;
c_{j\sigma}^+\;\;=\;\;{\ts{1\over 2}}\,(\,a_{j\sigma}\,-\,i\,b_{j\sigma}\,)   \pS 
\eeq
we have
\beq
\{\,a_{i\sigma}\,,\,a_{j\tau}\,\}\;\;=\;\;\{\,a_{i\sigma}\,,\,b_{j\tau}\,\}\;\;=\;\;\{\,b_{i\sigma}\,,\,b_{j\tau}\,\}\;\;=\;\;0   \pS  
\eeq
with the exception of  $\,(i,\sigma)\,=\,(j,\tau)\,$ where we have
\beq
(a_{i\sigma})^2\;\;=\;\;(b_{i\sigma})^2&=&1\,,\;\;\;\;\;\; \{ \,a_{i\sigma}\,,\,b_{i\sigma}\,\}\;\;=\;\; 0 \pS
\eeq
The density operator becomes 
\beq
c_{j\sigma}^+c_{j\sigma}&=&\ts {1\over 4}\,(\,a_{j\sigma}\,-\,i\,b_{j\sigma}\,) \,(\,a_{j\sigma}\,+\,i\,b_{j\sigma}\,) \;\;=\;\;
  \ts {i\over 2}\,a_{j\sigma}b_{j\sigma}\,+\,{1\over 2} \pS
\eeq
Thus, 
\beqn
C_{\rm spin}(i,j)&:=&\la\,(n_{i\su}-n_{i\sd})\,(n_{j\su}-n_{j\sd})\,\ra_\beta  \pS \nn \\ 
&=&\ts -\,{1\over 4}\;\la\,(a_{i\su}b_{i\su}-a_{i\sd}b_{i\sd})\,(a_{j\su}b_{j\su}-a_{j\sd}b_{j\sd})\,\ra    \pI \lbeq{A.8} \\ 
&=&\ts +\,{1\over 4}\;\Bigl[\, -\,\la\,a_{i\su}b_{i\su}\,a_{j\su}b_{j\su}\,\ra \,-\, \la\,a_{i\sd}b_{i\sd}\,a_{j\sd}b_{j\sd}\,\ra   
 \ts \,+\,\la\,a_{i\su}b_{i\su}\,a_{j\sd}b_{j\sd}\,\ra \,+\, \la\,a_{i\sd}b_{i\sd}\,a_{j\su}b_{j\su}\,\ra \,\Bigr] \pS  \nn
\eeqn
where for simplicity we omitted the subscript $\beta$ at the angular brackets for the expectations. Since $a_{i\sigma}$ does not anticommute with itself, 
we have $\,(a_{i\sigma})^2=1\,$ instead of $\,(a_{i\sigma})^2=0\,$, we have 
to distinguish the cases $i=j$ and $i\ne j$. We also recall that the $G^{aa}$ and $G^{ab}$ of the Main Theorem correspond to expectations $i\la aa\ra$ and $i\la ab\ra$, 
so there is an $i=\sqrt{-1}$ included. With that, we can write   
\beq
-\,\la\,a_{i\su}b_{i\su}\,a_{j\su}b_{j\su}\,\ra&\buildrel i\ne j\over =&+\,\bigl\la\;G^{ab}_{\su\su,ii}\,G^{ab}_{\su\su,jj}\;-\;G^{aa}_{\su\su,ij}\,G^{bb}_{\su\su,ij}
  \;+\;G^{ab}_{\su\su,ij}\,G^{ba}_{\su\su,ij} \;\bigr\ra \pS \\ 
-\,\la\,a_{i\sd}b_{i\sd}\,a_{j\sd}b_{j\sd}\,\ra&\buildrel i\ne j\over =&+\,\bigl\la\;G^{ab}_{\sd\sd,ii}\,G^{ab}_{\sd\sd,jj}\;-\;G^{aa}_{\sd\sd,ij}\,G^{bb}_{\sd\sd,ij}
  \;+\;G^{ab}_{\sd\sd,ij}\,G^{ba}_{\sd\sd,ij} \;\bigr\ra \pS 
\eeq
while for $\,i=j\,$ the above equations change to 
\beq
-\,\la\,a_{j\su}b_{j\su}\,a_{j\su}b_{j\su}\,\ra& =& +\,\la\,a_{j\su}a_{j\su}\,b_{j\su}b_{j\su}\,\ra \;\;=\;\; 1  \pS \\ 
-\,\la\,a_{j\sd}b_{j\sd}\,a_{j\sd}b_{j\sd}\,\ra& =& +\,\la\,a_{j\sd}a_{j\sd}\,b_{j\sd}b_{j\sd}\,\ra \;\;=\;\; 1   \pS 
\eeq
The last two terms in the last line of (\req{A.8}) we can evaluate for arbitrary $i,j$ equal or non equal. We obtain 
\beq
\la\,a_{i\su}b_{i\su}\,a_{j\sd}b_{j\sd}\,\ra&=&-\,\bigl\la\;G^{ab}_{\su\su,ii}\,G^{ab}_{\sd\sd,jj}\;-\;G^{aa}_{\su\sd,ij}\,G^{bb}_{\su\sd,ij}
  \;+\;G^{ab}_{\su\sd,ij}\,G^{ba}_{\su\sd,ij}\;\bigr\ra \pS \\ 
\la\,a_{i\sd}b_{i\sd}\,a_{j\su}b_{j\su}\,\ra&=&-\,\bigl\la\;G^{ab}_{\sd\sd,ii}\,G^{ab}_{\su\su,jj}\;-\;G^{aa}_{\sd\su,ij}\,G^{bb}_{\sd\su,ij}
  \;+\;G^{ab}_{\sd\su,ij}\,G^{ba}_{\sd\su,ij} \;\bigr\ra \pS 
\eeq
Using the skew symmetry of $G$, this can be rewritten as 
\beq
-\,\la\,a_{i\su}b_{i\su}\,a_{j\su}b_{j\su}\,\ra&\buildrel i\ne j\over =&+\,\bigl\la\;G^{ab}_{\su\su,ii}\,G^{ab}_{\su\su,jj}\;-\;G^{aa}_{\su\su,ij}\,G^{bb}_{\su\su,ij}
  \;-\;G^{ab}_{\su\su,ij}\,G^{ab}_{\su\su,ji}\;\bigr\ra \pS \\ 
-\,\la\,a_{i\sd}b_{i\sd}\,a_{j\sd}b_{j\sd}\,\ra&\buildrel i\ne j\over =&+\,\bigl\la\;G^{ab}_{\sd\sd,ii}\,G^{ab}_{\sd\sd,jj}\;-\;G^{aa}_{\sd\sd,ij}\,G^{bb}_{\sd\sd,ij}
  \;-\;G^{ab}_{\sd\sd,ij}\,G^{ab}_{\sd\sd,ji}\;\bigr\ra  \pS\\ 
+\,\la\,a_{i\su}b_{i\su}\,a_{j\sd}b_{j\sd}\,\ra&=&-\,\bigl\la\;G^{ab}_{\su\su,ii}\,G^{ab}_{\sd\sd,jj}\;-\;G^{aa}_{\su\sd,ij}\,G^{bb}_{\su\sd,ij}
  \;-\;G^{ab}_{\su\sd,ij}\,G^{ab}_{\sd\su,ji}\;\bigr\ra \pS \\ 
+\,\la\,a_{i\sd}b_{i\sd}\,a_{j\su}b_{j\su}\,\ra&=&-\,\bigl\la\;G^{ab}_{\sd\sd,ii}\,G^{ab}_{\su\su,jj}\;-\;G^{aa}_{\su\sd,ji}\,G^{bb}_{\su\sd,ji}
  \;-\;G^{ab}_{\sd\su,ij}\,G^{ab}_{\su\sd,ji}\;\bigr\ra \pS 
\eeq
With the notation of Theorem 3.1, 
\beq
\rho&:=&G^{ab} \\
i\,F^a&:=&G^{aa} \\ 
i\,F^b&:=&G^{bb}
\eeq
this looks as follows: 
\beqn
-\,\la\,a_{i\su}b_{i\su}\,a_{j\su}b_{j\su}\,\ra&\buildrel i\ne j\over =&+\,\bigl\la\;\rho_{\su\su,ii}\,\rho_{\su\su,jj}\;+\;F^{a}_{\su\su,ij}\,F^{b}_{\su\su,ij}
  \;-\;\rho_{\su\su,ij}\,\rho_{\su\su,ji}\;\bigr\ra \pS  \nn \\ 
-\,\la\,a_{i\sd}b_{i\sd}\,a_{j\sd}b_{j\sd}\,\ra&\buildrel i\ne j\over =&+\,\bigl\la\;\rho_{\sd\sd,ii}\,\rho_{\sd\sd,jj}\;+\;F^{a}_{\sd\sd,ij}\,F^{b}_{\sd\sd,ij}
  \;-\;\rho_{\sd\sd,ij}\,\rho_{\sd\sd,ji}\;\bigr\ra  \pS  \nn \\ 
+\,\la\,a_{i\su}b_{i\su}\,a_{j\sd}b_{j\sd}\,\ra&=&-\,\bigl\la\;\rho_{\su\su,ii}\,\rho_{\sd\sd,jj}\;+\;F^{a}_{\su\sd,ij}\,F^{b}_{\su\sd,ij}
  \;-\;\rho_{\su\sd,ij}\,\rho_{\sd\su,ji}\;\bigr\ra \pS  \nn \\ 
+\,\la\,a_{i\sd}b_{i\sd}\,a_{j\su}b_{j\su}\,\ra&=&-\,\bigl\la\;\rho_{\sd\sd,ii}\,\rho_{\su\su,jj}\;+\;F^{a}_{\su\sd,ji}\,F^{b}_{\su\sd,ji}
  \;-\;\rho_{\sd\su,ij}\,\rho_{\su\sd,ji}\;\bigr\ra \pS  \lbeq{A.9}
\eeqn
We can evaluate further in the $w_1=1$ representation or in the $w_2=1$ representation. 

\medskip
\bigskip
{\bf Evaluation for} $\bm{w_1=1:}$ In this case, we have $F^b=F^a$ and from part (a) of Theorem 3.2 we get with $F\equiv F^a$ 
\beq
-\,\la\,a_{i\su}b_{i\su}\,a_{j\su}b_{j\su}\,\ra&\buildrel i\ne j\over =&+\,\bigl\la\;\rho_{\su\su,ii}\,\rho_{\su\su,jj}\;+\;F_{\su\su,ij}\,F_{\su\su,ij}
  \;-\;\rho_{\su\su,ij}\,\rho_{\su\su,ji}\;\bigr\ra  \\ 
-\,\la\,a_{i\sd}b_{i\sd}\,a_{j\sd}b_{j\sd}\,\ra&\buildrel i\ne j\over =&+\,\bigl\la\;\rho_{\sd\sd,ii}\,\rho_{\sd\sd,jj}\;+\;F_{\sd\sd,ij}\,F_{\sd\sd,ij}
  \;-\;\rho_{\sd\sd,ij}\,\rho_{\sd\sd,ji}\;\bigr\ra  \pS\\ 
+\,\la\,a_{i\su}b_{i\su}\,a_{j\sd}b_{j\sd}\,\ra&=&-\,\bigl\la\;\rho_{\su\su,ii}\,\rho_{\sd\sd,jj}\;+\;0
  \;-\;0\;\bigr\ra  \\ 
+\,\la\,a_{i\sd}b_{i\sd}\,a_{j\su}b_{j\su}\,\ra&=&-\,\bigl\la\;\rho_{\sd\sd,ii}\,\rho_{\su\su,jj}\;+\;0
  \;-\;0\;\bigr\ra \pS 
\eeq
Using also $\rho_{\sigma\sigma}^T=\rho_{\sigma\sigma}$, we obtain 
\beq
C_{\rm spin}(i,j)&\buildrel i\ne j\over =&\ts +\,{1\over 4}\;\Bigl\la\, (\rho_{\su\su,ii}-\rho_{\sd\sd,ii})\,(\rho_{\su\su,jj}-\rho_{\sd\sd,jj}) 
  \,+\, F_{\su\su,ij}^2\,+\,F_{\sd\sd,ij}^2 \,-\,\rho_{\su\su,ij}^2 \,-\,\rho_{\sd\sd,ij}^2 \,\Bigr\ra \pS
\eeq
and for $i=j$ 
\beq
C_{\rm spin}(j,j)&=&\ts +\,{1\over 4}\;\bigl\la\, 2  \,-\,\rho_{\su\su,jj}\,\rho_{\sd\sd,jj}\,-\,\rho_{\sd\sd,jj}\,\rho_{\su\su,jj} \,\bigr\ra 
  \;\;=\;\; \ts {1\over 2}\,\bigl\la\, 1  \,-\,\rho_{\su\su,jj}\,\rho_{\sd\sd,jj} \,\bigr\ra  \pS 
\eeq
Finally, with part (a) of Theorem 4.2, we arrive at 
\beq
C_{\rm spin}(i,j)&\buildrel i\ne j\over =&\ts +\,{1\over 4}\;\Bigl\la\, (\rho_{\su\su,ii}+\eps_u\,\rho_{\su\su,ii})\,(\rho_{\su\su,jj}+\eps_u\,\rho_{\su\su,jj}) 
  \,+\, 2\,F_{\su\su,ij}^2 \,-\,2\,\rho_{\su\su,ij}^2  \,\Bigr\ra \pS \\ 
& =&\ts \,{1\over 2}\;\Bigl\la\, F_{\su\su,ij}^2 \,-\,\rho_{\su\su,ij}^2 
  \,+\,  {1\over 2}\,(1+\eps_u)^2\,\rho_{\su\su,ii}\,\rho_{\su\su,jj}  \,\Bigr\ra \pS
\eeq
and for $i=j$ 
\beq
C_{\rm spin}(j,j)&=& \ts {1\over 2}\,\bigl\la\, 1  \,+\,\eps_u\,\rho_{\su\su,jj}^2 \,\bigr\ra   \pS
\eeq
This proves part (a) of Theorem 4.3\,.   

\medskip
\bigskip
{\bf Evaluation for} $\bm{w_2=1:}$ In that case, recall from Theorem 3.2 that with $\,F\equiv F^a\,$ we have 
\beqn
\bmat F^{b}_{\su\su} &F^{b}_{\su\sd} \\ F^{b}_{\sd\su} & F^{b}_{\sd\sd} \emat  \;\;=\;\;  
  \bmat F^{a}_{\sd\sd} & F^{a}_{\sd\su}  \\  F^{a}_{\su\sd} & F^{a}_{\su\su} \emat \;\;=\;\;  
  \bmat F_{\sd\sd} & F_{\sd\su}  \\  F_{\su\sd} & F_{\su\su} \emat  \lbeq{A.10n}
\eeqn
Thus we get 
\beq
-\,\la\,a_{i\su}b_{i\su}\,a_{j\su}b_{j\su}\,\ra&\buildrel i\ne j\over =&+\,\bigl\la\;\rho_{\su\su,ii}\,\rho_{\su\su,jj}\;+\;F_{\su\su,ij}\,F_{\sd\sd,ij}
  \;-\;\rho_{\su\su,ij}\,\rho_{\su\su,ji}\;\bigr\ra \pS \\ 
-\,\la\,a_{i\sd}b_{i\sd}\,a_{j\sd}b_{j\sd}\,\ra&\buildrel i\ne j\over =&+\,\bigl\la\;\rho_{\sd\sd,ii}\,\rho_{\sd\sd,jj}\;+\;F_{\sd\sd,ij}\,F_{\su\su,ij}
  \;-\;\rho_{\sd\sd,ij}\,\rho_{\sd\sd,ji}\;\bigr\ra  \pS\\ 
+\,\la\,a_{i\su}b_{i\su}\,a_{j\sd}b_{j\sd}\,\ra&=&-\,\bigl\la\;\rho_{\su\su,ii}\,\rho_{\sd\sd,jj}\;+\;F_{\su\sd,ij}\,F_{\sd\su,ij}
  \;-\;\rho_{\su\sd,ij}\,\rho_{\sd\su,ji}\;\bigr\ra \pS \\ 
+\,\la\,a_{i\sd}b_{i\sd}\,a_{j\su}b_{j\su}\,\ra&=&-\,\bigl\la\;\rho_{\sd\sd,ii}\,\rho_{\su\su,jj}\;+\;F_{\su\sd,ji}\,F_{\sd\su,ji}
  \;-\;\rho_{\sd\su,ij}\,\rho_{\su\sd,ji}\;\bigr\ra \pS 
\eeq
From part (b) of Theorem 4.2, we have the symmetry relations 
\beq
\rho_{\sd\sd}\;\;=\;\;\rho_{\su\su}\;,&&\;\;\;\;F_{\sd\sd}\;\;=\;\;F_{\su\su}  \pS \\ 
\rho_{\sd\su}\;\;=\;\;\eps_u\,\rho_{\su\sd}\;,&&\;\;\;\;F_{\sd\su}\;\;=\;\;\eps_u\,F_{\su\sd}  
\eeq
such that 
\beq
-\,\la\,a_{i\su}b_{i\su}\,a_{j\su}b_{j\su}\,\ra&\buildrel i\ne j\over =&+\,\bigl\la\;\rho_{\su\su,ii}\,\rho_{\su\su,jj}\;+\;F_{\su\su,ij}^2
  \;-\;\rho_{\su\su,ij}\,\rho_{\su\su,ji}\;\bigr\ra \pS \\ 
-\,\la\,a_{i\sd}b_{i\sd}\,a_{j\sd}b_{j\sd}\,\ra&\buildrel i\ne j\over =&+\,\bigl\la\;\rho_{\su\su,ii}\,\rho_{\su\su,jj}\;+\;F_{\su\su,ij}^2
  \;-\;\rho_{\su\su,ij}\,\rho_{\su\su,ji}\;\bigr\ra  \pS\\ 
+\,\la\,a_{i\su}b_{i\su}\,a_{j\sd}b_{j\sd}\,\ra&=&-\,\bigl\la\;\rho_{\su\su,ii}\,\rho_{\su\su,jj}\;+\;\eps_u\,F_{\su\sd,ij}^2
  \;-\;\eps_u\,\rho_{\su\sd,ij}\,\rho_{\su\sd,ji}\;\bigr\ra \pS \\ 
+\,\la\,a_{i\sd}b_{i\sd}\,a_{j\su}b_{j\su}\,\ra&=&-\,\bigl\la\;\rho_{\su\su,ii}\,\rho_{\su\su,jj}\;+\;\eps_u\,F_{\su\sd,ji}^2
  \;-\;\eps_u\,\rho_{\su\sd,ij}\,\rho_{\su\sd,ji}\;\bigr\ra \pS 
\eeq
We also recall from part (b) of Theorem 3.2 that 
\beq
\rho_{\su\su}^T&=&\rho_{\sd\sd}\;\;=\;\;\rho_{\su\su}\,,\;\;\;\;\;\rho_{\su\sd}^T\;\;=\;\;\rho_{\su\sd} \pS
\eeq
Also, since $F^T=-F$, we have $F_{\su\sd}^T=-F_{\sd\su}=-\eps_u\,F_{\su\sd}$. Thus we get
\beq
C_{\rm spin}(i,j)&\buildrel i\ne j\over =&\ts +\,{1\over 4}\;\Bigl\la\, 2\,F_{\su\su,ij}^2\,-\,2\,\rho_{\su\su,ij}^2
 \ts \,-\,\eps_u\,(\,F_{\su\sd,ij}^2\,+\,F_{\su\sd,ji}^2\,) \,+\,2\,\eps_u\,\rho_{\su\sd,ij}^2 \,\Bigr\ra \pI \\ 
&=&\ts +\,{1\over 2}\;\bigl\la\, F_{\su\su,ij}^2\,-\,\rho_{\su\su,ij}^2
 \ts \,-\,\eps_u\,F_{\su\sd,ij}^2 \,+\,\eps_u\,\rho_{\su\sd,ij}^2 \,\bigr\ra \pS
\eeq
and with Theorem 4.1,
\beq
C_{\rm spin}(j,j)&=&\ts {1\over 2}\;\Bigl\la\, 1 \,-\,\rho_{\su\su,jj}^2 
 \ts \,-\,\eps_u\,F_{\su\sd,jj}^2 \,+\,\eps_u\,\rho_{\su\sd,jj}^2 \,\Bigr\ra \;\;=\;\; \ts {1\over 2}\;\bigl\la\, 1 \,-\,0
 \ts \,-\,0 \,+\,\eps_u\,\rho_{\su\sd,jj}^2 \,\bigr\ra \pI
\eeq
Finally, again with Theorem 4.1, for lattice sites $i\ne j$, 
\beq
C_{\rm spin}^{\,\rm on}(i,j)& =&\ts +\,{1\over 2}\;\bigl\la\, F_{\su\su,ij}^2 \,+\,\eps_u\,\rho_{\su\sd,ij}^2 \,\bigr\ra 
  \;\;\buildrel u>0 \over =\;\; {1\over 2}\;\bigl\la\, F_{\su\su,ij}^2 \,+\,\rho_{\su\sd,ij}^2 \,\bigr\ra \;\;\ge \;\; 0 \pI \\
C_{\rm spin}^{\,\rm off}(i,j)& =&\ts +\,{1\over 2}\;\bigl\la\, -\,\rho_{\su\su,ij}^2
 \ts \,-\,\eps_u\,F_{\su\sd,ij}^2 \,\bigr\ra  
\;\;\buildrel u>0 \over =\;\; \ts -\,{1\over 2}\,\bigl\la\, \rho_{\su\su,ij}^2 \,+\,F_{\su\sd,ij}^2 \,\bigr\ra \;\;\le \;\;0  \pS
\eeq
which completes the proof of Theorem 4.3\,. \;\;$\blacksquare$

\medskip
\bigskip
\bigskip
\bigskip
{\bf A.4 \;Proof of Theorem 4.4} 

\bigskip
\bigskip
In terms of Majorana fermion operators, we have 
\beq
c_{i\su}^+c_{i\sd}^+&=&\ts {1\over 4}\,\bigl[ \,+\,(\,a_{i\su}a_{i\sd}\,-\,b_{i\su}b_{i\sd}\,)\,-\,i\,(\, a_{i\su}b_{i\sd} \,-\,a_{i\sd}b_{i\su}\,)\,\bigr] \pS\\
c_{j\sd}c_{j\su}&=&\ts {1\over 4}\,\bigl[ \,-\,(\,a_{j\su}a_{j\sd}\,-\,b_{j\su}b_{j\sd}\,)\,-\,i\,(\, a_{j\su}b_{j\sd} \,-\,a_{j\sd}b_{j\su}\,)\,\bigr] \pS 
\eeq
Thus, 
\beqn
\la\,c_{i\su}^+c_{i\sd}^+c_{j\sd}c_{j\su}\,\ra&=&\ts -{1\over 16}\,\Bigl\la\,a_{i\su}a_{i\sd}a_{j\su}a_{j\sd}\,+\,b_{i\su}b_{i\sd}b_{j\su}b_{j\sd}
    \,-\,a_{i\su}a_{i\sd}b_{j\su}b_{j\sd}\,-\,a_{j\su}a_{j\sd}b_{i\su}b_{i\sd}  \pS  \nn \\ 
&&\phantom{mmm} \,-\,a_{i\su}a_{j\su}b_{i\sd}b_{j\sd}\,-\,a_{i\sd}a_{j\sd}b_{i\su}b_{j\su}
    \,+\,a_{i\su}a_{j\sd}b_{i\sd}b_{j\su}\,+\,a_{i\sd}a_{j\su}b_{i\su}b_{j\sd} \,\Bigr\ra \pS  \nn \\ 
&&-\;\ts {i\over 16}\,\Bigl\la\,a_{i\su}a_{i\sd} a_{j\su}b_{j\sd}\,+\,b_{i\su}b_{i\sd} a_{j\sd}b_{j\su} \,-\,b_{i\su}b_{i\sd} a_{j\su}b_{j\sd} \,-\,
  a_{i\su}a_{i\sd}a_{j\sd}b_{j\su}\,    \pS \nn  \\ 
&&\phantom{mmm} \,-\,a_{i\su}b_{i\sd} a_{j\su}a_{j\sd} \,-\,a_{i\sd}b_{i\su}b_{j\su}b_{j\sd} \,+\,a_{i\sd}b_{i\su}a_{j\su}a_{j\sd}\,+\,a_{i\su}b_{i\sd} b_{j\su}b_{j\sd} 
 \,\Bigr\ra \pS \nn \\ 
&=:& \ts -{1\over 16}\,{\rm term}_1 \;-\;\ts {i\over 16}\,{\rm term}_2 \pS
\eeqn
Let us first consider the case

\medskip
\bigskip
$\bm{i\ne j\,:}$ 

\bigskip
We leave term$_1$ as above and move all the $b$ operators to the right in term$_2$,
\beqn
{\rm term}_1&=&\Bigl\la\,a_{i\su}a_{i\sd}a_{j\su}a_{j\sd}\,+\,b_{i\su}b_{i\sd}b_{j\su}b_{j\sd}
    \,-\,a_{i\su}a_{i\sd}b_{j\su}b_{j\sd}\,-\,a_{j\su}a_{j\sd}b_{i\su}b_{i\sd}  \pS  \nn \\ 
&&\phantom{mmm} \,-\,a_{i\su}a_{j\su}b_{i\sd}b_{j\sd}\,-\,a_{i\sd}a_{j\sd}b_{i\su}b_{j\su}
    \,+\,a_{i\su}a_{j\sd}b_{i\sd}b_{j\su}\,+\,a_{i\sd}a_{j\su}b_{i\su}b_{j\sd} \,\Bigr\ra \pS\pS \\ 
{\rm term}_2&=&\Bigl\la\,a_{i\su}a_{i\sd} a_{j\su}b_{j\sd}\,+\, a_{j\sd}b_{j\su}b_{i\su}b_{i\sd}
   \,-\, a_{j\su}b_{j\sd}b_{i\su}b_{i\sd} \,-\,a_{i\su}a_{i\sd}a_{j\sd}b_{j\su}\,    \pS  \nn \\ 
&&\phantom{mmm} \,-\,a_{j\su}a_{j\sd}a_{i\su}b_{i\sd}  \,-\,a_{i\sd}b_{i\su}b_{j\su}b_{j\sd} \,+\,a_{j\su}a_{j\sd}a_{i\sd}b_{i\su}\,+\,a_{i\su}b_{i\sd} b_{j\su}b_{j\sd} 
 \,\Bigr\ra \pS\pS  \lbeq{A.11}
\eeqn
We can evaluate in the $w_1=1$ representation or in the $w_2=1$ representation. 

\goodbreak
\medskip
\bigskip
{\bf Evaluation for} $\bm{w_1=1:}$ 

\bigskip
From part (a) of Theorem 3.2 we have $\,G^{aa}_{\su\sd}=G^{ab}_{\su\sd}=G^{bb}_{\su\sd}=0\,$. Thus,  
recalling again that the $G^{aa}$ and $G^{ab}$ of the Main Theorem correspond to expectations $i\la aa\ra$ and $i\la ab\ra$ which accounts to the first 
overall minus sign in the next equations,  
\beq
\la\,a_{i\su}a_{i\sd}a_{j\su}a_{j\sd}\,\ra&=&\,-\,\bigl\la\; 0\, -\,G^{aa}_{\su\su,ij}\, G^{aa}_{\sd\sd,ij}\,+\,0\; \bigr\ra\;\;=\;\;
  +\,\bigl\la\,G^{aa}_{\su\su,ij}\, G^{aa}_{\sd\sd,ij}\, \bigr\ra \pS \\ 
\la\,b_{i\su}b_{i\sd}b_{j\su}b_{j\sd}\,\ra&=&\,-\,\bigl\la\; 0\, -\,G^{bb}_{\su\su,ij}\, G^{bb}_{\sd\sd,ij}\,+\,0\; \bigr\ra\;\;=\;\;
  +\,\bigl\la\,G^{bb}_{\su\su,ij}\, G^{bb}_{\sd\sd,ij}\, \bigr\ra \pS
\eeq
\beq
\la\,a_{i\su}a_{i\sd}b_{j\su}b_{j\sd}\,\ra&=&-\,\bigl\la\;0\,-\,G^{ab}_{\su\su,ij}\,G^{ab}_{\sd\sd,ij}\,+\,0\;\bigr\ra \;\;=\;\;
  +\,\bigl\la\;\,G^{ab}_{\su\su,ij}\,G^{ab}_{\sd\sd,ij}\,\bigr\ra \pS \\ 
\la\,a_{j\su}a_{j\sd}b_{i\su}b_{i\sd}\,\ra&=&-\,\bigl\la\;0\,-\,G^{ab}_{\su\su,ji}\,G^{ab}_{\sd\sd,ji}\,+\,0\;\bigr\ra \;\;=\;\;
  +\,\bigl\la\;\,G^{ab}_{\su\su,ji}\,G^{ab}_{\sd\sd,ji}\,\bigr\ra\pS 
\eeq
\beq
\la\,a_{i\su}a_{j\su}b_{i\sd}b_{j\sd}\,\ra&=&-\,\bigl\la \;G^{aa}_{\su\su,ij}\,G^{bb}_{\sd\sd,ij}\,-\,0\,+\,0\; \bigr\ra \;\;=\;\; 
  -\,\bigl\la \,G^{aa}_{\su\su,ij}\,G^{bb}_{\sd\sd,ij}\, \bigr\ra  \pS \\ 
\la\,a_{i\sd}a_{j\sd}b_{i\su}b_{j\su}\,\ra&=&-\,\bigl\la \;G^{aa}_{\sd\sd,ij}\,G^{bb}_{\su\su,ij}\,-\,0\,+\,0\; \bigr\ra \;\;=\;\; 
  -\,\bigl\la \,G^{aa}_{\sd\sd,ij}\,G^{bb}_{\su\su,ij}\, \bigr\ra\pS
\eeq
\beq
\la\,a_{i\su}a_{j\sd}b_{i\sd}b_{j\su}\,\ra&=&-\,\bigl\la\;0\,-\,0\,+\,G^{ab}_{\su\su,ij}\,G^{ab}_{\sd\sd,ji}\;\bigr\ra \;\;=\;\; 
  -\,\bigl\la\,G^{ab}_{\su\su,ij}\,G^{ab}_{\sd\sd,ji}\,\bigr\ra   \pS \\ 
\la\,a_{i\sd}a_{j\su}b_{i\su}b_{j\sd} \,\ra&=&-\,\bigl\la\;0\,-\,0\,+\,G^{ab}_{\sd\sd,ij}\,G^{ab}_{\su\su,ji}\;\bigr\ra \;\;=\;\; 
  -\,\bigl\la\,G^{ab}_{\sd\sd,ij}\,G^{ab}_{\su\su,ji}\,\bigr\ra\pS 
\eeq
With the notation of Theorem 3.1, 
\beq
\rho&:=&G^{ab} \\
i\,F^a&:=&G^{aa} \\ 
i\,F^b&:=&G^{bb}
\eeq
and recalling $\,F^b=F^a\equiv F\,$ from part (a) of Theorem 3.2, we get 
\beq
{\rm term}_1&=&\ts -\,4\,\bigl\la\,F_{\su\su,ij}F_{\sd\sd,ij}\,+\,\rho_{\su\su,ij}\rho_{\sd\sd,ij} \,\bigr\ra \pI
\eeq

Furthermore, 
\beq
\bigl\la\; a_{i\su}a_{i\sd} a_{j\su}b_{j\sd}\,+\, a_{j\sd}b_{j\su}b_{i\su}b_{i\sd} \;\bigr\ra &=&
  -\,\bigl\la\,-\,i\,F_{\su\su,ij}\rho_{\sd\sd,ij} \,+\, \rho_{\sd\sd,ji}\,iF_{\su\su,ji} \,\bigr\ra
  \;\;=\;\; +\,2i\,\bigl\la\,\rho_{\sd\sd,ij}F_{\su\su,ij}\,\bigr\ra \pS \\ 
\bigl\la\; a_{j\su}b_{j\sd}b_{i\su}b_{i\sd} \,+\,a_{i\su}a_{i\sd}a_{j\sd}b_{j\su}\;\bigr\ra &=&
  -\,\bigl\la\,-\,\rho_{\su\su,ji}\,iF_{\sd\sd,ji} \,+\,\rho_{\su\su,ij}\,iF_{\sd\sd,ij} \,\bigr\ra
 \;\;=\;\; -\,2i\,\bigl\la\,\rho_{\su\su,ij}F_{\sd\sd,ij}\,\bigr\ra \pS \\ 
\bigl\la\; a_{j\su}a_{j\sd}a_{i\su}b_{i\sd}  \,+\,a_{i\sd}b_{i\su}b_{j\su}b_{j\sd}\;\bigr\ra &=&
  -\,\bigl\la\,-\,iF_{\su\su,ji}\rho_{\sd\sd,ji}\,+\,\rho_{\sd\sd,ij}\,iF_{\su\su,ij} \,\bigr\ra
\;\;=\;\; -\,2i\,\bigl\la\,\rho_{\sd\sd,ij}F_{\su\su,ij}\,\bigr\ra  \pS \\
\bigl\la\; a_{j\su}a_{j\sd}a_{i\sd}b_{i\su}\,+\,a_{i\su}b_{i\sd} b_{j\su}b_{j\sd}\;\bigr\ra &=&
  -\,\bigl\la\,+\,\rho_{\su\su,ji}\,iF_{\sd\sd,ji} \,-\,\rho_{\su\su,ij}\,iF_{\sd\sd,ij} \,\bigr\ra
\;\;=\;\;+\,2i\,\bigl\la\,\rho_{\su\su,ij}F_{\sd\sd,ij}\,\bigr\ra \pS
\eeq
which gives 
\beq
{\rm term}_2&=&\Bigl\la\,\,+\,2i\,\rho_{\sd\sd,ij}F_{\su\su,ij}\,+\,2i\,\rho_{\su\su,ij}F_{\sd\sd,ij}\,+\,2i\,\rho_{\sd\sd,ij}F_{\su\su,ij} 
  \,+\,2i\,\rho_{\su\su,ij}F_{\sd\sd,ij} \,\Bigr\ra \pS \\ 
&=&\ts 4i\,\bigl\la\, F_{\su\su,ij}\,\rho_{\sd\sd,ij}\,+\, F_{\sd\sd,ij}\,\rho_{\su\su,ij}  \,\ra  \pS
\eeq
Alltogether, 
\beq
\la\,c_{i\su}^+c_{i\sd}^+c_{j\sd}c_{j\su}\,\ra&=&  \ts -{1\over 16}\,{\rm term}_1 \;-\;\ts {i\over 16}\,{\rm term}_2 \pS \\ 
&=& \ts +\,{1\over 4}\,\bigl\la\,F_{\su\su,ij}F_{\sd\sd,ij}\,+\,\rho_{\su\su,ij}\rho_{\sd\sd,ij} 
  \,+\, F_{\su\su,ij}\,\rho_{\sd\sd,ij}\,+\, F_{\sd\sd,ij}\,\rho_{\su\su,ij}  \,\bigr\ra \pI \\ 
&=&\ts +\,{1\over 4}\,\bigl\la\;(F_{\su\su,ij}+\rho_{\su\su,ij})\,(F_{\sd\sd,ij}+\rho_{\sd\sd,ij})\; \bigr\ra \pS
\eeq
Recall the notation (\req{4.4n}) to make the $\,(\GA,\GA)\,,\,(\GB,\GB)\,$, the on-matrix elements, or the $\,(\GA,\GB)\,,\,(\GB,\GA)\,$, the off-matrix elements, 
more explicit. With part (a) of Theorem 4.2, we get 
\beq
\la\,c_{i\su}^+c_{i\sd}^+c_{j\sd}c_{j\su}\,\ra&=&\ts {1\over 4}\,\Bigl\la\,(F_{\su\su,ij}+\rho_{\su\su,ij})^{\rm on}\,(F_{\sd\sd,ij}+\rho_{\sd\sd,ij})^{\rm on} \, \Bigr\ra \pI \\ 
&&\;+\;\ts {1\over 4}\,\Bigl\la\,(F_{\su\su,ij}+\rho_{\su\su,ij})^{\rm off}\,(F_{\sd\sd,ij}+\rho_{\sd\sd,ij})^{\rm off} \, \Bigr\ra \pS \\
&=&\ts {1\over 4}\,\Bigl\la\,(F_{\su\su,ij}^{\rm on}+\rho_{\su\su,ij}^{\rm on})\,
  (F_{\su\su,ij}^{\rm on}-\eps_u\,\rho_{\su\su,ij}^{\rm on}) \, \Bigr\ra \pI \\ 
&&\;+\;\ts {1\over 4}\,\Bigl\la\,(F_{\su\su,ij}^{\rm off}+\rho_{\su\su,ij}^{\rm off})\,(-\eps_u\,F_{\su\su,ij}^{\rm off}+\rho_{\su\su,ij}^{\rm off}) \, \Bigr\ra \pS\\
&=&\ts {1\over 4}\, (\chi_{ij}^{\rm on}-\eps_u \chi_{ij}^{\rm off})\;
  \Bigl\la\,(F_{\su\su,ij}+\rho_{\su\su,ij})(F_{\su\su,ij}-\,\eps_u \,\rho_{\su\su,ij}) \, \Bigr\ra  \pI  
\eeq
which proves part (a) of Theorem 4.4 for $i\ne j\,$.

\medskip
\bigskip
{\bf Evaluation for} $\bm{w_2=1:}$ 

\bigskip
Recall the expression for term$_2$ given by (\req{A.11}). Each of the eight expectations is zero, since from Theorem 4.1 we have 
\beq
\rho_{\su\su,ij}\cdot F_{\su\su,ij}\;\;=\;\;\rho_{\su\su,ij}\cdot F_{\sd\sd,ij}&=& 0 \;\;\;\;\; \forall \;i,j  \pS \\ 
\rho_{\su\sd,ij}\cdot F_{\su\sd,ij}\;\;=\;\;\rho_{\su\sd,ij}\cdot F_{\sd\su,ij}&=& 0 \;\;\;\;\; \forall \;i,j  \pS 
\eeq
and the expectations are given by, with $F^a\equiv F$ and recalling (\req{A.10n}),
\beq
\la a_{i\su}a_{i\sd}a_{j\su}b_{j\sd}\ra &\sim& \pm F_{\su\sd,ii}\rho_{\su\sd,jj}\; \pm\; F_{\su\su,ij}\rho_{\sd\sd,ij}\;\pm\; \rho_{\su\sd,ij}F_{\sd\su,ij} \;\;=\;\; 0 \pS \\ 
\la a_{j\sd}b_{j\su}b_{i\su}b_{i\sd}\ra &\sim& \pm \rho_{\sd\su,jj}F^b_{\su\sd,ii}\; \pm\; \rho_{\sd\su,ji}F^b_{\su\sd,ji}\;\pm\; \rho_{\sd\sd,ji}F^b_{\su\su,ij} \;\;=\;\; 0 \pS \\ 
\la a_{j\su}b_{j\sd}b_{i\su}b_{i\sd}\ra &\sim& \pm \rho_{\su\sd,jj}F^b_{\su\sd,ii}\; \pm\; \rho_{\su\su,ji}F^b_{\sd\sd,ji}\;\pm\; \rho_{\su\sd,ji}F^b_{\sd\su,ij} \;\;=\;\; 0 \pS \\
\la a_{i\su}a_{i\sd}a_{j\sd}b_{j\su}\ra &\sim& \pm F_{\su\sd,ii}\rho_{\sd\su,jj}\; \pm\; F_{\su\sd,ij}\rho_{\sd\su,ij}\;\pm\; \rho_{\su\su,ij}F_{\sd\sd,ij} \;\;=\;\; 0 \pS 
\eeq
with similar expressions for the expectations in the second line of (\req{A.11}). For the expectations of term$_1$, we find, using again (\req{A.10n}),  
\beq
\la\,a_{i\su}a_{i\sd}a_{j\su}a_{j\sd}\,\ra&=&\,-\,\la\, -F_{\su\sd,ii}F_{\su\sd,jj}\, +\,F_{\su\su,ij}F_{\sd\sd,ij}\,-\,F_{\su\sd,ij}F_{\sd\su,ij}\, \ra \pS \\ 
&=&\,-\,\la\, 0\, +\,F_{\su\su,ij}^2\,-\,\eps_u\,F_{\su\sd,ij}^2\, \ra \pS \\ 
\la\,b_{i\su}b_{i\sd}b_{j\su}b_{j\sd}\,\ra&=&\,-\,\la\, -F_{\sd\su,ii}F_{\sd\su,jj}\, +\,F_{\sd\sd,ij}F_{\su\su,ij}\,-\,F_{\sd\su,ij}F_{\su\sd,ij}\, \ra \pS \\ 
&=&\,-\,\la\, 0\, +\,F_{\su\su,ij}^2\,-\,\eps_u\,F_{\su\sd,ij}^2\, \ra \pS \\
\la\,a_{i\su}a_{i\sd}a_{j\su}a_{j\sd}\,\ra\,+\,\la\,b_{i\su}b_{i\sd}b_{j\su}b_{j\sd}\,\ra&=&\,-\,2\,\la\,F_{\su\su,ij}^2\,-\,\eps_u\,F_{\su\sd,ij}^2\, \ra \pI
\eeq
and
\beq
\la\,a_{i\su}a_{i\sd}b_{j\su}b_{j\sd}\,\ra&=&-\,\la\,-\,F_{\su\sd,ii}F_{\sd\su,jj}\,-\,\rho_{\su\su,ij}\rho_{\sd\sd,ij}\,+\,\rho_{\su\sd,ij}\rho_{\sd\su,ij}\,\ra \pS \\ 
&=&-\,\la\,-\,\rho_{\su\su,ij}^2\,+\,\eps_u\,\rho_{\su\sd,ij}^2\,\ra \pS \\ 
\la\,a_{j\su}a_{j\sd}b_{i\su}b_{i\sd}\,\ra&=&-\,\la\,-\,\rho_{\su\su,ji}^2\,+\,\eps_u\,\rho_{\su\sd,ji}^2\,\ra 
  \;\;=\;\; -\,\la\,-\,\rho_{\su\su,ij}^2\,+\,\eps_u\,\rho_{\su\sd,ij}^2\,\ra  \pS \\ 
\la\,a_{i\su}a_{i\sd}b_{j\su}b_{j\sd}\,\ra\,+\,\la\,a_{j\su}a_{j\sd}b_{i\su}b_{i\sd}\,\ra&=&\,+\,2\,\la\,\rho_{\su\su,ij}^2\,-\,\eps_u\,\rho_{\su\sd,ij}^2\,\ra \pI 
\eeq
and
\beq
\la\,a_{i\su}a_{j\su}b_{i\sd}b_{j\sd}\,\ra&=&-\,\la \,-\,F_{\su\su,ij}F_{\su\su,ij}\,-\,\rho_{\su\sd,ii}\rho_{\su\sd,jj}\,+\,\rho_{\su\sd,ij}\rho_{\su\sd,ji}\, \ra \pS \\ 
&=&-\,\la \,-\,F_{\su\su,ij}^2\,-\,\rho_{\su\sd,ii}\rho_{\su\sd,jj}\,+\,\rho_{\su\sd,ij}^2\, \ra \pS \\ 
\la\,a_{i\sd}a_{j\sd}b_{i\su}b_{j\su}\,\ra&=&-\,\la \,-\,F_{\su\su,ij}^2\,-\,\rho_{\su\sd,ii}\rho_{\su\sd,jj}\,+\,\rho_{\su\sd,ij}^2\, \ra \pS \\ 
\la\,a_{i\su}a_{j\su}b_{i\sd}b_{j\sd}\,\ra\,+\,\la\,a_{i\sd}a_{j\sd}b_{i\su}b_{j\su}\,\ra&=&
  +\,2\,\la \,F_{\su\su,ij}^2\,+\,\rho_{\su\sd,ii}\rho_{\su\sd,jj}\,-\,\rho_{\su\sd,ij}^2\, \ra \pI
\eeq
and
\beq
\la\,a_{i\su}a_{j\sd}b_{i\sd}b_{j\su}\,\ra&=&-\,\la\,-\,F_{\su\sd,ij}F^b_{\sd\su,ij}\,-\,\rho_{\su\sd,ii}\rho_{\sd\su,jj}\,+\,\rho_{\su\su,ij}\rho_{\sd\sd,ji}\,\ra \pS \\ 
&=&-\,\la\,-\,F_{\su\sd,ij}^2\,-\,\eps_u\,\rho_{\su\sd,ii}\rho_{\su\sd,jj}\,+\,\rho_{\su\su,ij}^2\,\ra \pS \\ 
\la\,a_{i\sd}a_{j\su}b_{i\su}b_{j\sd} \,\ra&=&-\,\la\,-\,F_{\su\sd,ij}^2\,-\,\eps_u\,\rho_{\su\sd,ii}\rho_{\su\sd,jj}\,+\,\rho_{\su\su,ij}^2\,\ra \pS \\ 
\la\,a_{i\su}a_{j\sd}b_{i\sd}b_{j\su}\,\ra\,+\,\la\,a_{i\sd}a_{j\su}b_{i\su}b_{j\sd}\,\ra&=&
  +\,2\,\la\,F_{\su\sd,ij}^2\,+\,\eps_u\,\rho_{\su\sd,ii}\rho_{\su\sd,jj}\,-\,\rho_{\su\su,ij}^2\,\ra \pI 
\eeq

\medskip
In total,
\beq
\la\,c_{i\su}^+c_{i\sd}^+c_{j\sd}c_{j\su}\,\ra&=&\ts -{1\over 16}\,\Bigl\la\,a_{i\su}a_{i\sd}a_{j\su}a_{j\sd}\,+\,b_{i\su}b_{i\sd}b_{j\su}b_{j\sd}
    \,-\,a_{i\su}a_{i\sd}b_{j\su}b_{j\sd}\,-\,a_{j\su}a_{j\sd}b_{i\su}b_{i\sd}  \pI \\ 
&&\phantom{mmm} \,-\,a_{i\su}a_{j\su}b_{i\sd}b_{j\sd}\,-\,a_{i\sd}a_{j\sd}b_{i\su}b_{j\su}
    \,+\,a_{i\su}a_{j\sd}b_{i\sd}b_{j\su}\,+\,a_{i\sd}a_{j\su}b_{i\su}b_{j\sd} \,\Bigr\ra \pS \\ 
&=&\ts +\,{2\over 16}\,\Bigl\la\,\,F_{\su\su,ij}^2\,-\,\eps_u\,F_{\su\sd,ij}^2\, + \,\rho_{\su\su,ij}^2\,-\,\eps_u\,\rho_{\su\sd,ij}^2\,   \pI \\ 
&&\phantom{mmm} \,+\,F_{\su\su,ij}^2\,+\,\rho_{\su\sd,ii}\rho_{\su\sd,jj}\,-\,\rho_{\su\sd,ij}^2\, 
   \,-\,F_{\su\sd,ij}^2\,-\,\eps_u\,\rho_{\su\sd,ii}\rho_{\su\sd,jj}\,+\,\rho_{\su\su,ij}^2\, \;\Bigr\ra \pS
\eeq
or
\beq
\la\, c_{i\su}^+c_{i\sd}^+c_{j\sd}c_{j\su}\,\ra &=&\ts +\,{1\over 4}\,\Bigl\la\,F_{\su\su,ij}^2\, + \,\rho_{\su\su,ij}^2 \,-\,{1+\eps_u\over 2}\,\bigl(\,F_{\su\sd,ij}^2
  \,+\,\rho_{\su\sd,ij}^2\,\bigr) \,+\,{1-\eps_u\over 2}\,\rho_{\su\sd,ii}\,\rho_{\su\sd,jj}  \;\Bigr\ra \pI
\eeq
This proves part (b) of Theorem 4.4 for $i\ne j$. Finally, let's consider the case 

\medskip
\bigskip
$\bm{i= j\,:}$ 

\bigskip
We write
\beq
\la\, c_{j\su}^+c_{j\sd}^+c_{j\sd}c_{j\su}\,\ra &=& \la\, c_{j\su}^+c_{j\su}\,c_{j\sd}^+c_{j\sd}\,\ra \;\;=\;\;\la\, n_{j\su}\,n_{j\sd}\,\ra 
\eeq
and recall that in terms of Majorana operators 
\beq
n_{j\sigma}\;\;=\;\;c_{j\sigma}^+c_{j\sigma}&=&\ts {1\over 4}\,(\,a_{j\sigma}\,-\,i\,b_{j\sigma}\,) \,(\,a_{j\sigma}\,+\,i\,b_{j\sigma}\,) \;\;=\;\;
  \ts {i\over 2}\,a_{j\sigma}b_{j\sigma}\,+\,{1\over 2} \pS
\eeq
Thus, 
\beq
\lefteqn{
\la\, c_{j\su}^+c_{j\sd}^+c_{j\sd}c_{j\su}\,\ra \;\;=\;\;\ts +\,{1\over 4}\, \la\, a_{j\su}a_{j\sd}b_{j\su}b_{j\sd}\,\ra 
\;+\; {i\over 4}\,\la\, a_{j\su}b_{j\su}\,+\,a_{j\sd}b_{j\sd} \,\ra \;+\; {1\over 4}   } \pS \\ 
&=&\ts -\,{1\over 4}\,\Bigl\la\, G^{aa}_{\su\sd,jj}\,G^{bb}_{\su\sd,jj}\,-\,G^{ab}_{\su\su,jj}\,G^{ab}_{\sd\sd,jj}\,+\,G^{ab}_{\su\sd,jj}\,G^{ab}_{\sd\su,jj}\,\Bigr\ra 
   \;+\;{1\over 4}\,\bigl\la\, G^{ab}_{\su\su,jj}\,+\,G^{ab}_{\sd\sd,jj}\,\bigr\ra \;+\;{1\over 4} \pI
\eeq
In the $w_1=1$ representation, this becomes 
\beq
\la\, c_{j\su}^+c_{j\sd}^+c_{j\sd}c_{j\su}\,\ra
&=&\ts -\,{1\over 4}\,\bigl\la\, 0\,-\,G^{ab}_{\su\su,jj}\,G^{ab}_{\sd\sd,jj}\,+\,0\,\bigr\ra 
   \;+\;{1\over 4}\,\bigl\la\, G^{ab}_{\su\su,jj}\,+\,G^{ab}_{\sd\sd,jj}\,\bigr\ra \;+\;{1\over 4} \pI \\ 
&=&\ts {1\over 4}\,\bigl\la \, \rho_{\su\su,jj}\,\rho_{\sd\sd,jj} \,+\,\rho_{\su\su,jj}\,+\,\rho_{\sd\sd,jj} \,+\, 1\,\bigr\ra \pS \\ 
&=&\ts {1\over 4}\,\bigl\la\;(\, 1\,+\,\rho_{\su\su,jj}\,)(\,1\,+\, \rho_{\sd\sd,jj}\,)\;\bigr\ra \pI
\eeq
and with part (a) of Theorem 4.2, we end up with 
\beq
\la\, c_{j\su}^+c_{j\sd}^+c_{j\sd}c_{j\su}\,\ra&=&\ts {1\over 4}\,\bigl\la\;(\, 1\,+\,\rho_{\su\su,jj}\,)(\,1\,-\,\eps_u\, \rho_{\su\su,jj}\,)\;\bigr\ra \pS
\eeq

\medskip
In the $w_2=1$ representation, we obtain 
\beq
\la\, c_{j\su}^+c_{j\sd}^+c_{j\sd}c_{j\su}\,\ra &=&\ts -\,{1\over 4}\,\Bigl\la\, iF^a_{\su\sd,jj}\,iF^{b}_{\su\sd,jj}\,-\,\rho_{\su\su,jj}\,\rho_{\sd\sd,jj}\,+\,\rho_{\su\sd,jj}\,\rho_{\sd\su,jj}\,\Bigr\ra 
   \;+\;{1\over 4}\,\bigl\la\, \rho_{\su\su,jj}\,+\,\rho_{\sd\sd,jj}\,\bigr\ra \;+\;{1\over 4} \pI
\eeq
With Theorem 4.1, this reduces to 
\beq
\la\, c_{j\su}^+c_{j\sd}^+c_{j\sd}c_{j\su}\,\ra &=&\ts +\,{1\over 4}\,\bigl\la \,1\,-\, \rho_{\su\sd,jj}\,\rho_{\sd\su,jj}\,\bigr\ra \pS 
\eeq
and with part (b) of Theorem 4.2, we end up with 
\beq
\la\, c_{j\su}^+c_{j\sd}^+c_{j\sd}c_{j\su}\,\ra &=&\ts {1\over 4}\,\bigl\la\,1\,-\, \eps_u\,\rho_{\su\sd,jj}^2 \,\bigr\ra \pI 
\eeq
which completes the proof of Theorem 4.4\,. \;\;$\blacksquare$

\newpage

\noindent{\bf\large References} 

{\small

\bigskip
\bigskip
\begin{itemize}
\item[{\rm[1]}] Detlef Lehmann, {\sl The Dynamics of the Hubbard Model through Stochastic Calculus and Girsanov Transformation}, International Journal of Theoretical 
  Physics 63, 139, May 2024.
\item[{\rm[2]}] Ze-Yao Han, Zhou-Quan Wan and Hong Yao, {\sl Pfaffian Quantum Monte Carlo: Solution to Majorana Sign Ambiguity and Applications}, 
\,{\footnotesize \url{https://arxiv.org/abs/2408.10311}}, August 2024. 
\item[{\rm[3]}] J.~P.~F.~LeBlanc, Andrey E.~Antipov, Federico Becca, Ireneusz W.~Bulik, Garnet Kin-Lic Chan, Chia-Min Chung, Youjin Deng, Michel Ferrero, 
  Thomas M.~Henderson, Carlos A.~Jimenez-Hoyos, E.~Kozik, Xuan-Wen Liu, Andrew J.~Millis, N.~V.~Prokof'ev, Mingpu Qin, Gustavo E.~Scuseria, Hao Shi, B.~V.~Svistunov, 
  Luca F.~Tocchio, I.~S.~Tupitsyn, Steven R.~White, \,Shiwei Zhang, \,Bo-Xiao Zheng, \,Zhenyue Zhu \,and \,Emanuel Gull \;(The Simons Collaboration on the 
  Many-Electron Problem), 
  \;{\sl Solutions of the Two Dimensional Hubbard Model: Benchmarks and Results from a Wide Range of Numerical Algorithms}, \;Physical Review X 5, 041041, December 2015. 
\item[{\rm[4]}] Elliott H.~Lieb, Michael Loss and Robert J.~McCann, {\sl Uniform Density Theorem for the Hubbard Model}, Journal of Mathematical Physics, Vol.34, No.3, 
  p.891-898, March 1993.
\item[{\rm[5]}] Shun-Qing Shen, {\sl Strongly Correlated Electron Systems: Spin-Reflection Positivity and Some Rigorous Results}, International Journal of Modern Physics B, 
  Vol.12, Nos.7+8, p.709-779, 1998.
\item[{\rm[6]}] Hal Tasaki, {\sl Physics and Mathematics of Quantum Many-Body Systems}, Graduate Texts in Physics Series, Springer, May 2020. 
\item[{\rm[7]}] S.~L.~Cotter, G.~O.~Roberts, A.~M.~Stuart and D.~White, {\sl MCMC Methods for Functions: Modifying Old Algorithms to Make Them Faster}, 
  Statistical Science, Vol.28, No.3, p.424-446, 2013. 
\item[{\rm[8]}] Martin Hairer, Andrew M.~Stuart and Sebastian J.~Vollmer, {\sl Spectral Gaps for a Metropolis-Hastings Algorithm in Infinite Dimensions}, 
  The Annals of Applied Probability, Vol.24, No.6, p.2455-2490, 2014. 
\item[{\rm[9]}] Jorge E.~Hirsch, {\sl Discrete Hubbard-Stratonovich Transformation for Fermion Lattice Models}, Physical Review B, Vol.28, No.7, p.4059-4061, October 1983. 
\item[{\rm[10]}] Jorge E.~Hirsch, {\sl Two-Dimensional Hubbard Model: Numerical Simulation Study}, Physical Review B, Vol.31, No.7, p.4403-4419, April 1985.
\item[{\rm[11]}] S.~R.~White, D.~J.~Scalapino, R.~L.~Sugar, E.~Y.~Loh, J.~E.~Gubernatis and R.~T.~Scalettar, {\sl Numerical Study of the Two-Dimensional Hubbard Model}, 
  Physical Review B, Vol.40, No.1, p.506-516, July 1989. 
\item[{\rm[12]}] A.~Moreo, D.~J.~Scalapino, R.~L.~Sugar, S.~R.~White and N.~E.~Bickers, {\sl Numerical Study of the Two-Dimensional Hubbard Model for Various Band Fillings}, 
  Physical Review B, Vol.41, No.4, p.2313-2320, February 1990. 
\item[{\rm[13]}] Richard T.~Scalettar, Reinhard M.~Noack and Rajiv R.~P.~Singh, {\sl Ergodicity at Large Couplings with the Determinant Monte Carlo Algorithm}, Physical Review B, 
  Vol.44, No.19, p.10502-10507, November 1991. 
\item[{\rm[14]}] Thereza Paiva, Raimundo R.~dos Santos, Richard T.~Scalettar and P.~J.~H.~Denteneer, {\sl Critical Temperature for the Two-Dimensional Attractive Hubbard 
  Model}, Physical Review B, Vol.69, No.18, p.184501, March 2004. 
\item[{\rm[15]}] E.~Y.~Loh, J.~E.~Gubernatis, R.~T.~Scalettar, S.~R.~White, D.~J.~Scalapino and R.~L.~Sugar, {\sl Numerical Stability and the Sign Problem in the 
  Determinant Quantum Monte Carlo Method}, International Journal of Modern Physics C, Vol.16, No.8, p.1319-1327, 2005. 
\item[{\rm[16]}] Zhaojun Bai, Wenbin Chen, Richard Scalettar and Ichitaro Yamazaki, {\sl Numerical Methods for Quantum Monte Carlo Simulations of the Hubbard Model}, 
  in: {\sl Multi-Scale Phenomena in Complex Fluids}, edited by Thomas Y.~Hou, Chun Liu and Jian-Guo Liu, Series in Contemporary Applied Mathematics CAM 12, Higher Education 
  Press Beijing and World Scientific Publishing Singapore, 2009. 
\item[{\rm[17]}] C.~N.~Varney, C.-R.~Lee, Z.~J.~Bai, S.~Chiesa, M.~Jarrell and Richard T. Scalettar, {\sl Quantum Monte Carlo Study of the Two-Dimensional Fermion 
  Hubbard Model}, Physical Review B, Vol.80, p.075116, August 2009.  
\item[{\rm[18]}] Tong Shen, Hartem Barghathi, Jiangjong Yu, Adrian Del Maestro and Brenda M.~Rubenstein, {\sl Stable Recursive Auxiliary Quantum Monte Carlo Algorithm 
  in the Canonical Ensemble: Applications to Thermometry and the Hubbard Model}, Physical Review E, Vol.107, p.055302, May 2023. 
\item[{\rm[19]}] Yu-Feng Song, Youjin Deng and Yuan-Yao He, {\sl Magnetic, Thermodynamic and Dynamical Properties of the Three-Dimensional Fermionic Hubbard-Model: 
  A Comprehensive Monte Carlo Study}, Physical Review B, Vol.111, p.035123, January 2025. 
\item[{\rm[20]}] Zhuotao Xie, Yu-Feng Song and Yuan-Yao He, {\sl Ising Phase Transitions and Thermodynamics of Correlated Fermions in a Two-Dimensional Spin-Dependent 
  Lattice Potential}, Physical Review B, Vol.111, p.125105, March 2025.

\item[{\rm[21]}] Guang-Shan Tian, {\sl Rigorous Theorems on Off-Diagonal Long-Range Order in the Negative-U Hubbard Model}, Physical Review B, Vol.45, No.6, p.3145-3148, 
  February 1992. 

\item[{\rm[22]}] Tohru Koma and Hal Tasaki, {\sl Decay of Superconducting and Magnetic Correlations in One- and Two-Dimensional Hubbard Models}, Physical Review Letters, 
  Vol.68, No.21, p.3248-3251, May 1992.  

\item[{\rm[23]}] Shun-Qing Shen and Zhao-Ming Qiu, {\sl Exact Demonstration of Off-Diagonal Long-Range Order in the Ground State of a Hubbard Model}, Physical Review Letters, 
  Vol.71, No.25, p.4238- 4240, December 1993. 

\item[{\rm[24]}] Shun-Qing Shen, Zhao-Ming Qiu and Guang-Shan Tian, {\sl Ferrimagnetic Long-Range Order of the Hubbard Model}, Physical Review Letters, Vol.72, No.8, 
  p.1280-1282, February 1994. 

\item[{\rm[25]}] Guang-Shan Tian, {\sl Antiferromagnetic Correlation in the Half-Filled Stronly Correlated Electron Models at Nonzero Temperature: A Rigorous Result}, 
  Physical Review B, Vol.63, p.224413, May 2001.  

\item[{\rm[26]}] Guang-Shan Tian, {\sl Lieb's Spin-Reflection-Positivity Method and Its Applications to Strongly Correlated Electron Systems}, Journal of Statistical 
  Physics, Vol.116, p.629-680, August 2004. 

\item[{\rm[27]}] Andreas Mielke, {\sl The Hubbard Model and its Properties}, in: {\sl Many-Body Physics: From Kondo to Hubbard}, edited by Eva Pavarini, Erik Koch and 
  Piers Coleman, Lecture Notes of the Autumn School on Correlated Electrons 2015, Schriften des Forschungszentrum J\"ulich, Reihe Modeling and Simulation Vol.5, 
  \,{\footnotesize \url{https://www.cond-mat.de/events/correl15/manuscripts/correl15.pdf}}, \,September 2015.

\item[{\rm[28]}] Yukimi Goto, Tohru Koma and Hironobu Yoshida, {\sl Superconductivity and Low Energy Excitations in an Attractive Hubbard Model}, 
  RIKEN-iTHEMS-Report-25, {\footnotesize \url{https://arxiv.org/pdf/2509.19780}}, September 2025.

\end{itemize}

}

\vfill

\end{document}